\documentclass[9pt, amssymb, preprint, superscriptaddress, nobibnotes, nofootinbib, aps]{revtex4-2}

\usepackage{xcolor}

\usepackage{xspace}

\newcommand{\hide}[1]{}
\newcommand{\cro}{Ca$_{3}$Ru$_{2}$O$_{7}$\xspace}
\newcommand{\ticrox}{Ca$_{3}$(Ru$_{1-x}$Ti$_{x}$)$_{2}$O$_{7}$\xspace}
\newcommand{\ticro}{Ca$_{3}$(Ru$_{0.99}$Ti$_{0.01}$)$_{2}$O$_{7}$\xspace}

\newcommand{\ruo}{RuO$_{6}$\xspace}
\newcommand{\de}{$^{\circ}$\xspace}

\newcommand{\Pn}{$Pn2_{1}a$\xspace}
\newcommand{\Bb}{$Bb2_{1}m$\xspace}
\newcommand{\Neel}{N\'eel\xspace} 
\newcommand{\tT}{$\tau \mathcal{T}$\xspace}
\newcommand{\FF}{$|F|^{2}$\xspace}

\usepackage{amsmath}
\usepackage{physics}
\usepackage{amssymb}
\usepackage{mathtools}
\mathtoolsset{showonlyrefs}                
\usepackage{booktabs}                      
\usepackage{url}                         
\usepackage{caption}                     
\usepackage{comment}                     
\usepackage{multirow}
\usepackage{graphicx}
\usepackage[hidelinks]{hyperref}         
\usepackage{subcaption}
\usepackage{wrapfig}

\usepackage[hidelinks]{hyperref}         
\usepackage{bookmark}                  

\usepackage{titletoc}

\begin{document}

\title{Evaluating the Structural Basis for Polar Altermagnet Candidate Ca$_{3}$(Ru,Ti)$_{2}$O$_{7}$}

\author{Akash Saha}
\thanks{This author contributed equally to this work}
\affiliation{Department of Materials Science and Engineering, The Pennsylvania State University, USA}

\author{Yihuang Xiong}
\thanks{This author contributed equally to this work}
\altaffiliation[Present address: ]{Research Institute of Superconductor Electronics, Nanjing University, China}
\affiliation{Thayer School of Engineering, Dartmouth College, USA}

\author{Vladimir A. Stoica} 
\affiliation{Department of Materials Science and Engineering, The Pennsylvania State University, USA}

\author{Subin Mali} 
\affiliation{Department of Physics, The Pennsylvania State University, USA}

\author{Aaron Pearre} 
\affiliation{Department of Physics, The Pennsylvania State University, USA}

\author{Saugata Sarker}
\affiliation{Department of Materials Science and Engineering, The Pennsylvania State University, USA}

\author{Huaiyu Wang}
\affiliation{Stanford Institute for Materials and Energy Sciences, SLAC National Accelerator Laboratory, USA}

\author{Yufei Zhao} 
\affiliation{Department of Physics, The Pennsylvania State University, USA}

\author{Evguenia Karapetrova} 
\affiliation{Advanced Photon Source, Argonne National Lab, USA}

\author{Yu Wang}
\affiliation{Department of Physics, The Pennsylvania State University, USA}

\author{Jadupati Nag} 
\affiliation{Department of Materials Science and Engineering, The Pennsylvania State University, USA}

\author{Zachary W. Hazenstab} 
\affiliation{Department of Physics, The Pennsylvania State University, USA}

\author{Seng Huat Lee} 
\affiliation{Department of Physics, The Pennsylvania State University, USA}

\author{Long-Qing Chen}
\affiliation{Department of Materials Science and Engineering, The Pennsylvania State University, USA}

\author{Geoffroy Hautier}
\affiliation{Thayer School of Engineering, Dartmouth College, USA}

\author{Binghai Yan}
\affiliation{Department of Physics, The Pennsylvania State University, USA}

\author{Zhiqiang Mao}
\affiliation{Department of Physics, The Pennsylvania State University, USA}

\author{Venkatraman Gopalan}
\thanks{Corresponding author: vgopalan@psu.edu}
\affiliation{Department of Materials Science and Engineering, The Pennsylvania State University, USA}
\affiliation{Department of Physics, The Pennsylvania State University, USA}

\begin{abstract}

   The interplay between polar and altermagnetic orders remains largely unexplored in the broader landscape of correlated electron systems. Density functional theory (DFT) within the PBEsol + $U$ treatment propose Ca$_{3}$Ru$_{2}$O$_{7}$ as a polar altermagnet, reliant on the transformation of the experimentally reported $Bb2_{1}m$ phase to a lower symmetry $Pn2_{1}a$ structure. Here, we perform a targeted search for the $Pn2_{1}a$ phase using synchrotron X-ray diffraction on single crystals of Ca$_{3}$Ru$_{2}$O$_{7}$ and Ca$_{3}$(Ru$_{0.99}$Ti$_{0.01}$)$_{2}$O$_{7}$. No diffraction signature of the $Pn2_{1}a$ structure is detected down to 20 K within experimental limits. Through diffraction simulations, we translate this result into a quantitative upper bound on the symmetry breaking distortion amplitude, which we find to be significantly smaller than the DFT predicted distortion. Our DFT calculations further motivate dilute Ti substitution ($\lesssim 2\%$) as an alternative route to stabilize the low symmetry phase, for which we again find no clear evidence for this putative \Pn structure that hosts altermagnetism. Our findings indicate that PBEsol + $U$ over-stabilizes a \Pn distortion, and the measured bounds provide benchmark for future calculations. Ti substitution exceeding $\sim$3\% reveals a chemically tunable polar altermagnetic phase within the $Bb2_{1}m$ structure and we theoretically demonstrate that the spin-splitting can be tuned by Ti substitution. By combining DFT, diffraction simulations and synchrotron X-ray measurements, our study demonstrates a general methodology to quantify subtle structural distortions that underlie emergent electronic and magnetic phases. 

\end{abstract}

\maketitle

\clearpage

\section{Introduction}

Traditionally, magnetic materials have been broadly categorized into ferromagnets and antiferromagnets. These two magnetic classes were thought to be mutually exclusive, until the recent advent of altermagnetism as a new class of magnetic ordering \cite{alm1, alm2, alm3, alm4, alm5}. Similar to antiferromagnets, they possess a collinear spin structure without any net magnetic moment, while the momentum space exhibits non-degenerate spin polarized bands akin to a ferromagnet. These characteristics make them useful for exploring unconventional transport phenomena and spintronic functionality while remaining robust to stray magnetic fields. The interplay between the magnetic structure and the crystallographic symmetries is fundamental to altermagnetic classification \cite{alm2, almcrystal, alm6, alm7}. Collinear compensated magnets for which a fractional lattice translation $\tau$ or inversion $\mathcal{P}$ symmetry operation relates the two opposite spin sub-lattices enforces spin degeneracy of the electronic bands resulting in a conventional antiferromagnet. In the absence of time reversal symmetry $\mathcal{T}$ combined with fractional translation or inversion symmetry operations ($\tau\mathcal{T}$ or $\mathcal{P}\mathcal{T}$ respectively), the resulting magneto-crystallographic order is defined as an altermagnet and will display a spin-split band structure \cite{nrss1, nrss2, nrss3, nrss4, nrss5, nrss6, nrss7, nrss8}. Importantly, the spin splitting is non-relativistic in origin and momentum dependent, directly related to the crystallographic symmetries. This symmetry based classification provides the basis for the verification of altermagnetic ordering in any material with known magnetic and crystal structures \cite{ssg1, ssg2, ssg3}. \\

In strongly correlated materials where altermagnetism may coexist with other ferroic order parameters, such as a polar distortion or spontaneous strain, the synergy between the crystal lattice and the magnetic degrees of freedom become even richer, enabling tunable altermagnetic functionality such as the altermagnetoelectric \cite{ame1, ame2, ame3, ame4, ame5, ame6} and altermagnetoelastic effects \cite{afe1, afe2, afe3}, where the spin-splitting in momentum space can be switched by external controls such as electric fields or mechanical strain respectively, without requiring a reorientation of the \Neel vector. This deterministic control and non-volatile switching are promising for spintronics applications and such altermagnetic switching pathways have been theoretically demonstrated in Ca$_{3}$Mn$_{2}$O$_{7}$ for example \cite{ame1, ame2}. The closely related bilayer Ruddlesden-Popper ruthenate \cro is a rich system with multiple coexisting ferroic orders such as antiferromagnetism and ferroelasticity which stabilizes in a polar lattice. While ferroelastic domain switching has been experimentally demonstrated, the reversal of the polar order is challenging due to the presence of free carriers in the system, rendering electric field biasing ineffective \cite{lei18}. Substituting the Ru site with Ti opens an electronic bandgap \cite{ke11, peng13} that makes it insulating and could allow for such electrical biasing. The coexistence and coupling of the structural, electronic and magnetic order parameters in this system provide multiple pathways to control the underlying symmetry landscape, making Ca$_{3}$(Ru,Ti)$_{2}$O$_{7}$ a promising platform for tunable polar altermagnetic functionality. \\

In this work, we reexamine the crystal structure of single crystals of \cro using temperature dependent synchrotron X-ray diffraction in light of recent density functional theory (DFT) calculations \cite{puggioni20, zhao25} and nonlinear transport measurements \cite{mali26} that indicate a low temperature symmetry lowering that places the ground state of \cro in a polar altermagnetic phase. This predicted structural phase has remained elusive in prior diffraction experiments, and to date, no study has systematically quantified its absence. In addition, our DFT calculations motivate dilute Ti substitution ($\lesssim$ 2\%) as an alternative route to stabilize this low symmetry phase. For Ti concentrations exceeding $\sim$3\%, a different route towards polar altermagnetism emerges within the \Bb structure enabled through a change in the magnetic order to a G-type antiferromagnet. Compared to the parent compound, this phase realizes a slightly larger spin splitting ($\sim$ few meV), and our DFT calculations demonstrate that the spin splitting can be tuned by Ti substitution while being insulating and ferroelastic.

\section{Methods}

\subsection*{Sample Preparation}

Single crystals of \cro and \ticro were grown using the floating zone method. The Ti concentration for \ticro was confirmed using Energy Dispersive X-ray Spectroscopy. The absence of secondary Ruddlesden-Popper phases was validated by laboratory based X-ray $\theta$-$2\theta$ scans. Temperature dependent linear transport and SQUID magnetometry measurements were performed to verify the electronic and magnetic transition temperatures. Refer to Supplementary Note \ref{sample_char} for details on sample characterization \cite{suppm}. All the measurements for this study were performed on predominantly single domain crystals measuring roughly $\sim$2 mm $\times \sim$2 mm $\times \sim$0.5 mm.  

\subsection*{Synchrotron X-ray diffraction measurements}

Experiments were performed at the Sector 33 bending magnet beamline of the Advanced Photon Source at Argonne National Laboratory. The incident X-ray energy was 15.5 keV (wavelength 0.799 \AA) essentially probing the bulk of the crystal at typical incidence angles. Beam defining slits were adjusted to ensure that the X-ray footprint remained within the sample dimensions, minimizing beam overspill and associated background scattering. All the reported intensities in this study are presented in normalized units, correcting for incident beam intensity fluctuations and attenuation filters. The samples were mounted on a copper rod using silver epoxy and were oriented such that the sample surface was the (001) plane with the [100] crystal direction lying along the incident beam. The sample space was enclosed by a hemispherical beryllium dome and cooling was achieved using an Advanced Research Systems Displex cryostat capable of reaching a base temperature of 20 K. A Pilatus 100K detector positioned 1.253 metres from the sample was used to collect the scattering patterns. The three dimensional reciprocal space was mapped by rocking the sample on a Huber six circle diffractometer, and rsMap3D \cite{rsmap3d} was used for reconstruction. 

\subsection*{Density Functional Theory calculations}

All first-principles calculations were performed using the Vienna ab initio Simulation Package (VASP) with the projector augmented-wave (PAW) method \cite{dft1, dft2}. Spin-orbit coupling was included throughout unless otherwise stated. The on-site Coulomb interaction was applied to the Ru $4d$ orbital manifold within Dudarev’s DFT+$U$ formalism \cite{dft3}. The application was to mitigate self-interaction errors associated with localized $d$ states and to effectively simulate the gap opening effect due to Ti substitution. The lattice parameters were fixed to the previously reported experimental structure with space group \Bb \cite{yoshida05}. Bi-layer AFM ordering and \Neel-type AFM ordering were imposed for the AFM-$b$ magnetic phase, with the easy axis along [010], and the G-AFM phase, respectively. The PBEsol exchange-correlation functional was employed to describe the structural and electronic properties \cite{puggioni20}. Structural optimizations were carried out using a plane-wave cutoff energy of 520 eV. The Brillouin zone was sampled with a $\Gamma$-centered 10 $\times$ 10 $\times$ 6 $\vb{k}$-point mesh for static calculations, while a reduced 5 $\times$ 5 $\times$ 3 mesh was used for structural relaxations. A Gaussian smearing width of 20 meV was applied. The total energy and atomic forces were converged to within $10^{-7}$ eV and 0.1 meV \AA$^{-1}$, respectively.

\section{Results and Discussion}

\subsection{Prediction of structural symmetry lowering in \cro}

 \cro crystallizes in a polar orthorhombic lattice under the point group $m2m$ (Schoenflies notation $C_{2v}$ with the $C_{2}$ axis along the $b$ axis). Reported diffraction studies have assigned it to the space group \Bb (space group 36, $a<b<c$ setting), which remains unchanged over a temperature range from 8 K to 300 K and under applied magnetic fields of up to 5 T \cite{yoshida05, petkov23}. At room temperature, \cro is a paramagnetic metal, which then transitions into an antiferromagnet (AFM) below $T_{N} = 56$ K with the spins on the Ru sites pointing along the crystallographic $a$ axis (AFM-$a$ magnetic structure). On further cooling below $T_{S} = 48$ K, the Fermi surface becomes partially gapped resulting in a pseudogap phase \cite{yoshida04}. Concurrent with the presence of the pseudogap, the easy axis of the spins changes from the $a$ to the $b$ axis, resulting in the AFM-$b$ magnetic structure. Both the AFM-$a$ and AFM-$b$ magnetic structures show A-type antiferromagnetic ordering where the spins are ferromagnetically ordered within the \ruo octahedral bilayer, while neighboring bilayers are coupled antiferromagnetically along the $c$ axis. Crystallographic investigations have found \cro to be isostructural across the various electronic and magnetic phases induced through temperature, while the lattice constants show an abrupt change at the spin-reorientation transition at 48 K \cite{yoshida05, nelson07}. \\ 

\begin{figure*}[!htb]
    \centering
    \includegraphics[scale=0.95]{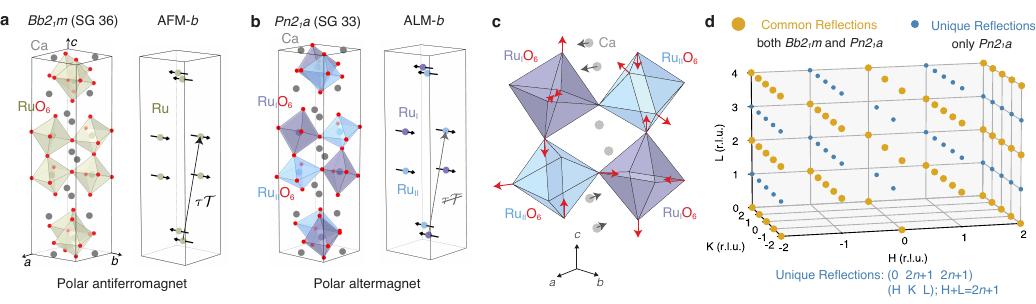}
    \caption{\textbf{Candidate crystal structures of \cro and their associated diffraction signatures.} (a) Crystal structure in the \Bb space group. All the \ruo octahedra are equivalent due to the $\tau = B$ lattice centering operation. This translation combined with the time reversal operator $\tau \mathcal{T}$ for the AFM-$b$ magnetic structure makes it a conventional antiferromagnet. (b) Proposed \Pn space group where the broken translation symmetry allows neighboring octahedra to have different local environments. This crystal structure combined with the AFM-$b$ magnetic ordering makes it an altermagnet ALM-$b$. (c) Visualization of the atomic displacements that modify the \ruo octahedra thereby driving the \Pn structural transition. (d) Overview of the reciprocal space for the two crystal structures. The different systematic absences allows the differentiation of the two crystal structures.}
    \label{fig1}
\end{figure*}

Our DFT calculations performed with on-site Coulomb repulsion $U$ applied in the Dudarev scheme, spin-orbit interactions, and using the PBEsol functional predict a structural transition with increasing Hubbard $U$ values (detailed later in section \ref{dftsection}). This drives the system from space group \Bb ($U = 1.2$ eV) to \Pn (space group 33, $a<b<c$ setting) for $U > 1.2$ eV, consistent with several other DFT reports on this material \cite{puggioni20, zhao25, lda}. While all the \ruo octahedra in the unit cell are equivalent in the \Bb structure (Fig. \ref{fig1}a), the \Pn phase breaks this equivalence and stabilizes a structure characterized by the spatial variation of the \ruo octahedral sizes as shown in Fig. \ref{fig1}b. This structural distortion from \Bb is mainly driven by atomic displacements of the oxygen and calcium atoms as illustrated in Fig. \ref{fig1}c\ (refer to Supplementary Material Figs. \ref{proj}-\ref{RuO} for the projected displacements \cite{suppm}). A symmetry mode decomposition of this distortion using AMPLIMODES \cite{amplimodes} shows that it comprises of a zone boundary $Y_{2}$ mode coupled to a zone center $\Gamma_{1}$ mode, with the $Y_{2}$ mode acting as the primary symmetry breaking order parameter (details in Supplementary Note \ref{samdist} \cite{suppm}). The distortion results in an alternating pattern of the \ruo octahedral deformations where the apical oxygen atoms move towards the ruthenium center in one octahedron and away from it in the neighboring octahedron, producing a staggered compression -- elongation of the Ru-O$_{\text{apical}}$ bonds. Concurrently, the in-plane Ru-O bonds exhibit a similar staggering, with the neighboring octahedra undergoing Ru-O$_{\text{in-plane}}$ shortening and elongation. This produces a cooperative checkerboard-like modulation of the \ruo octahedral volumes, which is accommodated by corresponding displacements of the Ca atoms. Overall, this leads to a loss of the fractional translation symmetry $\tau = B : (x,y,z) \rightarrow (x+\frac{1}{2}, y, z+\frac{1}{2})$ operation that reduces the symmetry of the crystal to \Pn. This structural transition renders \cro a rare polar altermagnet for the same AFM-$b$ magnetic structure \cite{zhao25} (see Supplementary Fig. \ref{spingroup} for an illustration of the spin group operations \cite{suppm}). \\ 

Puggioni et al. interpret the variation of Hubbard $U$ as a proxy for temperature, arising from narrowing of the electronic bandwidth on cooling, which effectively enhances $U$ \cite{puggioni20}. Within this framework, larger $U$ values corresponds to lower temperatures, leading to the proposal that the low temperature crystal structure of \cro is \Pn instead of the experimentally observed \Bb. They further suggest that the transition from the high temperature \Bb to low temperature \Pn phase occurs at $\sim$30 K. Recent transport calculations and measurements identify the emergence of a nonlinear response at low temperatures that is symmetry forbidden for the \Bb phase while being allowed for \Pn, suggesting the \Pn structure as the ground state for \cro \cite{mali26, zhao25}. While the signatures for the low symmetry \Pn phase observed in nonlinear transport measurements are compelling, it is important to validate them with other complementary probes. The anomalous Hall effect, for example, is a hallmark of time-reversal symmetry breaking and has been demonstrated for several altermagnets \cite{ahe1, ahe2, ahe3}. However, the crystal symmetries of both \Pn and \Bb forbid an anomalous Hall response \cite{mali26}.  Moreover, DFT calculations predict a non-relativistic spin splitting of $\sim$ 0.1 meV \cite{zhao25}, which is too small to be experimentally detected using angle resolved photoemission spectroscopy (ARPES). Notably, this low symmetry phase has not been observed in reported X-ray and neutron diffraction studies \cite{yoshida05, resonantxray, rashba24, dashwood23, nscat}. Given the symmetry restrictions and limited sensitivity to detect this putative \Pn structure, a targeted crystallographic search seeking direct structural evidence, supplemented with quantitative bound estimates on the symmetry breaking distortions, is warranted.  

\subsection{Survey for \Pn space group in \cro}

The two candidate space groups \Bb and \Pn belong to the same polar point group $m2m$ ($C_{2v}$). Consequently, they possess identical symmetry allowed forms for the property tensors, making them indistinguishable through probes sensitive to the point group symmetry such as pyroelectric measurements or second harmonic generation. The presence of the fractional lattice translation $B$ differentiates between these two space groups (see Supplementary Fig. \ref{symm} for an analysis of the symmetry operations \cite{suppm}), making X-ray diffraction a suitable probe for distinguishing between the candidate phases through their unique selection rules (also see Supplementary Note \ref{shgsec} \cite{suppm}). The reflections permitted for the higher symmetry \Bb structure are a subset of those allowed for symmetry lowered \Pn as shown in Table \ref{tab1}. Hence, measuring reflections unique to the \Pn phase would provide direct structural evidence for its presence. These reflections appear as half-order satellite peaks between the primary reflections allowed for the \Bb phase (also allowed in \Pn) as shown in Fig. \ref{fig1}d. We index all the reflections in this article based on the conventional unit cell of the \Bb structure in the $a<b<c$ orthorhombic setting. We note that symmetry breaking X-ray reflections are insensitive to atomic displacements orthogonal to the scattering vector. As shown in Table \ref{tab1}, the reflections unique to \Pn fall into three types: $0 K L$, $H 0 L$ and $H K L$ with $H+L=$odd. The first two reflection types are not sensitive to the symmetry breaking atomic displacements along the crystallographic $a$ and $b$ directions respectively. On the other hand, the reflection $HKL$ with $H+L=$odd and $K \neq 0$ is sensitive to any general symmetry breaking atomic displacements. Based on the DFT predictions shown in Supplementary Figs. \ref{proj} and \ref{dist_xyz} \cite{suppm}, the symmetry-breaking distortion has finite components along all three crystallographic axes, making these better candidate reflections to probe the proposed distortion. Based on nonlinear transport measurements, the structural transition from the \Bb to \Pn is anticipated at 48 K \cite{mali26}. Given that prior diffraction studies were unable to resolve this phase, it is likely that the associated symmetry breaking distortions are subtle. To overcome this limited sensitivity, we employ high flux synchrotron X-rays to probe the three dimensional reciprocal space around the reflections of interest. \\

\begin{table*}[!htb]
\begin{center}
\begin{tabular}{|c|c|c|c|c|}
\hline
Reflection & \Bb           & \Pn           & Reflections unique to \Pn & Examples              \\
\hline
H K L & H + L = even   & all allowed  & H + L = odd        & 1 1 8,\ 2 2 7 \\
0 K L & K, L = even  & K + L = even & K, L = odd         & 0 1 1,\ 0 1 3 \\
H 0 L & H + L = even & all allowed     & H + L = odd  & 1 0 2,\ 2 0 5 \\
H K 0 & H = even     & H = even  & none               & -                     \\
H 0 0 & H = even     & H = even     & none               & -                     \\
0 K 0 & K = even     & K = even     & none               & -                     \\
0 0 L & L = even     & L = even     & none               & -                    \\
\hline
\end{tabular}
\caption{\textbf{Conditions for allowed reflections for the two space groups.} All peaks are indexed in the $a<b<c$ setting of the conventional unit cell of the \Bb structure. The reflections permitted for \Pn include all the reflections allowed for \Bb and also contain additional reflections unique only to \Pn.}
\label{tab1}
\end{center}
\end{table*}

\begin{figure}[!htb]
    \centering
    \includegraphics[scale=1.8]{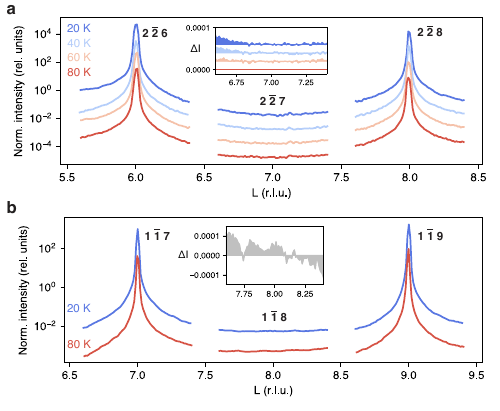}
    \caption{\textbf{Absence of diffraction signatures associated with \Pn structure in \cro.} (a) Integrated peak profiles of consecutive 2 $\bar{2}$ $L$ reflections between 20 K and 80 K. The 2 $\bar{2}$ 6 and 2 $\bar{2}$ 8 reflections are allowed for both the candidate structures while the 2 $\bar{2}$ 7 reflection is forbidden for the \Bb space group while being allowed for \Pn structure. No peak is observed at 2 $\bar{2}$ 7 within the detection limit. The data at different temperatures are shifted for visual clarity. Inset shows the differential intensities relative to the 80 K data near the 2 $\bar{2}$ 7 reflection. (b) Complementary set of 1 $\bar{1}$ $L$ reflections at 20 K and 80 K. No intensity is detected at 1 $\bar{1}$ 8 peak above the noise floor. Inset shows the difference between the 20 K and 80 K dataset near the 1 $\bar{1}$ 8 reflection condition.}
    \label{fig2}
\end{figure}

At 20 K, the lattice parameters of \cro were determined to be $a = 5.370(4)$ \AA, $b = 5.523(2)$ \AA, and $c = 19.535(1)$ \AA, in good agreement with the values obtained from earlier studies \cite{yoshida05}. The small deviations may be attributed to the orientation accuracy as well as the nature of the sample used, namely single crystals in the present work versus powder measurements in Ref. \cite{yoshida05}. Tracking the temperature dependence of the $c$-axis lattice constant revealed a sharp decrease at the spin reorientation transition at 48 K (Supplementary Material Fig. \ref{clattice} \cite{suppm}), consistent with prior reports \cite{nelson07, yoshida05}. The extra peaks due to the \Pn structure are expected to appear below this temperature. However, no signs of additional peaks appearing below 48 K were detected down to a base temperature of 20 K. Fig. \ref{fig2}a shows the $L$ scans for consecutive 2 $\bar{2}$ $L$ reflections at four representative temperatures between 20 K and 80 K. The 2 $\bar{2}$ 6 and 2 $\bar{2}$ 8 reflections are allowed for both the \Bb as well as the \Pn structures, while a peak at 2 $\bar{2}$ 7 is expected only for the low temperature \Pn phase. No peak corresponding to the 2 $\bar{2}$ 7 reflection was observed above the experimental background at all temperatures. To improve sensitivity for detecting a possibly diffuse \Pn signal, we examined the differential intensities relative to the 80 K null data, but found no credible evidence of any feature above the noise floor. Fig. \ref{fig2}b presents another related set of reflections, where the 1 $\bar{1}$ 8 reflection which is expected for the \Pn structure shows no detectable intensity. Several additional \Pn allowed reflections measured at 20 K are provided in the Supplementary Fig. \ref{11L_all} \cite{suppm}. \\

For the measured reflections, we can quantify the ``sensitivity'' by the ratio $Q_{\text{out-of-plane}}/Q_{\text{in-plane}} = L c^{*}/\sqrt{(H a^{*})^{2} + (K b^{*})^{2}}$ where $a^{*}, b^{*}, c^{*}$ represent the magnitude of the reciprocal lattice vectors. This sensitivity ranges from $\sim0.6$ for the 2 2 6 reflection to $\sim1.8$ for 1 1 9, demonstrating that we do sample a range of reflections with comparable sensitivity to both in-plane and out-of-plane components. Furthermore, our high flux synchrotron X-ray measurements revealed a small fraction of residual twin domains ($\approx 0.3$\%) that were not detected by lab based X-ray Laue and polarized light microscopy (Supplementary Note \ref{sample_char} \cite{suppm}). Due to the similar in-plane lattice constants $a \approx b$, reflections from the twin may nearly coincide with that of the primary domain. We hence limit our analysis to reflections satisfying $|H| = |K|$ as for these reflections, the twin domain contribution satisfies the same selection rules as the primary domain, so they are both either simultaneously allowed or simultaneously forbidden, eliminating ambiguities arising from the twinning (refer to Supplementary Note \ref{multidomain} for a detailed explanation \cite{suppm}). Across the multiple measurements, no evidence for the proposed \Pn phase is observed. 

\subsection{Constraints on symmetry breaking structural distortions}

Although no clear evidence for a low temperature \Pn phase in \cro is observed, our measurements can be used to place an upper bound on the magnitude of the structural distortion. We use the $3 \sigma$ fluctuations in the measured background as the detection threshold for a Bragg peak, where $\sigma$ represents the standard deviation of the background counts estimated from a region in reciprocal space around the reflection of interest. Any signal below this level is indistinguishable from noise. This threshold directly translates to an upper bound on the integrated intensity $I_{HKL}$, which is linearly related to the square of the structure factor $F_{HKL}$, that is $I_{HKL} \propto |F_{HKL}|^{2}$. The structure factor squared, rather than the raw integrated intensity, is the appropriate quantity for comparison across various reflections since it represents the intrinsic scattering efficiency of a given reflection independent of the experimental geometry and thus enables quantitative bound estimation on the underlying symmetry breaking distortions. Details on  extracting the integrated intensities and its consequent conversion to structure factor squared are discussed in the appendix. \\

To determine the upper limit for the structural distortion that would produce a Bragg peak at the $3 \sigma$ detection level, we calculate the structure factor squared as a function of the amplitude of the \Pn distortion. For this, we consider the relaxed crystal structures obtained from DFT corresponding to the \Bb ($U = 1.2$ eV) and \Pn ($U > 1.2$ eV) space groups, and calculate the displacement vectors between the equivalent atomic positions in the two structures. The resulting displacements are then linearly scaled by a factor $\lambda$, and applied to the \Bb structure, thereby generating a family of distorted configurations that all belong to the \Pn space group, but span a range of distortion amplitudes characterized by $\lambda$. Justification of a linear scaling using a single scalar parameter $\lambda$ as well as a more rigorous symmetry based distortion mode analysis is presented in Supplementary Note \ref{samdist} \cite{suppm}. Concretely, if $\vb{r}_{B}$ ($\vb{r}_{P}$) represents the atomic position of a particular atom in the \Bb (\Pn) unit cell, then the atomic position in the distorted unit cell is given by $\vb{r}_{\lambda} = \vb{r}_{B} + \lambda (\vb{r}_{P} - \vb{r}_{B})$. In this framework, $\lambda=0$ ($\lambda=1$) corresponds to the relaxed \Bb (\Pn) structure obtained from DFT. The structures for other $\lambda$ values are linearly interpolated/extrapolated between these limits. Based on comparison of the DFT derived band structures to experimental ARPES data, the electronic structure of \cro at 16 K is seemingly well described by $U = 1.6$ eV \cite{puggioni20}. Accordingly, the relaxed structure obtained at $U = 1.6$ eV is used as a representative \Pn structure at 20 K for the calculation of the distortion amplitude from the \Bb space group. The square of the structure factors for the resulting distorted configurations are then calculated using \textsc{xrayutilitites} \cite{xu} at an X-ray energy of 15.5 keV with lattice constants fixed to those determined from our experiments at 20 K. \\

\begin{figure}[]
    \centering
    \includegraphics[scale=1.5]{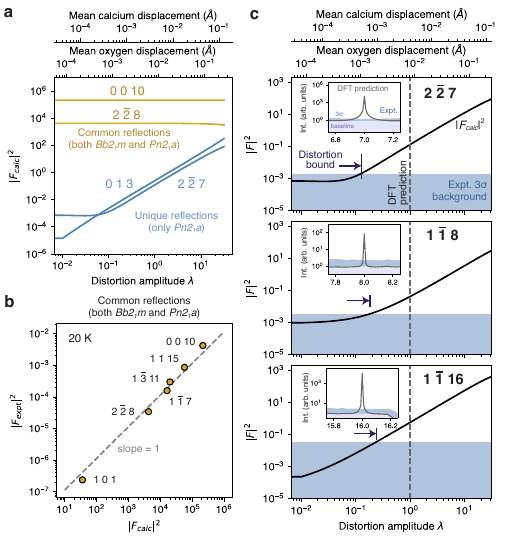}
    \caption{\textbf{Estimation of an upper bound associated with the \Pn distortion from diffraction data.} (a) Calculated square of structure factors of selected reflections obtained by scaling the amplitude $\lambda$ of the \Pn distortion. The reflections unique to the \Pn structure show an increasing structure factor squared with the distortion amplitude while the common reflections show negligible change. (b) Comparison of the experimental \FF at 20 K to the model calculated values for reflections common to both the structures. The dashed line denotes the line of best fit for calibrating the experimental \FF where the slope is fixed to unity. (c) Experimentally derived constraints on the distortion. The solid black line represents the calculated \FF as a function of distortion amplitude while the shaded region denotes the experimentally obtained value based on the $3 \sigma$ background level at the various reflection conditions, placing an upper bound on the \Pn distortion denoted by the arrow. The DFT predicted distortion magnitude for the relaxed \Pn structure denoted by the vertical dashed line is larger than the estimate obtained from experiments. Insets show the experimental $L$ scan at the respective reflection conditions where only the background is observed. The $3 \sigma$ estimate is also indicated. The gray peak represents the expected scan profile if the distortion predicted by DFT was experimentally realized.}
    \label{fig3}
\end{figure}

The variation of the calculated square of the structure factor \FF for a few different reflections as a function of the distortion amplitude $\lambda$ is shown in Fig. \ref{fig3}a. The reflections common to both \Bb and \Pn are largely insensitive to the magnitude of the structural distortion. In contrast, the reflections unique to \Pn are forbidden for the \Bb structure ($\lambda = 0$), and acquire a finite intensity with \FF increasing with the scaling factor $\lambda$. Hence, probing the unique reflections allow us to place experimentally derived constraints on the distortion amplitude characterizing the \Pn structure. To assess the agreement between the calculated \FF to the experimental data taken at 20 K, we examine their scaling for the reflections common to both the structural phases, since the intensities for these reflections do not depend strongly on the distortion. Fig. \ref{fig3}b shows excellent agreement between the experimental and calculated values of the structure factors squared for these reflections. Since we expect $|F_{\text{expt}}|^{2} \propto |F_{\text{calc}}|^{2}$, we fit a line with unit slope to determine the proportionality constant. This calibration then enables quantitative comparison for the weaker reflections unique to \Pn. \\

The integrated intensity threshold for detecting a Bragg peak unique to the \Pn structure is related to the $3 \sigma$ background level at the reflection of interest. Using the established calibration, this limit can be re-expressed in terms of the structure factor squared, enabling a quantitative estimate of the atomic displacements as shown in Fig. \ref{fig3}c. For instance, the $3 \sigma$ background level for the 2 $\bar{2}$ 7 reflection corresponds to $\lambda \approx 0.12$. This upper bound is much smaller than $\lambda=1$, which corresponds to the distortion obtained from the relaxed \Pn structure predicted by DFT. Further, the experimentally derived upper bound on the structure factor squared for the 2 $\bar{2}$ 7 reflection is around two orders of magnitude lower than the value calculated for the DFT predicted structure. In other words, if the distortion amplitude predicted by DFT was experimentally realized, then we should have observed a Bragg peak corresponding to the \Pn phase at 2 $\bar{2}$ 7 with an integrated intensity at least two orders higher than the baseline (insets in Fig. \ref{fig3}c). However this peak was not experimentally observed, thereby restricting the amplitude of the \Pn distortion to $\lambda \lesssim 0.12$. Similar analysis for other reflections unique to \Pn structure, as shown in Fig. \ref{fig3}c, lead to the same conclusion that the \Pn structure predicted by DFT should lead to an observable Bragg peak much higher than the $3 \sigma$ background uncertainty. Provided such peaks are absent in our experiment for all cases, we conclude that DFT over-estimates the amplitude of the \Pn distortion. Our experiments therefore place a strong upper bound of $\lambda \leq 0.22$ (averaged over several reflections) on the symmetry breaking structural distortions driving the transition from the \Bb to the \Pn structure. This corresponds to mean oxygen and calcium distortions of 1.8 $\times 10^{-3}$ \AA \ and 8.9 $\times 10^{-4}$ \AA \ respectively. It is to be noted that oxygen is a light element and interacts weakly with X-rays. However, the oxygen displacements are accompanied by displacements of the calcium atoms to accommodate the \ruo octahedral changes, and the change in the structure factor is dominated by these calcium displacements as demonstrated in Supplementary Note \ref{F2Ca} \cite{suppm}. Expressed in terms of the primary order parameter $Y_{2}$ (see Supplementary Note \ref{samdist} for details \cite{suppm}), these $\lambda$ bounds translate to a mode amplitude equal to 1.3 $\times 10^{-2}$ \AA, while the most restrictive single reflection bound (for 2 $\bar{2}$ 7) corresponds to a $Y_{2}$ amplitude of 8 $\times 10^{-3}$ \AA. These experimental constraints provide a quantitative benchmark for future DFT calculations on this material.

\subsection{Potential alternative to stabilize putative \Pn structure through Ti substitution} \label{dftsection}

Our experimental results from the previous section highlighted that the amplitude of the symmetry breaking distortion responsible for polar altermagnetism in \cro is significantly smaller than that predicted by the theoretical calculations. The evolution of the ground states with increasing Hubbard $U$ was interpreted in terms of a temperature induced structural transition \cite{puggioni20}. This naturally raises the question if alternate external tuning parameters could stabilize this low symmetry \Pn phase. Given that the \Pn structure emerges for higher on-site Coulomb repulsion strengths, a plausible route is to explore perturbations that enhance the electronic correlations. One such route is offered by isovalent chemical substitution using Ti \cite{ke11, peng13}. With dilute Ti substitution at the Ru site, the electronic ground state changes from the quasi two-dimensional metal to a Mott insulator with as little as 0.3\% Ti doping. The magnetic ground state also evolves from the AFM-$b$ magnetic structure in the parent compound to a G-AFM magnetic ordering as the Ti doping concentration is above 3\% \cite{peng13}. The G-AFM magnetic ordering will be discussed in detail later. A summary of the ground states with Ti doping is shown in Fig. \ref{fig4}a, with more details in the Supplementary Note \ref{Tiphasediagram} \cite{suppm}. Throughout this section, Hubbard $U$ is used purely as a phenomenological proxy for the electronic localization induced by Ti substitution, rather than an explicit microscopic simulation of individual Ti impurities in the lattice. \\

\begin{figure}[!htb]
    \centering
    \includegraphics[scale=1.8]{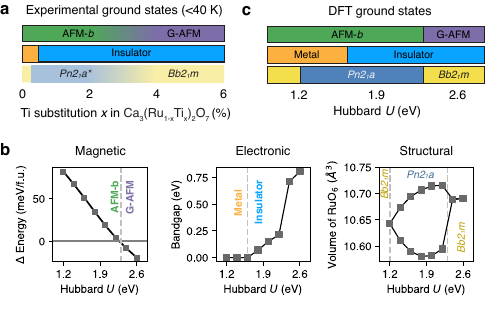}
    \caption{\textbf{Ti substitution as a proxy for Hubbard $U$.} (a) Experimentally observed phases in \cro with Ti substitution. The different colors represent the different phases. The color gradient for the magnetic phase represents a phase coexistence of AFM-$b$ and G-AFM structures between 2\% and 4\% Ti concentration. (b) DFT+$U$ calculations showing the effect of Hubbard $U$ on the formation energies of the AFM-$b$ and G-AFM magnetic structures, the electronic bandgap and the volume of the \ruo octahedra. Spin-orbit interactions have been included for these calculations. On increasing Hubbard $U$, we encounter a metal-insulator transition followed by a magnetic transition at higher $U$ values. The calculations predicts an intermediate \Pn structure, reflected by the two distinct \ruo volumes in this phase. (c) Summary of the DFT predicted magnetic, electronic and structural ground states. The experimentally observed low temperature electronic and magnetic phases on increasing Ti substitution qualitatively match those obtained from DFT, suggesting an alternate mechanism for stabilizing the \Pn structure. This is pending experimental validation and is hence represented as \Pn$^{*}$ in (a). The mapping between the Hubbard $U$ and experimental Ti concentration is qualitative only and no quantitative secondary axis between the two is implied.}
    \label{fig4}
\end{figure}

We perform first principles DFT calculations using the PBEsol functional, employing Hubbard $U$ correction (using the Dudarev scheme) as well as including spin-orbit interactions. We consider the two magnetic structures, AFM-$b$ and G-AFM, and systematically vary the on-site Hubbard $U$ correction term from 1.2 eV to 2.6 eV. Comparing the formation energy difference between the two magnetic structures as a function of Hubbard $U$, we find that the AFM-$b$ is energetically favored upto a critical $U$ value of about 2.3 eV, above which the G-AFM structure becomes more stable. The electronic band structures for the two magnetic phases are also calculated by varying the $U$ values (Supplementary Figs. \ref{BS_AFMb}, \ref{BS_GAFM} \cite{suppm}), revealing a band gap opening for the AFM-$b$ magnetic structure for $U$ values above 1.6 eV. On the other hand, the G-AFM phase remains insulating for the chosen $U$ range. Following the evolution of the most energetically preferred phase from our DFT calculations (see Supplementary Note \ref{dft_sm} for full details \cite{suppm}) with increasing Hubbard $U$ values, we observe that a lower $U \leq$ 1.2 eV favors the stabilization of an AFM-$b$ metallic \Bb phase. Upon increasing the on-site interaction from $U = $ 1.2 eV, we identify a structural symmetry lowering to \Pn while retaining the magnetic and electronic phases. With a further increase of the Hubbard $U$ beyond 1.6 eV, the gap is opened, achieving a metal-insulator transition in the low symmetry \Pn AFM-$b$ phase. For $U >$ 2.3 eV, we realize a change from the AFM-$b$ to G-AFM magnetic orders concurrent with a lattice symmetry change back to \Bb while remaining insulating. These $U$ dependent changes are summarized in Fig. \ref{fig4}b and a phase diagram tracking the changes in the ground state with the correlation strength $U$ is shown in Fig. \ref{fig4}c. The change in the electronic and magnetic ground states as $U$ is varied phenomenologically resembles the low temperature phases in \cro with Ti substitution, as can be seen by comparing the experimental phase diagram (Fig. \ref{fig4}a) to the DFT phase diagram obtained as a function of Hubbard $U$ (Fig. \ref{fig4}c), which we take as a qualitative validation of the proxy approach in the absence of explicit Ti substitution. Our DFT calculations suggest Hubbard $U$ to be an effective way to simulate dilute Ti substitution, providing a potential alternative pathway for stabilizing the low symmetry \Pn phase. We emphasize that this correspondence is qualitative rather than quantitative and a one-to-one comparison is not to be expected.  \\

The approach is physically motivated by the differing orbital character of the two magnetic configurations. The \Neel type G-AFM configuration opens a band gap at comparatively low $U$ ($\approx$1.0 eV), whereas the A-type AFM-$a/b$ phase remains metallic over a much wider $U$ range, reflecting the different bonding/antibonding character of the relevant Ru $4d$ orbitals in the two magnetic states \cite{lda, npj}. Because Ti induced localization is known to experimentally drive the same AFM-$b$ to G-AFM evolution, this orbital selective $U$ sensitivity provides a microscopic rationale for why increasing $U$ tracks the effect of increasing the Ti content, even though the two perturbations are not expected to map onto each other quantitatively. \\

\begin{figure}[!htb]
    \centering
    \includegraphics[scale=1.8]{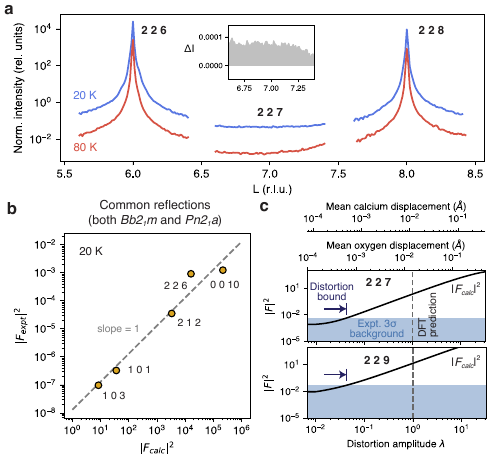}
    \caption{\textbf{Lack of structural evidence of \Pn phase in \ticro.} (a) Integrated line profiles of select 2 2 $L$  peaks. No detectable intensity associated with a Bragg peak is observed for the 2 2 7 reflection condition. Inset shows the difference between the low and high temperature data. (b) Scaling of the experimental \FF at 20 K compared to those from calculations for select reflections that are common to both \Bb and \Pn structures. (c) Constraints on the structural distortion set by experiments on \ticro. The experimental upper bound on this distortion amplitude (arrow) is almost an order of magnitude lower than the corresponding value from DFT denoted by the dashed vertical line.}
    \label{fig5}
\end{figure}

Qualitative comparison of the DFT derived ground states with the experimental low temperature phase diagram points to a compositional window between 0.3\% and 2\% Ti where the \Pn phase is most likely to be realized. We chose 1\% Ti doped \cro as a representative case for an insulating AFM-$b$ ground state which could potentially stabilize the \Pn order. Similar to the parent compound, we interrogate the reciprocal space for this composition to seek evidence of additional reflections being allowed due to possible symmetry lowering. At 20 K, the orthorhombic lattice parameters for \ticro were found to be $a = 5.406(6)$ \AA, $b = 5.516(9)$ \AA, and $c = 19.378(11)$ \AA. The shortening of the $c$ axis lattice constant at the metal-insulator transition was also observed (Supplementary Fig. \ref{clattice} \cite{suppm}), consistent with previous literature \cite{peng13}. At all measured temperatures between 20 K and 80 K, the reflections probed in reciprocal space are compatible with the \Bb space group. No additional peaks associated with \Pn structure were detected within the experimental background. Fig. \ref{fig5}a shows an $L$ scan for the 2 2 $L$ family of peaks at 20 K and 80 K, highlighting the absence of any symmetry change between the two temperatures. \\

Similar to \cro, these results can be translated to upper bounds on the amplitude of the symmetry breaking distortion. For generating the $\lambda$ scaled distortions, we consider the \Pn structure for $U = 2.0$ eV since the electronic and magnetic ground states from DFT are consistent with the experimentally reported phases in \ticro \cite{peng13}. Our structure factor calculations explicitly include Ti in Ru sites. The change in the \FF with the distortion amplitude displays a similar trend as before, where the reflections common to both the \Bb and \Pn structures are independent of the distortion magnitude, while the reflections unique to the \Pn phase show an increasing \FF as the distortion is varied. Fig. \ref{fig5}b compares the \FF obtained from the experimental intensities at 20 K to the calculated values for several reflections common to both the space groups, and we find that the values scale very well. With this calibration, we find that the DFT predicted \Pn phase (for $U = 2.0$ eV) should give rise to a detectable Bragg peak with an intensity almost two orders of magnitude higher than the $3\sigma$ background level as shown in Fig. \ref{fig5}c. While this is not observed, the $3 \sigma$ background level from our experiments on \ticro provide an upper bound of $\lambda \leq 0.08$ for this distortion, corresponding to oxygen and calcium atom displacements of 1.2 $\times 10^{-3}$ \AA \ and 9 $\times 10^{-4}$ \AA \ respectively. This translates to a maximum $Y_{2}$ mode amplitude of 1.9 $\times 10^{-2}$ \AA. We note that our targeted search for the \Pn distortion was limited to a single composition within the proposed 0.3\% to 2\% Ti substitution range. A composition dependent survey of the low temperature structural phase remains an important direction for future work. 

\subsection{G-type altermagnetism with Ti substitution}

\begin{figure}[!htb]
    \centering
    \includegraphics[scale=1.5]{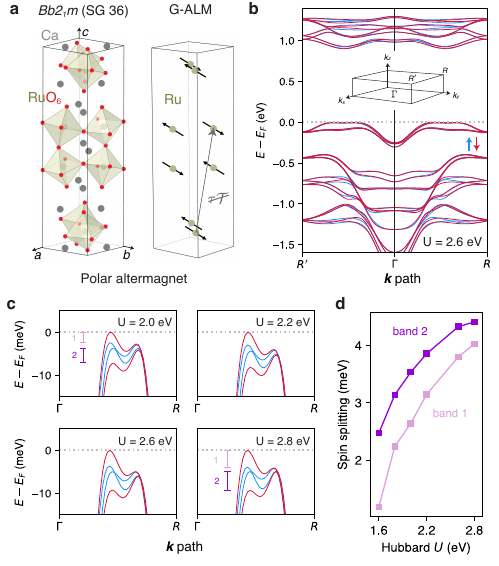}
    \caption{\textbf{Tuning of G-type altermagnetic spin-splitting with Hubbard $U$} (a) Illustration of the combined translation ($\tau = B$) and time-reversal symmetry operation $\tau \mathcal{T}$ that is broken for the G-type altermagnet in the \Bb crystal structure. (b) Spin resolved band structures calculated by excluding spin-orbit coupling along the $R^{\prime} - \Gamma - R$ $\vb{k}$-path over a large energy range. The colors represent opposite spin polarized bands. Inset shows a schematic of the Brillouin zone with the $\Gamma (0,0,0)$, $R (0.5, 0.5, 0.5)$ and $R^{\prime} (-0.5, 0.5, 0.5)$ high symmetry points. (c) Close-up of the bands near the Fermi level for the mentioned $U$ values. (d) Variation in the magnitude of spin-splitting with Hubbard $U$ tuning for the two bands closest to the Fermi level as shown in (c). The spin splitting increases with increasing Hubbard $U$, suggesting higher substitution levels of Ti in \cro are better experimental candidates for exploring polar altermagnetism.}
    \label{fig6}
\end{figure}

The structural transition from \Bb to the \Pn phase is essential for realizing altermagnetic ordering for the AFM-$b$ magnetic structure. At slightly higher Ti concentrations ($\gtrsim 3\%$), the G-type antiferromagnetic ordering emerges as discussed earlier. The G-AFM phase features antiferromagnetic coupling of the neighboring spins along both the in-plane and out-of-plane directions, along with a change in the magnetic easy axis direction which now lies approximately along the $[1 \ 1.732 \ 1]$ crystallographic direction \cite{ke11}. This magnetic structure, illustrated in Fig. \ref{fig6}a, breaks the combined translation and time-reversal \tT symmetry within the \Bb space group ($\tau = B$-centering operation), offering an alternative route to polar altermagnetism without requiring a structural change (refer to Supplementary Fig. \ref{spingroup} for the spin group operations \cite{suppm}). Further, the magnitude of spin-splitting near the Fermi level is significantly enhanced in the G-type altermagnet ($\sim$ few meV) compared to the AFM-$b$ order which showed a spin-splitting of $\sim$ 0.1 meV (Supplementary Fig. \ref{ALMcomp} \cite{suppm}). The G-type \Neel magnetic order gives rise to a P-2 $d_{xz}$-wave altermagnetic state, which when combined with the spin-orbit coupling induced spin-splitting of a $p$-wave character, results in a unique hybrid $d/p$ altermagnetic phase which has been recently reported in this material system \cite{npj}. \\

Using DFT, we theoretically investigate the non-relativistic spin splitting in the G-type altermagnet with increasing Ti substitution, using the Hubbard $U$ parameter as a proxy for the Ti concentration. The spin-orbit coupling is excluded to capture the spin polarized electronic bands arising solely due to the symmetries of the magnetic and crystal structures. Fig. \ref{fig6}b shows the spin-resolved band structure along the $R^{\prime} - \Gamma - R$ $\vb{k}$-path. In contrast to the bands near the Fermi energy, those well below the Fermi level exhibit a significantly larger spin-splitting, approximately 30 meV at about 0.8 eV below the Fermi level. This arises due to orbital selectivity, where the bands near the Fermi level predominantly possess a $d_{xy}$ orbital character, which is nearly filled and hence do not contribute strongly to the altermagnetic splitting \cite{npj}. Nevertheless, they still display a spin-splitting of a few meV and more importantly, this spin-splitting can be tuned via the Hubbard $U$. Fig. \ref{fig6}c shows the two highest occupied bands near the Fermi level for a series of $U$ values (the bands over a larger energy range are shown in Supplemental Material Fig. \ref{BS_noSOC} \cite{suppm}). The extracted spin-splitting for these bands, plotted in Fig. \ref{fig6}d, show a monotonous increase in the splitting with increasing Hubbard $U$. This has direct experimental implications, suggesting higher Ti substitution levels may provide more favorable conditions to realize a larger spin-splitting. \\

Although theoretical predictions indicate a small spin-splitting, the potential coupling of the altermagnetic order to the polar lattice offers a route to control, and possibly enhance altermagnetic functionality. Furthermore, the absence of inversion symmetry permits spin-orbit coupling induced effects, leading to an additional splitting of the bands with $p$-wave character \cite{npj}, whose momentum dependence is sensitive to the orientation of the \Neel vector. The experimentally determined \Neel vector points roughly along the body diagonal, in contrast to theoretical calculations where the \Neel vector is oriented primarily along the $b$ axis \cite{npj}. The enhanced projections of the experimental \Neel vector along the $a$ and $c$ axes could enable non-trivial momentum dependent relativistic band splitting arising from spin-orbit coupling coexisting with the crystal symmetry induced non-relativistic splitting. Strain engineering, especially in thin-film geometry, could provide a route to control this orientation and tune such unconventional band-splitting. Notably, the G-AFM order remains the magnetic ground state up to at least 15\% Ti-substitution, with the \Neel temperature increasing with Ti concentration and reaching $\sim$ 120 K at 15\% \cite{peng16}. This robust G-AFM region thus makes it a more experimentally accessible setting for exploring the coupling between the polar and altermagnetic order in this material system. Furthermore, the ferroelastic and insulating character of this phase offers an experimental platform to investigate chemically tunable altermagnetoelectric and altermagnetoelastic responses. 

\section{Conclusion}

Our low temperature X-ray diffraction measurements establish that the emergent \Pn structure predicted by DFT has a maximum $Y_{2}$ mode amplitude of  1.3 $\times 10^{-2}$ \AA, and a comparable bound of 1.9 $\times 10^{-2}$ \AA \ for the 1\% Ti substituted composition, where we have assumed a complete volume conversion from \Bb to \Pn. Within this assumption and corresponding sensitivity, we find no evidence for long-range static \Pn displacements, and the structures for both \cro and \ticro are consistent with \Bb at all temperatures. These results agree with all prior X-ray studies, which report no structural symmetry change \cite{yoshida05, resonantxray, rashba24, dashwood23}, and with neutron diffraction measurements in the $H$0$L$ scattering plane which detects no any additional symmetry lowering at low temperatures \cite{nscat}. The transition from \Bb to \Pn doubles the primitive unit cell and consequently halves the Brillouin zone (see Supplementary Fig. \ref{platt} \cite{suppm}). ARPES measurements on \cro reveal an electronic reconstruction while preserving the original Brillouin zone \cite{horio21}. The larger primitive unit cell of \Pn results in additional phonon modes relative to the \Bb structure, which polarized Raman scattering measurements do not observe \cite{raman05, hugo23}, supporting the absence of the \Pn phase. \\

The \Pn instability is highly sensitive to the computational treatment. For PBEsol+$U$ in the Dudarev scheme, \Pn becomes energetically favorable for $U \geq$ 1.4 eV, whereas LDA+$U$ retains the \Bb while still preserving the key low-temperature electronic features \cite{lda}. The \Pn instability is also influenced by spin-orbit coupling and the choice of Hubbard $U$ formulation -- it vanishes in the absence of spin-orbit coupling and shifts to higher $U$ values ($\geq$ 2.0 eV) under the Liechtenstein scheme \cite{lda}. Our experimental bound provides strong support for the view that the \Pn instability is over-stabilized within PBEsol+$U$. \\

Recent nonlinear transport measurements in \cro detect a low temperature response that is symmetry forbidden for \Bb structure \cite{mali26}. If the nonlinear transport represents \tT symmetry breaking that is intrinsic in origin, the associated long-range static ionic displacement is below the present experimental detection limit. In such a case, the lack of a detectable long-range structural symmetry change obtained from our X-ray measurements may suggest that the symmetry breaking is restricted to the electronic subsystem, albeit our diffraction results cannot directly establish this. While our results rule out long-range ordering of the \Pn phase, they do not exclude the possibility of \Pn that is ordered on much shorter length scales, and dynamic or strain-coupled lattice contributions. We consider short-range order unlikely to be the source of the bulk-sensitive nonlinear transport response, although testing it requires specialized X-ray techniques such as extended X-ray absorption fine structure (EXAFS) or X-ray diffuse scattering, which is a subject for future studies. Electronic Raman measurements find signatures of incoherent density wave fluctuations which inherit a similar distortion pattern of the \Pn structure, characterized by the spatial modulation of the \ruo octahedral volumes \cite{hugo23}. These density wave fluctuations have a characteristic lifetime of around a hundred femtoseconds \cite{hugo23} and phononic control via ultrafast light pulses seem like a potential route to stabilize this phase. \\

Considering the increasing interest and rapid growth in the number of predicted altermagnetic candidates, it is important to realize that these predictions can be highly sensitive to the specific choice of DFT parameters and methods, as illustrated, for instance, by the ongoing debate surrounding RuO$_{2}$. Given the importance of crystalline symmetries in altermagnetic classification, high sensitivity structural validation is an essential step in this identification. By combining DFT, diffraction simulations and synchrotron X-ray measurements, we present a general methodology to detect and place quantitative bounds on minute structural deviations. This approach can, in principle, be extended to scanning transmission electron microscopy, electron diffraction as well as laboratory-based X-ray experiments, which would enable the rapid screening of a large number of predicted altermagnet candidates. 

\section*{Acknowledgements}

 A.S. acknowledges fruitful discussions with John. W. Freeland, Ismaila Dabo, Maxwell Wetherington and Guo-Dong Zhao. A.S. also acknowledges the use of Penn. State Materials Characterization Lab (MCL) facilities. A.S., L.-Q.C. and V.G. acknowledge the primary support from Computational Mesoscale Science and Open Software for Quantum Materials, under award number DE-SC0020145 as part of the Computational Materials Sciences program of the US Department of Energy, Office of Science, Basic Energy Sciences. Y.X. and G.H. were supported by the U.S. Department of Energy, Office of Science, Basic Energy Sciences in Quantum Information Science under Award Number DESC0022289. This research used resources of the National Energy Research Scientific Computing Center, a DOE Office of Science User Facility supported by the Office of Science of the U.S. Department of Energy under Contract No. DE-AC02-05CH11231 using NERSC award BES-ERCAP0020966. V.A.S., H.W., J.N. and V.G. acknowledge support from the US Department of Energy X-ray scattering program grant number DE-SC0012375. S.M. and Z.M. acknowledge support from the US National Science Foundation under grant DMR-2211327. A.P. and Z.M. acknowledge the support from NSF through the Materials Research Science and Engineering Center DMR-2011839. S.S. and V.G. acknowledge the National Science Foundation grant number DMR-2011839. S.H.L. and Y.W. acknowledge the support of Penn State Two-Dimensional Crystal Consortium center under grant number NSF-DMR-2039351. This research was performed on APS beam time award \href{https://doi.org/10.46936/APS-192457/60015923}{192457} from the Advanced Photon Source, a U.S. Department of Energy (DOE) Office of Science user facility operated for the DOE Office of Science by Argonne National Laboratory under Contract No. DE-AC02-06CH11357.

\section*{Author Contributions}

A.S. and Y.X. contributed equally to this work. A.S. and V.G. conceived the project. A.S. and V.A.S. planned and performed the synchrotron X-ray measurements with assistance from E.K. and under supervision from V.G.. A.S. analyzed the synchrotron X-ray data with inputs from V.A.S., H.W. and V.G.. DFT calculations were performed by Y.X., with inputs from G.H., Y.Z. and B.Y.. Y.W. and Z.M. grew the single crystals for this study. S.M. and A.S. performed the transport measurements with assistance from S.S., J.N., S.H.L., Z.W.H. and were supervised by Z.M. and V.G.. A.P., S.M. and Z.M. performed the SQUID magnetometry measurements. All authors contributed to the discussion of the results. A.S. wrote the manuscript with inputs from all authors. The project was supervised by V.G. and L.-Q.C..

\section*{Data Availability}

The data that support the findings of this study are openly available in figshare at \url{https://figshare.com/s/2d5a6a9de2cafe1a67ca}.

\section*{Competing Interests}

The authors declare no competing interests.

\appendix

\section{Extraction of integrated intensities}

Intensity line profiles along the $L$ direction were obtained by integrating the reconstructed three dimensional intensities within a region of interest in the $HK$ plane centered on the reciprocal lattice point. For reflections where a clear peak is measured, the integrated intensity of the reflection is calculated by integrating the obtained line profile. For \Pn allowed reflections where a clear peak is not observed, we can only assign an upper bound to the integrated intensity. For this, we use the fluctuations in the measured background $b$ as the upper limit for the maximum intensity $I_{\text{max}}$ of the Bragg peaks. The integrated intensity $I_{\text{intg}}$ of the reflection is related to the maximum intensity $I_{\text{max}}$ and the peak widths by $I_{\text{intg}} = \eta I_{\text{max}} \Gamma_{H} \Gamma_{K} \Gamma_{L}$ where $\Gamma_H, \Gamma_K, \Gamma_L$ are the full-widths at half-maximum along the orthogonal $H, K, L$ reciprocal space directions, and $\eta$ is a numerical factor depending on the intensity profile model (see Supplementary Note \ref{profile} \cite{suppm}). The detection threshold for the integrated intensity can then be expressed as $b \cdot V_{HKL}$ where $b$ is the $3\sigma$ background level and $V_{HKL} = \eta \Gamma_{H} \Gamma_{K} \Gamma_{L}$ is related to the real-space correlation volume, which we assume it to be independent of the reflection being probed. Specifically, we take it to be the same for allowed reflections in \Bb and \Pn (see Supplementary Note \ref{fwhm} \cite{suppm}). We also assume a full volume conversion from \Bb to \Pn. The effect of a partial volume fraction conversion is discussed further in Supplementary Note \ref{partialvol} \cite{suppm} and it does not affect our conclusions. In our analysis, we extract $V_{HKL} = \eta \Gamma_{H} \Gamma_{K} \Gamma_{L} = I_{\text{intg}}/I_{\text{max}}$ by taking the ratio of the integrated experimental intensity to the experimental intensity maxima for the reflections where a clear peak is observed. The obtained values are comparable across multiple reflections. 

\section{Conversion of intensities to structure factor squared}

All intensities are already normalized to the incident synchrotron beam flux and attenuation filters between the sample and the detector. The square of the structure factor for the reflection being probed is related to the normalized intensity, but must be corrected for geometrical conditions of the sample and detector. This is expressed as $I_{HKL} \propto |F_{HKL}|^{2} \cdot L \cdot P \cdot A$ where $L$ refers to the Lorentz factor, $P$ is the polarization correction and $A$ accounts for the absorption effects. For the six circle (4S + 2D) diffractometer geometry, the Lorentz factor $L = 1/(\sin \delta \cos \beta_{in} \cos \gamma )$ and the polarization correction factor is given by $P = p_{h}P_{hor} + (1-p_{h})P_{vert}$ where $P_{hor} = 1 - \sin^{2}\gamma$ and $P_{vert} = 1 - \sin^{2} \delta \cos^{2} \gamma$ ($\alpha=0$ as in our experiments) \cite{vlieg}. $p_{h}$ is the horizontal polarization factor of the beam, which we set to 0.99. For the absorption factor correction, the penetration depth as a function of the incident angle was calculated at the X-ray energy using Ref. \cite{penetration}. 

\clearpage

\renewcommand{\bibname}{References}      
\bibliography{refs.bib}

@article{nrss1,
  title={{Three-dimensional mapping of the altermagnetic spin splitting in CrSb}},
  author={Yang, Guowei and Li, Zhanghuan and Yang, Sai and Li, Jiyuan and Zheng, Hao and Zhu, Weifan and Pan, Ze and Xu, Yifu and Cao, Saizheng and Zhao, Wenxuan and others},
  journal={Nature Communications},
  volume={16},
  number={1},
  pages={1442},
  year={2025},
  publisher={Nature Publishing Group UK London}
}

@article{nrss2,
  title={{Direct observation of altermagnetic band splitting in CrSb thin films}},
  author={Reimers, Sonka and Odenbreit, Lukas and {\v{S}}mejkal, Libor and Strocov, Vladimir N and Constantinou, Procopios and Hellenes, Anna B and Jaeschke Ubiergo, Rodrigo and Campos, Warlley H and Bharadwaj, Venkata K and Chakraborty, Atasi and others},
  journal={Nature Communications},
  volume={15},
  number={1},
  pages={2116},
  year={2024},
  publisher={Nature Publishing Group UK London}
}

@article{nrss3,
  title={{Observation of a giant band splitting in altermagnetic MnTe}},
  author={Osumi, T and Souma, S and Aoyama, T and Yamauchi, K and Honma, A and Nakayama, K and Takahashi, T and Ohgushi, K and Sato, T},
  journal={Physical Review B},
  volume={109},
  number={11},
  pages={115102},
  year={2024},
  publisher={APS}
}

@article{nrss4,
  title={{Large band splitting in g-wave altermagnet CrSb}},
  author={Ding, Jianyang and Jiang, Zhicheng and Chen, Xiuhua and Tao, Zicheng and Liu, Zhengtai and Li, Tongrui and Liu, Jishan and Sun, Jianping and Cheng, Jinguang and Liu, Jiayu and others},
  journal={Physical Review Letters},
  volume={133},
  number={20},
  pages={206401},
  year={2024},
  publisher={APS}
}

@article{nrss5,
  title={{Altermagnetic lifting of Kramers spin degeneracy}},
  author={Krempask{\`y}, Juraj and {\v{S}}mejkal, L and D’souza, SW and Hajlaoui, M and Springholz, G and Uhl{\'\i}{\v{r}}ov{\'a}, K and Alarab, F and Constantinou, PC and Strocov, V and Usanov, D and others},
  journal={Nature},
  volume={626},
  number={7999},
  pages={517--522},
  year={2024},
  publisher={Nature Publishing Group UK London}
}

@article{nrss6,
  title={{A metallic room-temperature d-wave altermagnet}},
  author={Jiang, Bei and Hu, Mingzhe and Bai, Jianli and Song, Ziyin and Mu, Chao and Qu, Gexing and Li, Wan and Zhu, Wenliang and Pi, Hanqi and Wei, Zhongxu and others},
  journal={Nature Physics},
  volume={21},
  number={5},
  pages={754--759},
  year={2025},
  publisher={Nature Publishing Group UK London}
}

@article{alm3,
  title={{Altermagnets as a new class of functional materials}},
  author={Song, Cheng and Bai, Hua and Zhou, Zhiyuan and Han, Lei and Reichlova, Helena and Dil, J Hugo and Liu, Junwei and Chen, Xianzhe and Pan, Feng},
  journal={Nature Reviews Materials},
  volume={10},
  number={6},
  pages={473--485},
  year={2025},
  publisher={Nature Publishing Group UK London}
}

@article{almcrystal,
  title={{Manipulation of the altermagnetic order in CrSb via crystal symmetry}},
  author={Zhou, Zhiyuan and Cheng, Xingkai and Hu, Mengli and Chu, Ruiyue and Bai, Hua and Han, Lei and Liu, Junwei and Pan, Feng and Song, Cheng},
  journal={Nature},
  volume={638},
  number={8051},
  pages={645--650},
  year={2025},
  publisher={Nature Publishing Group UK London}
}

@article{alm4,
  title={{Altermagnetism: Exploring new frontiers in magnetism and spintronics}},
  author={Bai, Ling and Feng, Wanxiang and Liu, Siyuan and {\v{S}}mejkal, Libor and Mokrousov, Yuriy and Yao, Yugui},
  journal={Advanced Functional Materials},
  volume={34},
  number={49},
  pages={2409327},
  year={2024},
  publisher={Wiley Online Library}
}

@article{alm1,
  title={{Emerging research landscape of altermagnetism}},
  author={{\v{S}}mejkal, Libor and Sinova, Jairo and Jungwirth, Tomas},
  journal={Physical Review X},
  volume={12},
  number={4},
  pages={040501},
  year={2022},
  publisher={APS}
}

@article{alm2,
  title={{Beyond conventional ferromagnetism and antiferromagnetism: A phase with nonrelativistic spin and crystal} rotation symmetry},
  author={{\v{S}}mejkal, Libor and Sinova, Jairo and Jungwirth, Tomas},
  journal={Physical Review X},
  volume={12},
  number={3},
  pages={031042},
  year={2022},
  publisher={APS}
}

@article{alm6,
  title={{Crystal chemistry and design principles of altermagnets}},
  author={Wei, Chao-Chun and Lawrence, Erick and Tran, Alyssa and Ji, Huiwen},
  journal={ACS Organic \& Inorganic Au},
  volume={4},
  number={6},
  pages={604--619},
  year={2024},
  publisher={ACS Publications}
}

@article{alm7,
  title={{Altermagnetism: A chemical perspective}},
  author={Fender, Shannon S and Gonzalez, Oscar and Bediako, D Kwabena},
  journal={Journal of the American Chemical Society},
  volume={147},
  number={3},
  pages={2257--2274},
  year={2025},
  publisher={ACS Publications}
}

@article{alm5,
  title={{Symmetry, microscopy and spectroscopy signatures of altermagnetism}},
  author={Jungwirth, Tomas and Sinova, Jairo and Fernandes, Rafael M and Liu, Qihang and Watanabe, Hikaru and Murakami, Shuichi and Nakatsuji, Satoru and {\v{S}}mejkal, Libor},
  journal={Nature},
  volume={649},
  number={8098},
  pages={837--847},
  year={2026},
  publisher={Nature Publishing Group UK London}
}

@article{ssg1,
  title={{Spin space groups: Full classification and applications}},
  author={Xiao, Zhenyu and Zhao, Jianzhou and Li, Yanqi and Shindou, Ryuichi and Song, Zhi-Da},
  journal={Physical Review X},
  volume={14},
  number={3},
  pages={031037},
  year={2024},
  publisher={APS}
}

@article{ssg2,
  title={{Enumeration and representation theory of spin space groups}},
  author={Chen, Xiaobing and Ren, Jun and Zhu, Yanzhou and Yu, Yutong and Zhang, Ao and Liu, Pengfei and Li, Jiayu and Liu, Yuntian and Li, Caiheng and Liu, Qihang},
  journal={Physical Review X},
  volume={14},
  number={3},
  pages={031038},
  year={2024},
  publisher={APS}
}

@article{ssg3,
  title={{Symmetry classification of magnetic orders using oriented spin space groups}},
  author={Liu, Yuntian and Chen, Xiaobing and Yu, Yutong and Etxebarria, Jes{\'u}s and Perez-Mato, J Manuel and Liu, Qihang},
  journal={Nature},
  volume={652},
  number={8111},
  pages={869--873},
  year={2026},
  publisher={Nature Publishing Group UK London}
}

@article{nrss7,
  title={{Broken Kramers degeneracy in altermagnetic MnTe}},
  author={Lee, Suyoung and Lee, Sangjae and Jung, Saegyeol and Jung, Jiwon and Kim, Donghan and Lee, Yeonjae and Seok, Byeongjun and Kim, Jaeyoung and Park, Byeong Gyu and {\v{S}}mejkal, Libor and others},
  journal={Physical review letters},
  volume={132},
  number={3},
  pages={036702},
  year={2024},
  publisher={APS}
}

@article{nrss8,
  title={{Observation of time-reversal symmetry breaking in the band structure of altermagnetic RuO$_{2}$}},
  author={Fedchenko, Olena and Min{\'a}r, Jan and Akashdeep, Akashdeep and D’souza, Sunil Wilfred and Vasilyev, Dmitry and Tkach, Olena and Odenbreit, Lukas and Nguyen, Quynh and Kutnyakhov, Dmytro and Wind, Nils and others},
  journal={Science advances},
  volume={10},
  number={5},
  pages={eadj4883},
  year={2024},
  publisher={American Association for the Advancement of Science}
}

@article{ame3,
  title={{Electrical switching of altermagnetism}},
  author={Chen, Yiyuan and Liu, Xiaoxiong and Lu, Hai-Zhou and Xie, XC},
  journal={Physical Review Letters},
  volume={135},
  number={1},
  pages={016701},
  year={2025},
  publisher={APS}
}

@article{ame2,
  title={{Altermagnetic multiferroics and altermagnetoelectric effect}},
  author={{\v{S}}mejkal, Libor},
  journal={arXiv preprint arXiv:2411.19928},
  year={2024}
}

@article{ame1,
  title={{Ferroelectric switchable altermagnetism}},
  author={Gu, Mingqiang and Liu, Yuntian and Zhu, Haiyuan and Yananose, Kunihiro and Chen, Xiaobing and Hu, Yongkang and Stroppa, Alessandro and Liu, Qihang},
  journal={Physical review letters},
  volume={134},
  number={10},
  pages={106802},
  year={2025},
  publisher={APS}
}

@article{afe1,
  title={{Ferroelastic altermagnetism}},
  author={Peng, Rui and Fang, Shibo and Ho, Pin and Liu, Fanxin and Zhou, Tong and Liu, Junwei and Ang, Yee Sin},
  journal={npj Quantum Materials},
  year={2025},
  publisher={Nature Publishing Group UK London}
}

@article{afe3,
  title={{Mechanically and electrically switchable triferroic altermagnet in a pentagonal FeO$_2$ monolayer}},
  author={Guo, Deping and Dai, Jiaqi and Wang, Renhong and Wang, Cong and Ji, Wei},
  journal={Physical Review B},
  volume={112},
  number={19},
  pages={195410},
  year={2025},
  publisher={APS}
}

@article{afe2,
  title={{Ferroelastically tunable altermagnets}},
  author={Ding, Ning and Ye, Haoshen and Wang, Shan-Shan and Dong, Shuai},
  journal={Physical Review B},
  volume={112},
  number={22},
  pages={L220410},
  year={2025},
  publisher={APS}
}

@article{ame4,
  title={{Antiferroelectric altermagnets: Antiferroelectricity alters magnets}},
  author={Duan, Xunkai and Zhang, Jiayong and Zhu, Ziye and Liu, Yuntian and Zhang, Zhenyu and {\v{Z}}uti{\'c}, Igor and Zhou, Tong},
  journal={Physical review letters},
  volume={134},
  number={10},
  pages={106801},
  year={2025},
  publisher={APS}
}

@article{ame5,
  title={{Proposing altermagnetic-ferroelectric type-III multiferroics with robust magnetoelectric coupling}},
  author={Sun, Wei and Yang, Changhong and Wang, Wenxuan and Liu, Ying and Wang, Xiaotian and Huang, Shifeng and Cheng, Zhenxiang},
  journal={Advanced Materials},
  volume={37},
  number={26},
  pages={2502575},
  year={2025},
  publisher={Wiley Online Library}
}

@article{ame6,
  title={{Designing spin symmetry for altermagnetism with strong magnetoelectric coupling}},
  author={Sun, Wei and Wang, Wenxuan and Yang, Changhong and Huang, Shifeng and Ding, Ning and Dong, Shuai and Cheng, Zhenxiang},
  journal={Advanced Science},
  volume={12},
  number={30},
  pages={e03235},
  year={2025},
  publisher={Wiley Online Library}
}

@article{npj,
  title={{Hybrid d/p-wave altermagnetism in Ca$_3$Ru$_2$O$_7$ and strain-controlled spin splitting}},
  author={Le{\'o}n, Andrea and Autieri, Carmine and Brumme, Thomas and Gonz{\'a}lez, Jhon W},
  journal={npj Quantum Materials},
  volume={10},
  number={1},
  pages={98},
  year={2025},
  publisher={Nature Publishing Group UK London}
}

@article{lei18,
  title={{Observation of quasi-two-dimensional polar domains and ferroelastic switching in a metal, Ca$_3$Ru$_2$O$_7$}},
  author={Lei, Shiming and Gu, Mingqiang and Puggioni, Danilo and Stone, Greg and Peng, Jin and Ge, Jianjian and Wang, Yu and Wang, Baoming and Yuan, Yakun and Wang, Ke and others},
  journal={Nano letters},
  volume={18},
  number={5},
  pages={3088--3095},
  year={2018},
  publisher={ACS Publications}
}

@article{yoshida05,
  title={{Crystal and magnetic structure of Ca$_3$Ru$_2$O$_7$}},
  author={Yoshida, Yoshiyuki and Ikeda, Shin-Ichi and Matsuhata, Hirofumi and Shirakawa, Naoki and Lee, CH and Katano, Susumu},
  journal={Physical Review B—Condensed Matter and Materials Physics},
  volume={72},
  number={5},
  pages={054412},
  year={2005},
  publisher={APS}
}

@article{peng13,
  title={{From quasi-two-dimensional metal with ferromagnetic bilayers to Mott insulator with G-type antiferromagnetic} order in Ca$_3$(Ru$_{1-x}$Ti$_x$)$_2$O$_7$},
  author={Peng, Jin and Ke, X and Wang, Gaochao and Ortmann, JE and Fobes, David and Hong, Tao and Tian, Wei and Wu, Xiaoshan and Mao, ZQ},
  journal={Physical Review B—Condensed Matter and Materials Physics},
  volume={87},
  number={8},
  pages={085125},
  year={2013},
  publisher={APS}
}

@article{ke11,
  title={{Emergent electronic and magnetic state in Ca$_3$Ru$_2$O$_7$ induced by Ti doping}},
  author={Ke, Xianglin and Peng, J and Singh, DJ and Hong, Tao and Tian, Wei and Dela Cruz, CR and Mao, ZQ},
  journal={Physical Review B—Condensed Matter and Materials Physics},
  volume={84},
  number={20},
  pages={201102},
  year={2011},
  publisher={APS}
}

@article{petkov23,
  title={{Lattice distortions and the metal--insulator transition in pure and Ti-substituted Ca$_3$Ru$_2$O$_7$}},
  author={Petkov, V and Rao, T Durga and Zafar, A and Abeykoon, AM Milinda and Fletcher, E and Peng, J and Mao, ZQ and Ke, X},
  journal={Journal of Physics: Condensed Matter},
  volume={35},
  number={1},
  pages={015402},
  year={2023},
  publisher={IOP Publishing}
}

@article{nelson07,
  title={{Spin-charge-lattice coupling near the metal-insulator transition in Ca$_3$Ru$_2$O$_7$}},
  author={Nelson, CS and Mo, H and Bohnenbuck, B and Strempfer, J and Kikugawa, N and Ikeda, SI and Yoshida, Y},
  journal={Physical Review B—Condensed Matter and Materials Physics},
  volume={75},
  number={21},
  pages={212403},
  year={2007},
  publisher={APS}
}

@article{puggioni20,
  title={{Cooperative interactions govern the fermiology of the polar metal Ca$_3$Ru$_2$O$_7$}},
  author={Puggioni, Danilo and Horio, M and Chang, J and Rondinelli, James M},
  journal={Physical Review Research},
  volume={2},
  number={2},
  pages={023141},
  year={2020},
  publisher={APS}
}

@article{horio21,
  title={{Electronic reconstruction forming a $C_2$-symmetric Dirac semimetal in Ca$_3$Ru$_2$O$_7$}},
  author={Horio, M and Wang, Q and Granata, V and Kramer, KP and Sassa, Y and J{\"o}hr, S and Sutter, D and Bold, A and Das, L and Xu, Y and others},
  journal={npj Quantum Materials},
  volume={6},
  number={1},
  pages={29},
  year={2021},
  publisher={Nature Publishing Group UK London}
}

@article{hugo23,
  title={{Strong electron-phonon coupling driven pseudogap modulation and density-wave fluctuations in a correlated polar} metal},
  author={Wang, Huaiyu and Xiong, Yihuang and Padma, Hari and Wang, Yi and Wang, Ziqi and Claes, Romain and Brunin, Guillaume and Min, Lujin and Zu, Rui and Wetherington, Maxwell T and others},
  journal={Nature communications},
  volume={14},
  number={1},
  pages={5769},
  year={2023},
  publisher={Nature Publishing Group UK London}
}

@article{raman05,
  title={{Raman spectroscopy of Ca$_3$Ru$_2$O$_7$: Phonon line assignment and electron scattering}},
  author={Iliev, MN and Jandl, S and Popov, VN and Litvinchuk, AP and Cmaidalka, J and Meng, RL and Meen, J},
  journal={Physical Review B—Condensed Matter and Materials Physics},
  volume={71},
  number={21},
  pages={214305},
  year={2005},
  publisher={APS}
}

@article{lda,
  title={{Ca$_3$Ru$_2$O$_7$: Interplay among degrees of freedom and the role of the exchange correlation}},
  author={Le{\'o}n, AM and Gonz{\'a}lez, JW and Rosner, Helge},
  journal={Physical Review Materials},
  volume={8},
  number={2},
  pages={024411},
  year={2024},
  publisher={APS}
}

@article{stone19,
  title={{Atomic and electronic structure of domains walls in a polar metal}},
  author={Stone, Greg and Puggioni, Danilo and Lei, Shiming and Gu, Mingqiang and Wang, Ke and Wang, Yu and Ge, Jianjian and Lu, Xue-Zeng and Mao, Zhiqiang and Rondinelli, James M and others},
  journal={Physical Review B},
  volume={99},
  number={1},
  pages={014105},
  year={2019},
  publisher={APS}
}

@article{zhao25,
  title={{Nonlinear transport signatures of hidden symmetry breaking in a Weyl altermagnet}},
  author={Zhao, Yufei and Mao, Zhiqiang and Yan, Binghai},
  journal={Physical Review B},
  volume={112},
  number={16},
  pages={165127},
  year={2025},
  publisher={APS}
}

@article{mali26,
  title={{Probing hidden symmetry via nonlinear transport in an altermagnet candidate Ca$_3$Ru$_2$O$_7$}},
  author={Mali, Subin and Zhao, Yufei and Wang, Yu and Sarker, Saugata and Chen, Yangyang and Li, Zixuan and Zhu, Jun and Liu, Ying and Gopalan, Venkatraman and Yan, Binghai and others},
  journal={Nature Communications},
  volume={17},
  number={1},
  pages={3074},
  year={2026},
  publisher={Nature Publishing Group UK London}
}

@article{resonantxray,
  title={{Magnetic structure and orbital state of Ca$_3$Ru$_2$O$_7$ investigated by resonant x-ray diffraction}},
  author={Bohnenbuck, B and Zegkinoglou, I and Strempfer, J and Sch{\"u}{\ss}ler-Langeheine, C and Nelson, CS and Leininger, Ph and Wu, H-H and Schierle, E and Lang, JC and Srajer, G and others},
  journal={Physical Review B—Condensed Matter and Materials Physics},
  volume={77},
  number={22},
  pages={224412},
  year={2008},
  publisher={APS}
}

@article{rashba24,
  title={{Pressure-induced enhancement of polar distortions in a metal and implications for the Rashba spin splitting}},
  author={Ladbrook, Evie and Dey, Urmimala and Bristowe, Nicholas C and Perry, Robin S and Daisenberger, Dominik and Warren, Mark R and Senn, Mark S},
  journal={Physical Review B},
  volume={111},
  number={20},
  pages={205110},
  year={2025},
  publisher={APS}
}

@article{nscat,
  title={{Spin valve effect and magnetoresistivity in single crystalline Ca$_3$Ru$_2$O$_7$}},
  author={Bao, Wei and Mao, Z Qu and Qu, Z and Lynn, JW},
  journal={Physical review letters},
  volume={100},
  number={24},
  pages={247203},
  year={2008},
  publisher={APS}
}

@article{cycloid2,
  title={{Magnetic structure and field dependence of the cycloid phase mediating the spin reorientation transition in} Ca$_3$Ru$_2$O$_7$},
  author={Faure, Q and Dashwood, CD and Colin, CV and Johnson, RD and Ressouche, E and Stenning, GBG and Spratt, J and McMorrow, DF and Perry, RS},
  journal={Physical Review Research},
  volume={5},
  number={1},
  pages={013040},
  year={2023},
  publisher={APS}
}

@article{dashwood23,
  title={{Strain control of a bandwidth-driven spin reorientation in Ca$_3$Ru$_2$O$_7$}},
  author={Dashwood, CD and Walker, AH and Kwasigroch, MP and Veiga, LSI and Faure, Q and Vale, JG and Porter, DG and Manuel, P and Khalyavin, DD and Orlandi, F and others},
  journal={Nature Communications},
  volume={14},
  number={1},
  pages={6197},
  year={2023},
  publisher={Nature Publishing Group UK London}
}

@article{xu,
  title={{Xrayutilities: A versatile tool for reciprocal space conversion of scattering data recorded with linear and area} detectors},
  author={Kriegner, Dominik and Wintersberger, Eugen and Stangl, Julian},
  journal={Applied Crystallography},
  volume={46},
  number={4},
  pages={1162--1170},
  year={2013},
  publisher={International Union of Crystallography}
}

@article{peng16,
  title={{Extremely large anisotropic transport caused by electronic phase separation in Ti-doped Ca$_3$Ru$_2$O$_7$}},
  author={Peng, Jin and Liu, JY and Gu, Xiaomin and Zhou, Guotai and Wang, Wei and Hu, J and Zhang, FM and Wu, XS},
  journal={Journal of Physics D: Applied Physics},
  volume={49},
  number={24},
  pages={245004},
  year={2016},
  publisher={IOP Publishing}
}

@article{vlieg,
  title={{Integrated intensities using a six-circle surface X-ray diffractometer}},
  author={Vlieg, Elias},
  journal={Applied Crystallography},
  volume={30},
  number={5},
  pages={532--543},
  year={1997},
  publisher={International Union of Crystallography}
}

@article{dft1,
  title={{Efficient iterative schemes for ab initio total-energy calculations using a plane-wave basis set}},
  author={Kresse, Georg and Furthm{\"u}ller, J{\"u}rgen},
  journal={Physical review B},
  volume={54},
  number={16},
  pages={11169},
  year={1996},
  publisher={APS}
}

@article{dft2,
  title={{Efficiency of ab-initio total energy calculations for metals and semiconductors using a plane-wave basis set}},
  author={Kresse, Georg and Furthm{\"u}ller, J{\"u}rgen},
  journal={Computational materials science},
  volume={6},
  number={1},
  pages={15--50},
  year={1996},
  publisher={Elsevier}
}

@article{dft3,
  title={{Electron-energy-loss spectra and the structural stability of nickel oxide: An LSDA+ U study}},
  author={Dudarev, Sergei L and Botton, Gianluigi A and Savrasov, Sergey Y and Humphreys, CJ and Sutton, Adrian P},
  journal={Physical Review B},
  volume={57},
  number={3},
  pages={1505},
  year={1998},
  publisher={APS}
}

@misc{rsmap3d,
  author = {Hammonds, John and Zhang, Zhan},
  year = {2024},
  month = {jun 4},
  title = {AdvancedPhotonSource/{rsMap3D}},
  url = {https://github.com/AdvancedPhotonSource/rsMap3D},
  howpublished = {https://github.com/AdvancedPhotonSource/rsMap3D},
}

@misc{penetration,
  title = {Penetration depth and optical properties for X-rays calculator},
  howpublished = {\url{https://gixa.ati.tuwien.ac.at/tools/penetrationdepth.xhtml}},
  note = {Accessed: 2026-04-27}
}

@article{amplimodes,
  title={{AMPLIMODES: symmetry-mode analysis on the Bilbao Crystallographic Server}},
  author={Orobengoa, Danel and Capillas, Cesar and Aroyo, Mois I and Perez-Mato, J Manuel},
  journal={Applied Crystallography},
  volume={42},
  number={5},
  pages={820--833},
  year={2009},
  publisher={International Union of Crystallography}
}

@article{yoshida04,
  title={{Quasi-two-dimensional metallic ground state of Ca$_3$Ru$_2$O$_7$}},
  author={Yoshida, Yoshiyuki and Nagai, Ichiro and Ikeda, Shin-Ichi and Shirakawa, Naoki and Kosaka, Masashi and M{\^o}ri, Nobuo},
  journal={Physical Review B—Condensed Matter and Materials Physics},
  volume={69},
  number={22},
  pages={220411},
  year={2004},
  publisher={APS}
}

@article{ahe1,
  title={{Spontaneous anomalous Hall effect arising from an unconventional compensated magnetic phase in a semiconductor}},
  author={Gonzalez Betancourt, RD and Zub{\'a}{\v{c}}, Jan and Gonzalez-Hernandez, R and Geishendorf, Kevin and {\v{S}}ob{\'a}{\v{n}}, Zbynek and Springholz, Gunther and Olejn{\'\i}k, Kamil and {\v{S}}mejkal, Libor and Sinova, Jairo and Jungwirth, Tomas and others},
  journal={Physical Review Letters},
  volume={130},
  number={3},
  pages={036702},
  year={2023},
  publisher={APS}
}

@article{ahe2,
  title={{An anomalous Hall effect in altermagnetic ruthenium dioxide}},
  author={Feng, Zexin and Zhou, Xiaorong and {\v{S}}mejkal, Libor and Wu, Lei and Zhu, Zengwei and Guo, Huixin and Gonz{\'a}lez-Hern{\'a}ndez, Rafael and Wang, Xiaoning and Yan, Han and Qin, Peixin and others},
  journal={Nature Electronics},
  volume={5},
  number={11},
  pages={735--743},
  year={2022},
  publisher={Nature Publishing Group UK London}
}

@article{ahe3,
  title={{Observation of a spontaneous anomalous Hall response in the Mn$_5$Si$_3$ d-wave altermagnet candidate}},
  author={Reichlova, Helena and Lopes Seeger, Rafael and Gonz{\'a}lez-Hern{\'a}ndez, Rafael and Kounta, Ismaila and Schlitz, Richard and Kriegner, Dominik and Ritzinger, Philipp and Lammel, Michaela and Leivisk{\"a}, Miina and Birk Hellenes, Anna and others},
  journal={Nature Communications},
  volume={15},
  number={1},
  pages={4961},
  year={2024},
  publisher={Nature Publishing Group UK London}
}

@misc{suppm,
  title = {Supplementary Material available at [URL will be inserted by publisher]},
}

@article{spatialticro,
  title={{Spatial inhomogeneity and the metal-insulator transition in Ca$_3$(Ru$_{1-x}$Ti$_x$)$_2$O$_7$}},
  author={Lechermann, Frank and Han, Qiang and Millis, Andrew J},
  journal={Physical review research},
  volume={2},
  number={3},
  pages={033490},
  year={2020},
  publisher={APS}
}

@article{percolation3,
  title={Bond and site percolation in three dimensions},
  author={Wang, Junfeng and Zhou, Zongzheng and Zhang, Wei and Garoni, Timothy M and Deng, Youjin},
  journal={Physical Review E—Statistical, Nonlinear, and Soft Matter Physics},
  volume={87},
  number={5},
  pages={052107},
  year={2013},
  publisher={APS}
}

@article{percolation2,
  title={The crossover from 2D to 3D percolation: theory and numerical simulations},
  author={Sotta, Paul and Long, Didier},
  journal={The European Physical Journal E},
  volume={11},
  number={4},
  pages={375--388},
  year={2003},
  publisher={Springer}
}
\nocite{stone19}
\nocite{percolation3}
\nocite{percolation2}


\newcommand{\beginsupplement}{
      \setcounter{equation}{0}
      \setcounter{figure}{0}
      \setcounter{table}{0}
      \setcounter{section}{0}
      \renewcommand{\theequation}{S\arabic{equation}}
      \renewcommand{\thefigure}{S\arabic{figure}}
      \renewcommand{\thetable}{S\arabic{table}}
      \renewcommand{\thesection}{\Alph{section}}
}

\clearpage

\beginsupplement

\begin{center}
    \textbf{\Large{Supplementary Material}}
\end{center}


\section*{Supplementary Figures}

\begin{figure}[!htb]
    \centering
    \includegraphics[scale=0.25]{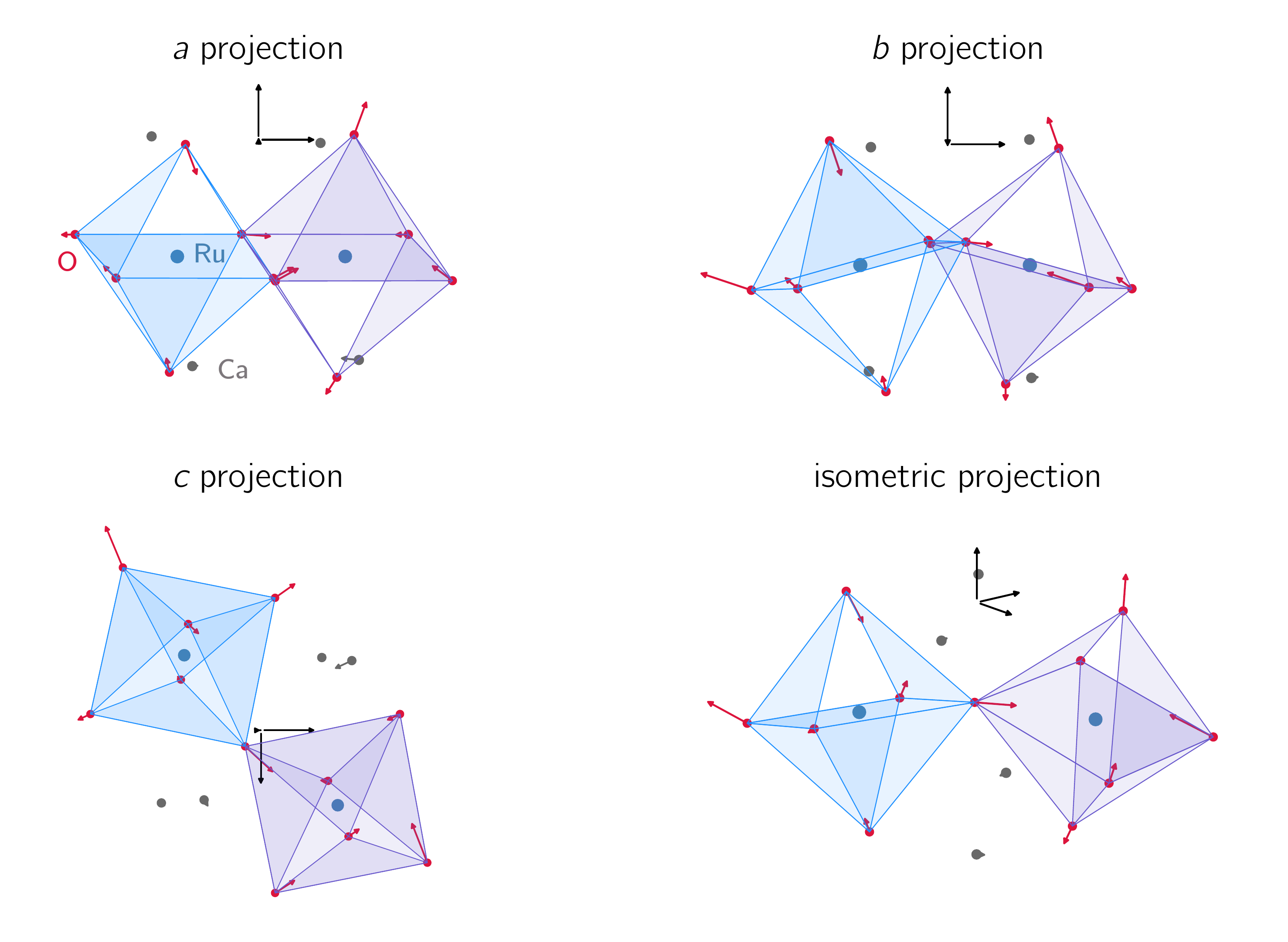}
    \caption{Projections of the distortion that drives the lattice from \Bb to \Pn viewed along various crystallographic axes. The distortion is obtained by computing the atomic displacement vectors between symmetry-equivalent sites in the two structures calculated from DFT. All the atomic displacements are uniformly scaled for visual clarity. A symmetry mode analysis using AMPLIMODES \cite{amplimodes} suggest this distortion is a combination of a $Y_{2}$ and $\Gamma_{1}$ modes (details in Supplementary Note \ref{samdist}).}
    \label{proj}
\end{figure}

\begin{figure}[!htb]
    \centering
    \includegraphics[scale=0.2]{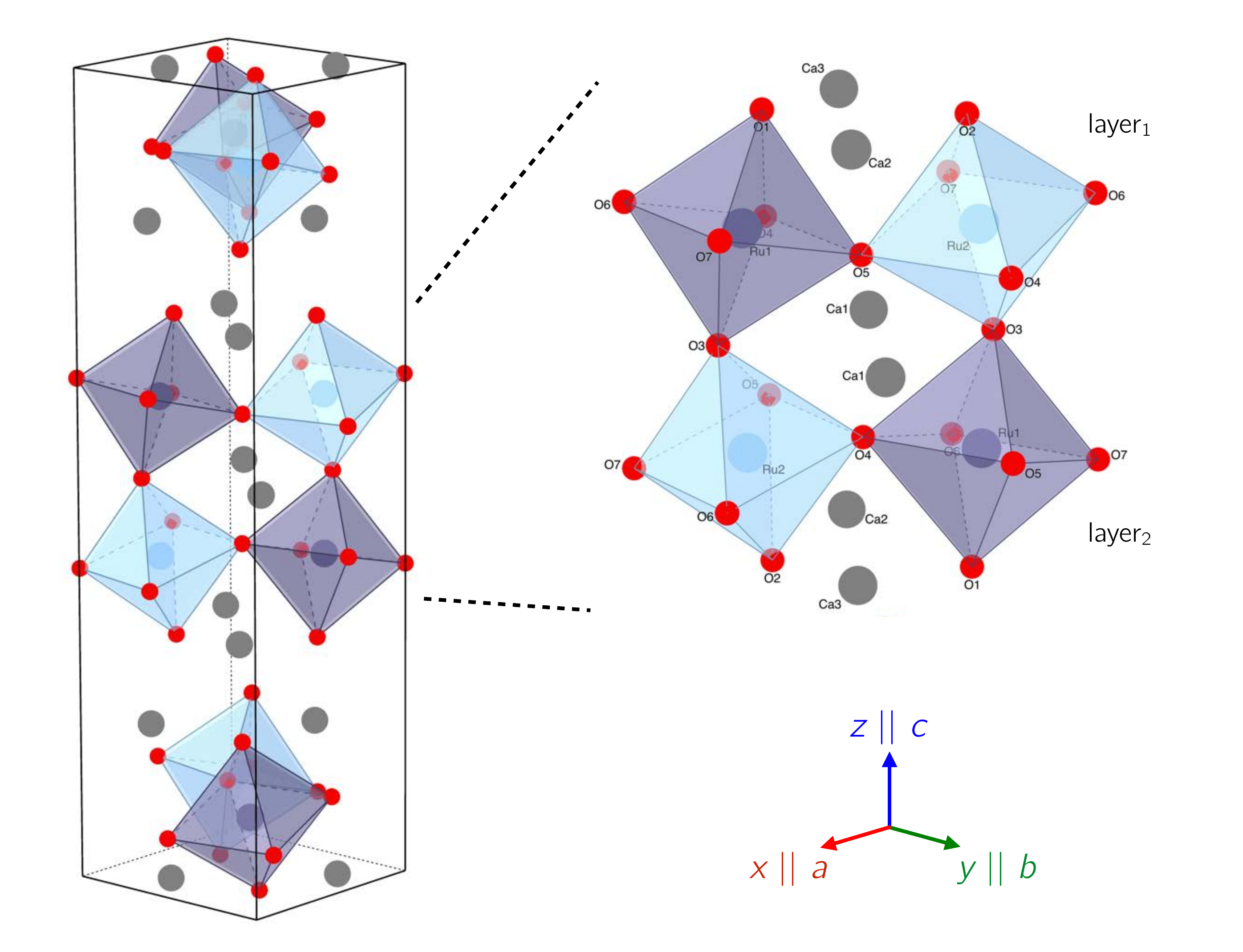}
    \caption{Definition of the unique atomic sites in a \ruo bi-layer of the \cro unit cell. The top and bottom \ruo layers are labelled as layers 1 and 2 respectively.}
    \label{defn}
\end{figure}

\begin{figure}[!htb]
    \centering
    \includegraphics[scale=0.3]{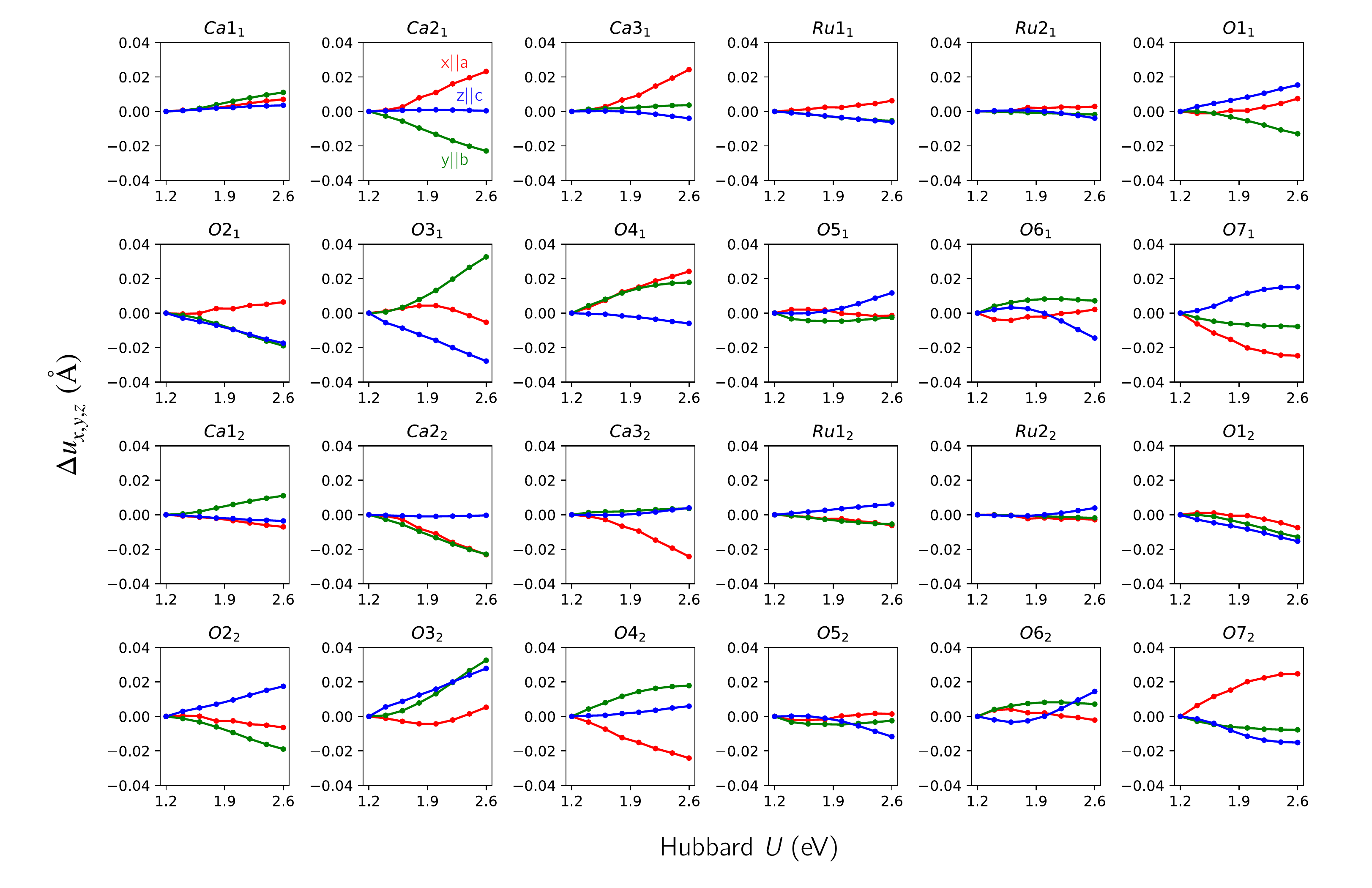}
    \caption{Atomic site resolved distortions as a function of the Hubbard $U$ projected along the crystallographic axes. The distortions are calculated with respect to the $U=1.2$ \Bb structure. Refer to Fig. \ref{defn} for the labeling of the atoms. The sub-scripts refer to the respective layers. The different colors represent the distortions along the different crystallographic axes.}
    \label{dist_xyz}
\end{figure}

\begin{figure}[!htb]
    \centering
    \includegraphics[scale=0.23]{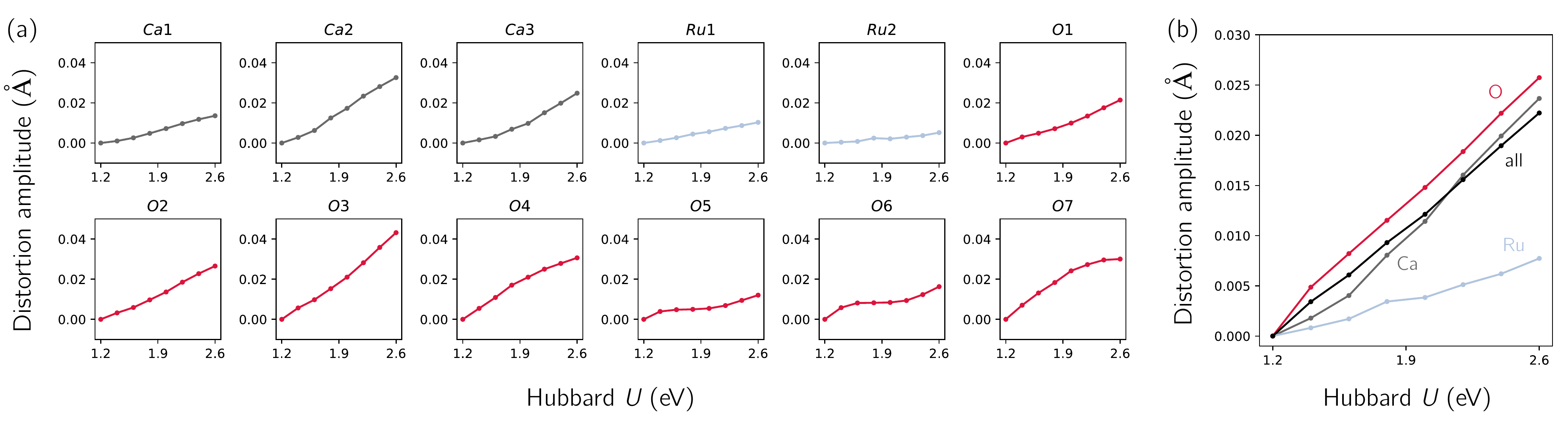}
    \caption{(a) Magnitude of the atomic site resolved distortions as a function of the Hubbard $U$. Refer to Fig. \ref{defn} for the labeling of the atoms. (b) Mean atomic distortion for Ca, Ru, O, obtained by averaging over all corresponding atomic sites, and across all atoms in the unit cell. The Ca and O atoms dominate the distortions.}
    \label{dist_mag}
\end{figure}

\begin{figure}[!htb]
    \centering
    \includegraphics[scale=0.7]{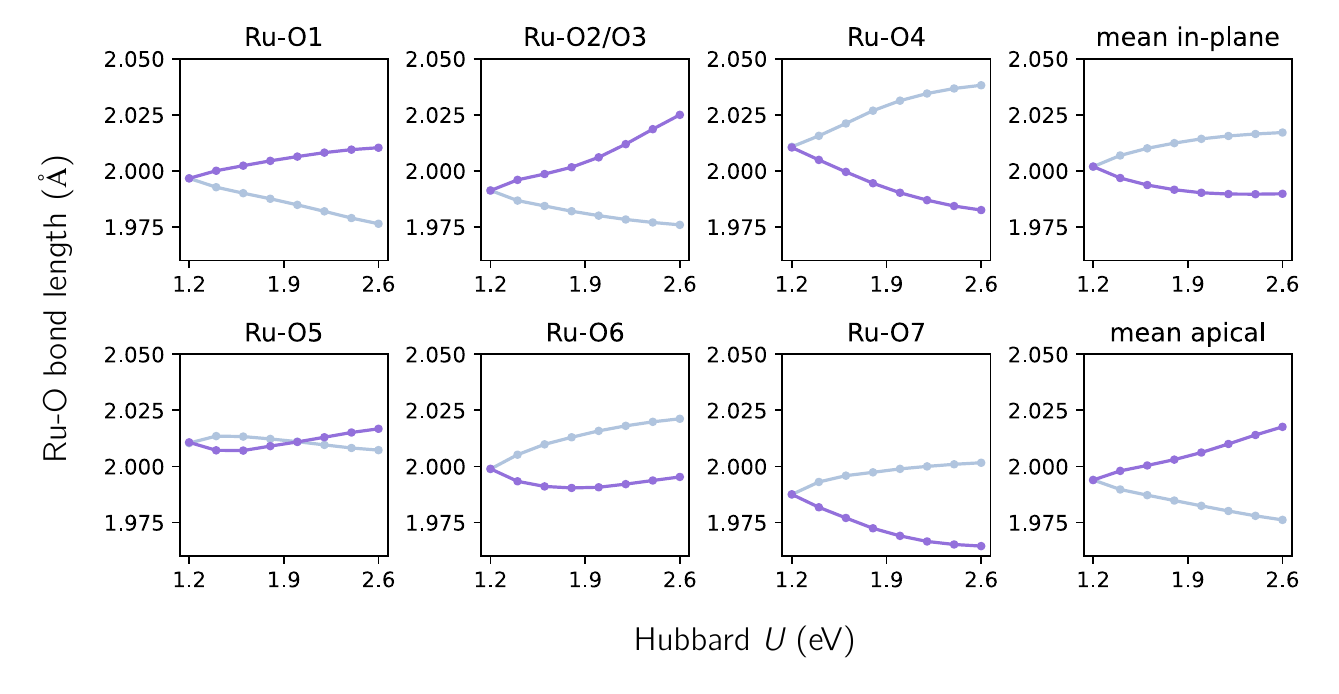}
    \caption{RuO bond lengths as a function of the Hubbard $U$. Refer to Fig. \ref{defn} for the labeling of the atoms. The distortion modifies the \ruo octahedra by shortening the Ru-O$_{\text{apical}}$ bond while elongating the Ru-O$_{\text{in-plane}}$ bonds and vice versa on the neighboring octahedra.}
    \label{RuO}
\end{figure}


\begin{figure}[!htb]
  \centering
  \includegraphics[scale=0.28]{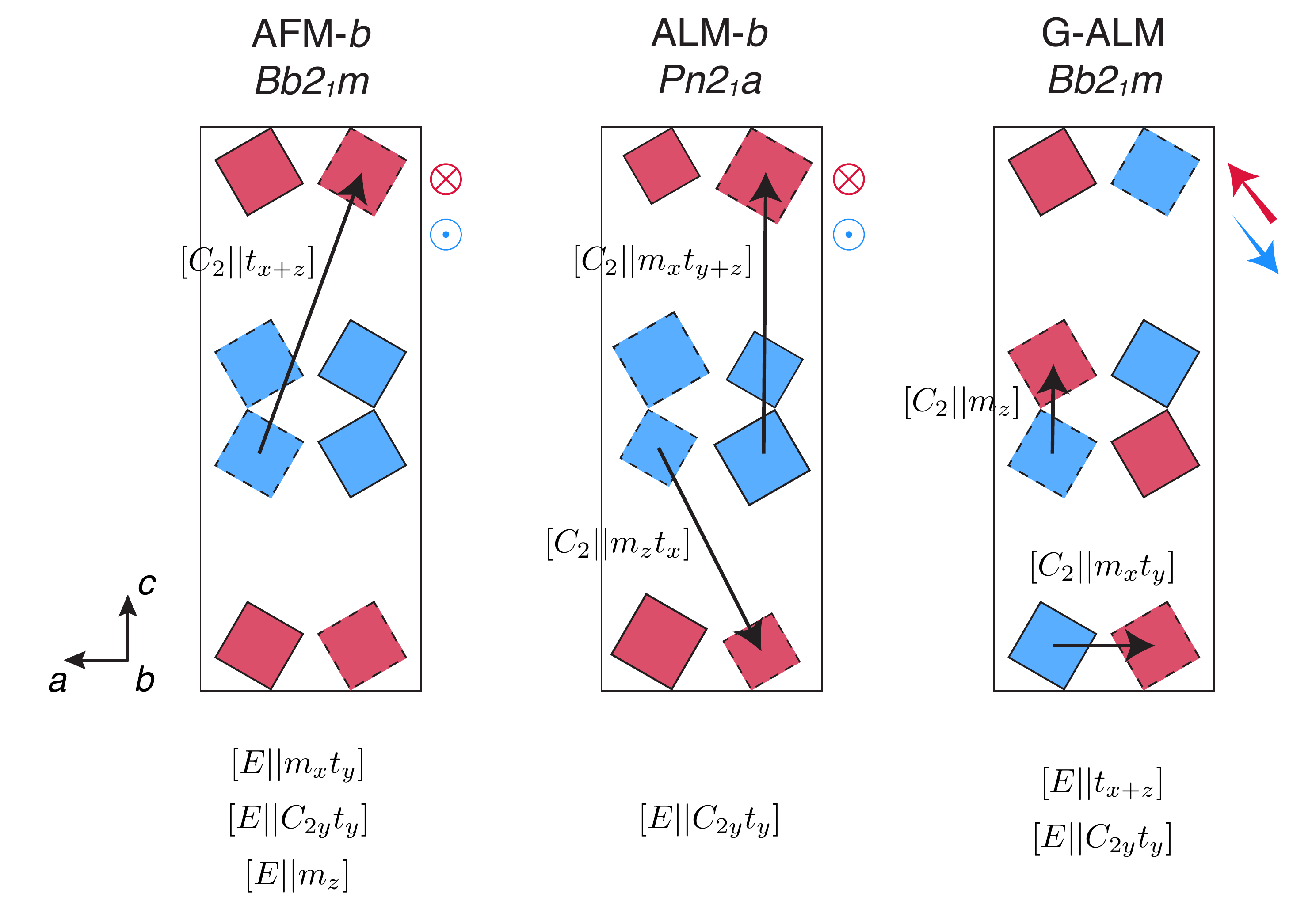}
  \caption{Illustration of the spin group symmetry operation that relate the two opposite spin sub-lattices in various phases considered: AFM-$b$ (non-altermagnetic AFM-$b$), ALM-$b$ (altermagnetic AFM-$b$) and G-ALM (altermagnetic G-AFM). Both the altermagnetic structures belong to the spin group 2, planar $d$-wave classification and have spin group $\ ^2m^2m^1m$. The arrows represent the operations that map the two opposite spin sub-lattices colored by blue and red. The other crystal operations that do not map the opposite spin sub-lattices are mentioned below each structure. The tilted squares represent the local \ruo octahedral environment at each Ru site while the size of the square represents the volume of the \ruo following the volume distortion in \Pn. Additionally, a solid (dotted) outline around the squares represents the Ru site being shifted towards (away from) the viewer.}
  \label{spingroup}
\end{figure}


\begin{figure}[!htb]
    \centering
    \includegraphics[scale=0.22]{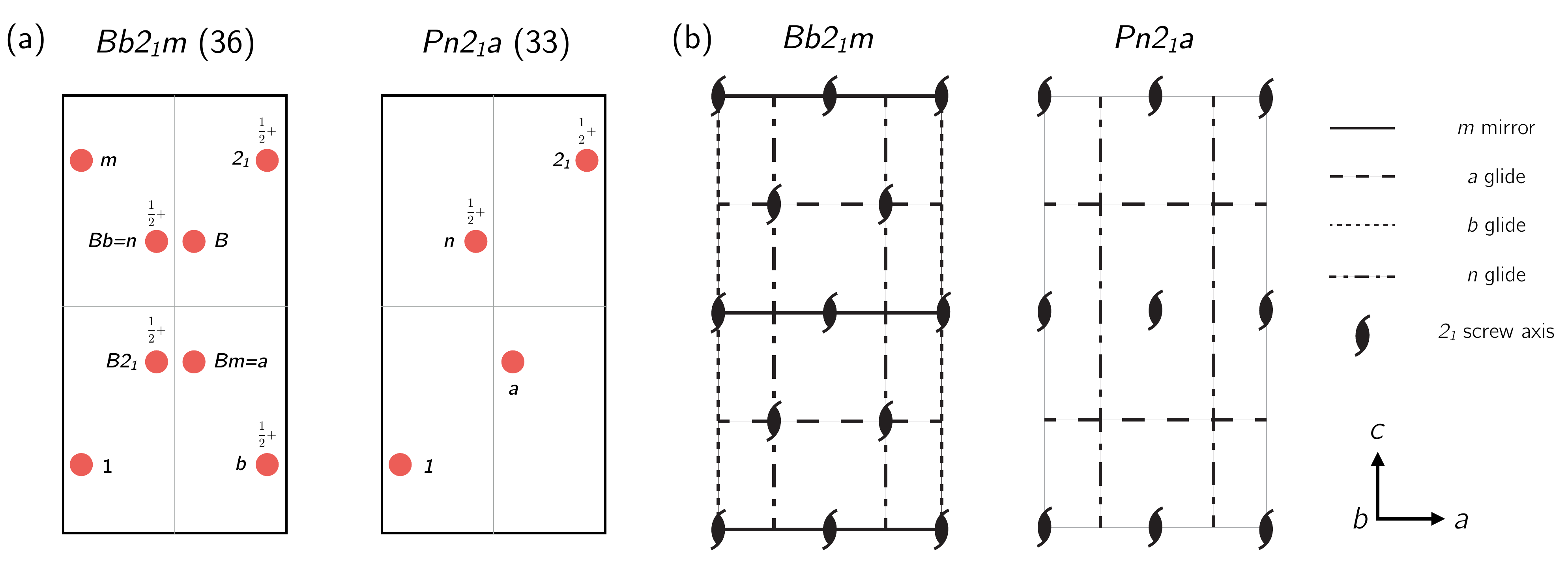}
    \caption{(a) Visualization of the symmetry related equivalent sites for the two considered crystal structures. The $B$ lattice centering operation, combined with the $b_{x}$ glide and mirror $m_{z}$ operations generate the $n_{x}$ and $a_{z}$ glide planes respectively as seen for the \Pn structure i.e. $B b_{x} = n_{x}$ and $B m_{z} = a_{z}$. (b) Illustrations of the various lattice symmetry operations in the two space groups. The presence (absence) of the fractional lattice translation $B$ differentiates these two space groups.}
    \label{symm}
\end{figure}


\begin{figure}[!htb]
    \centering
    \includegraphics[scale=0.6]{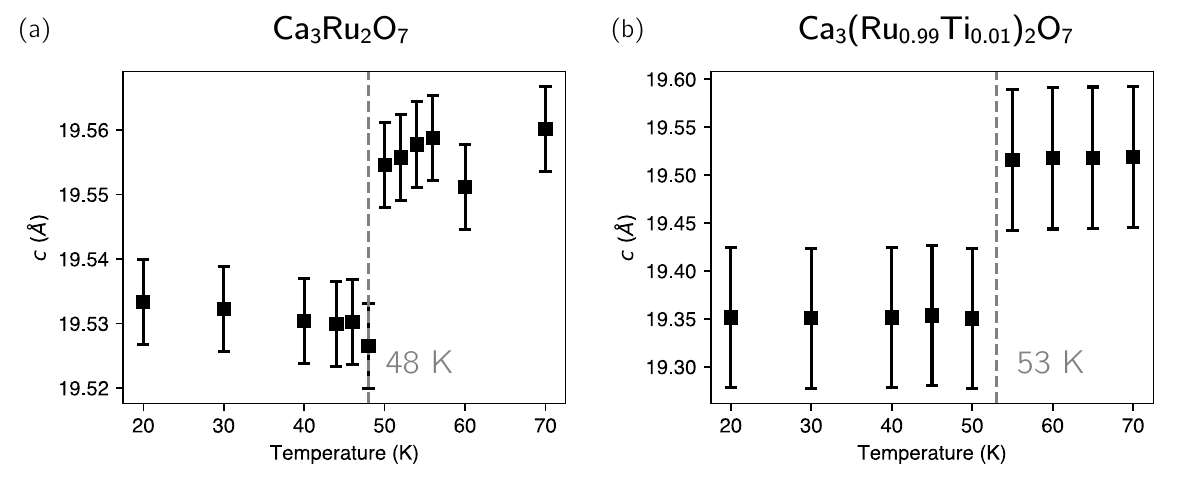}
    \caption{Temperature dependency of the $c$ lattice parameter for (a) \cro and (b) \ticro. The lattice constant is abruptly shortened on cooling at spin reorientation transition for both the compositions.}
    \label{clattice}
\end{figure}


\begin{figure}[!htb]
    \centering
    \includegraphics[scale=0.25]{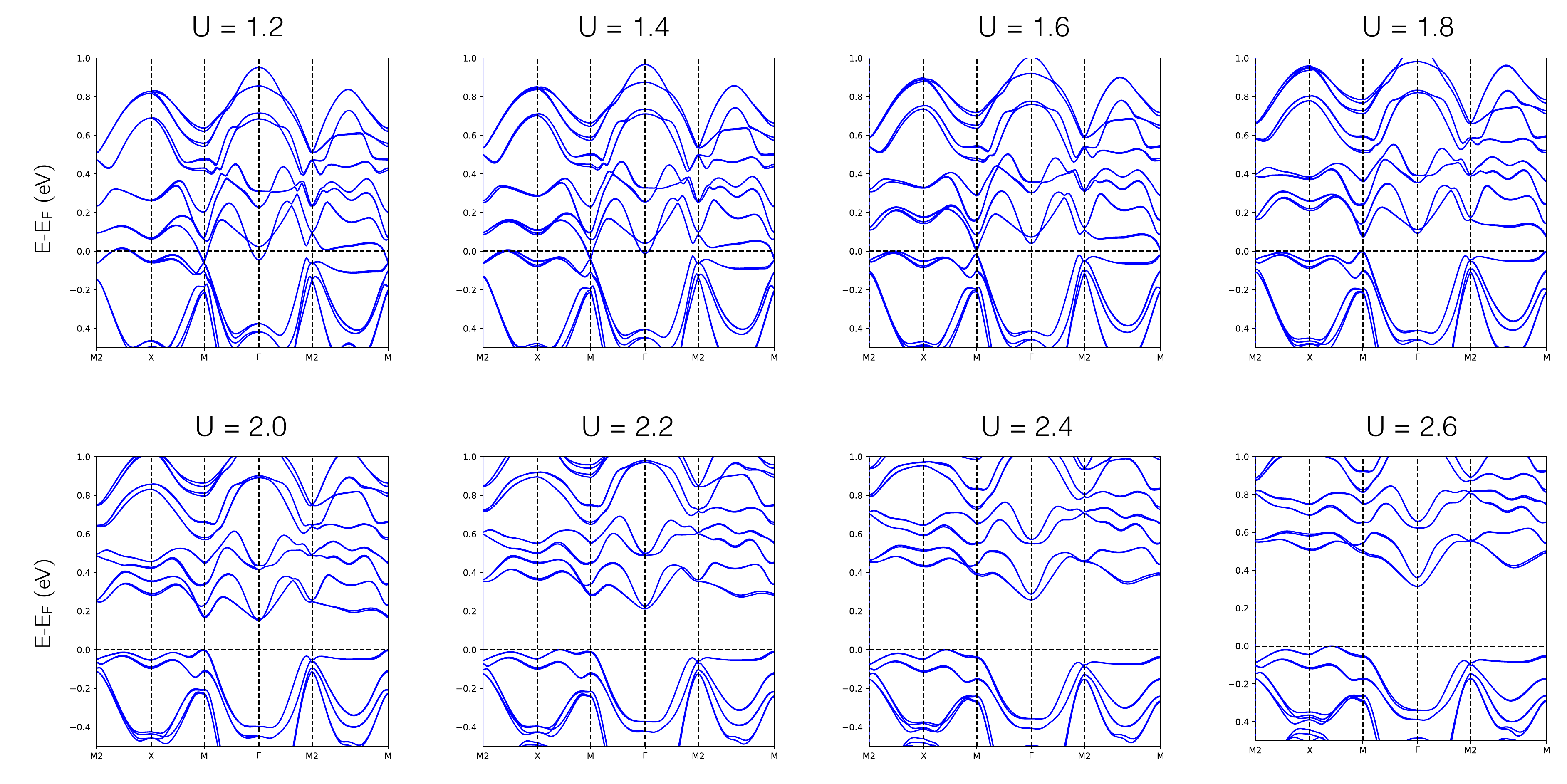}
    \caption{Hubbard $U$ dependent electronic band structures for the AFM-$b$ magnetic ordering. These calculations include spin-orbit coupling. A band gap opens up at higher values of $U$ ($U > 1.6$ eV). }
    \label{BS_AFMb}
\end{figure}

\begin{figure}[!htb]
    \centering
    \includegraphics[scale=0.25]{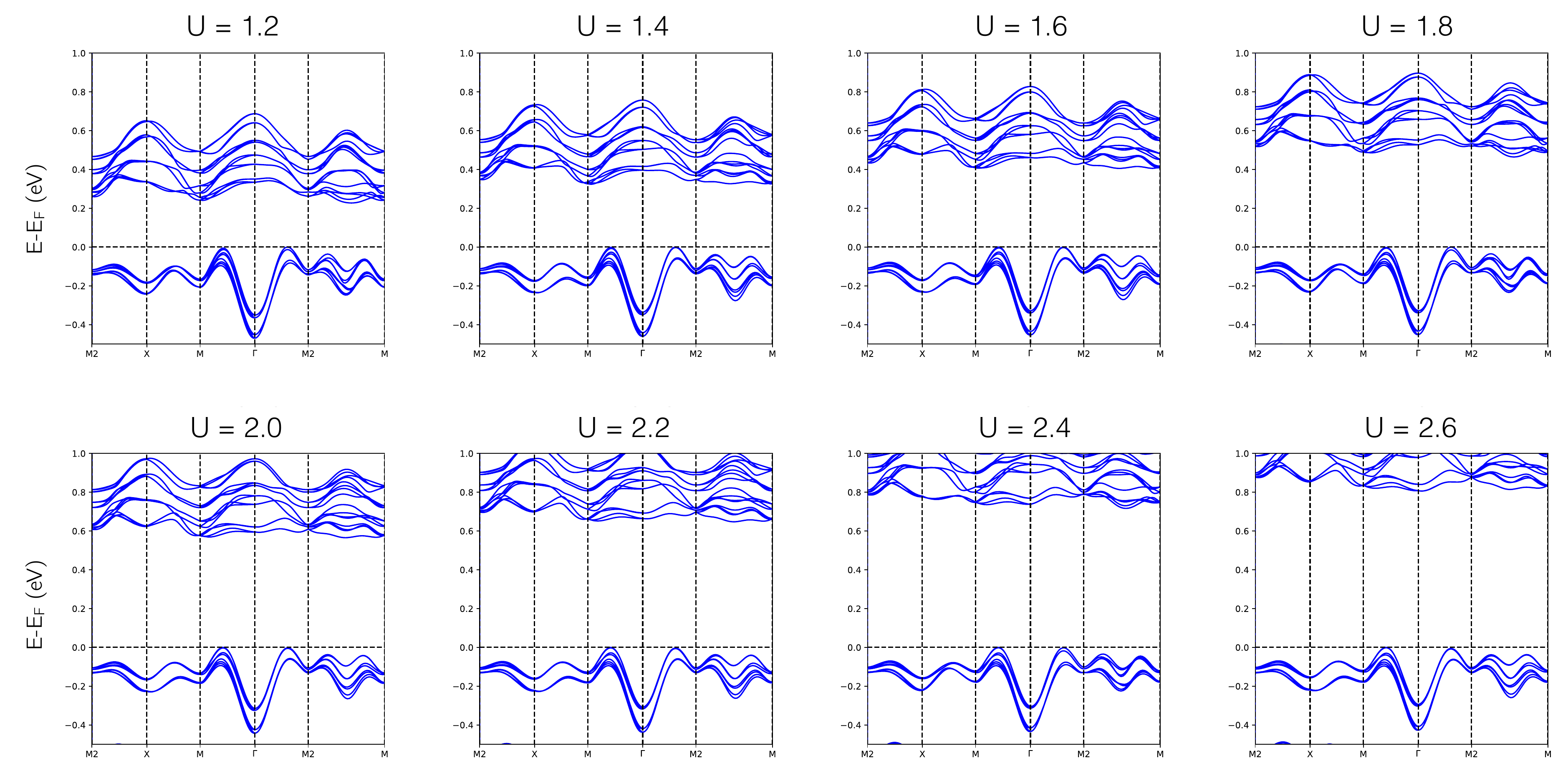}
    \caption{Hubbard $U$ dependent electronic band structures for the G-AFM magnetic ordering. These calculations include spin-orbit coupling. The electronic phase is already gapped at all considered $U$ values.}
    \label{BS_GAFM}
\end{figure}

 
\begin{figure}[!htb]
    \centering
    \includegraphics[scale=0.35]{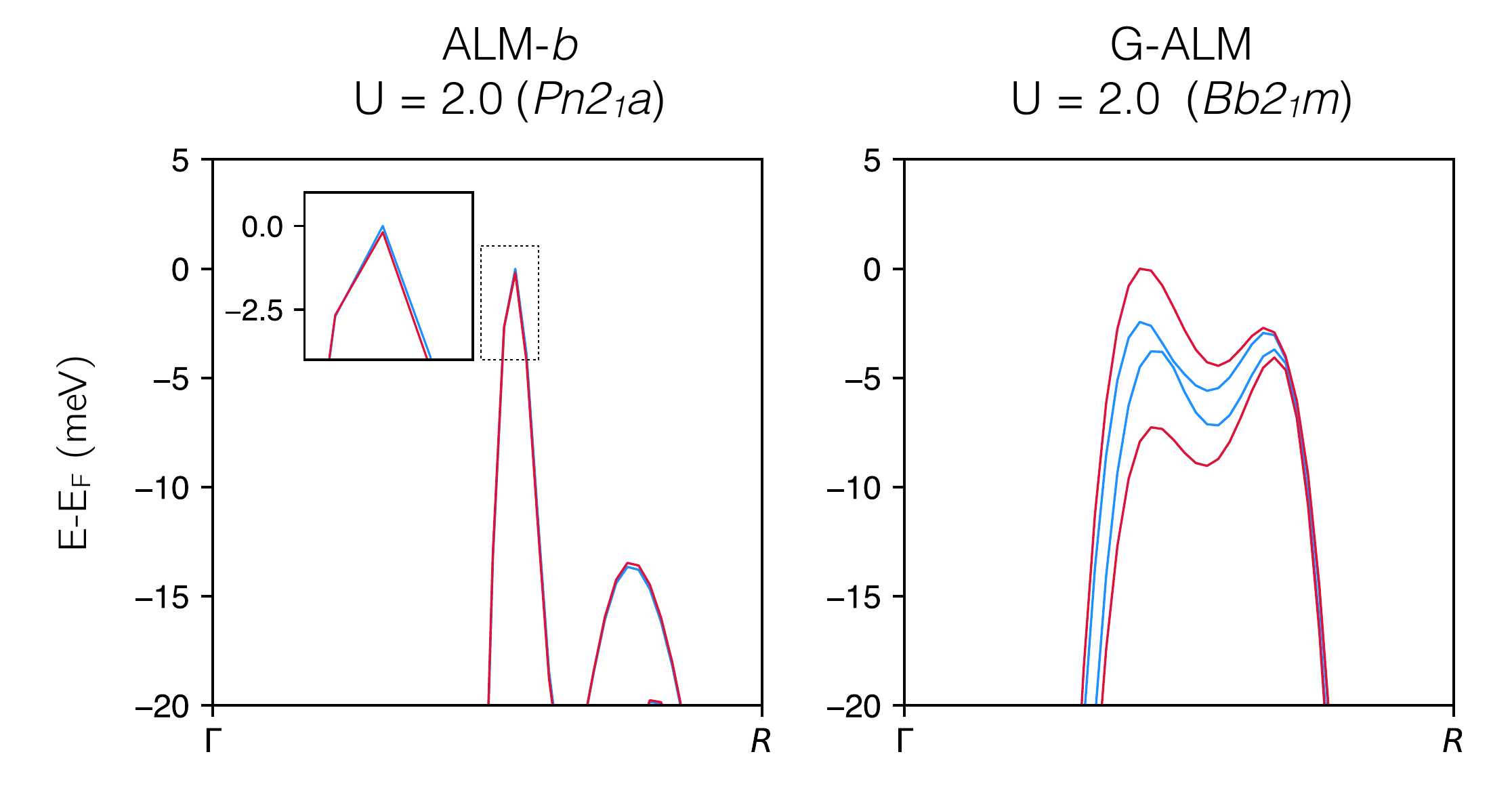}
    \caption{Comparison of spin-split band structure near the Fermi energy for the altermagnetic ALM-$b$ (AFM-$b$ \Pn) and the G-ALM (G-AFM \Bb) phases. Spin-orbit coupling has been excluded for these calculations. The magnitude of the spin-splitting is enhanced in the G-ALM phase compared to the ALM-$b$ phase.}
    \label{ALMcomp}
\end{figure}

\begin{figure}[!htb]
    \centering
    \includegraphics[scale=0.3]{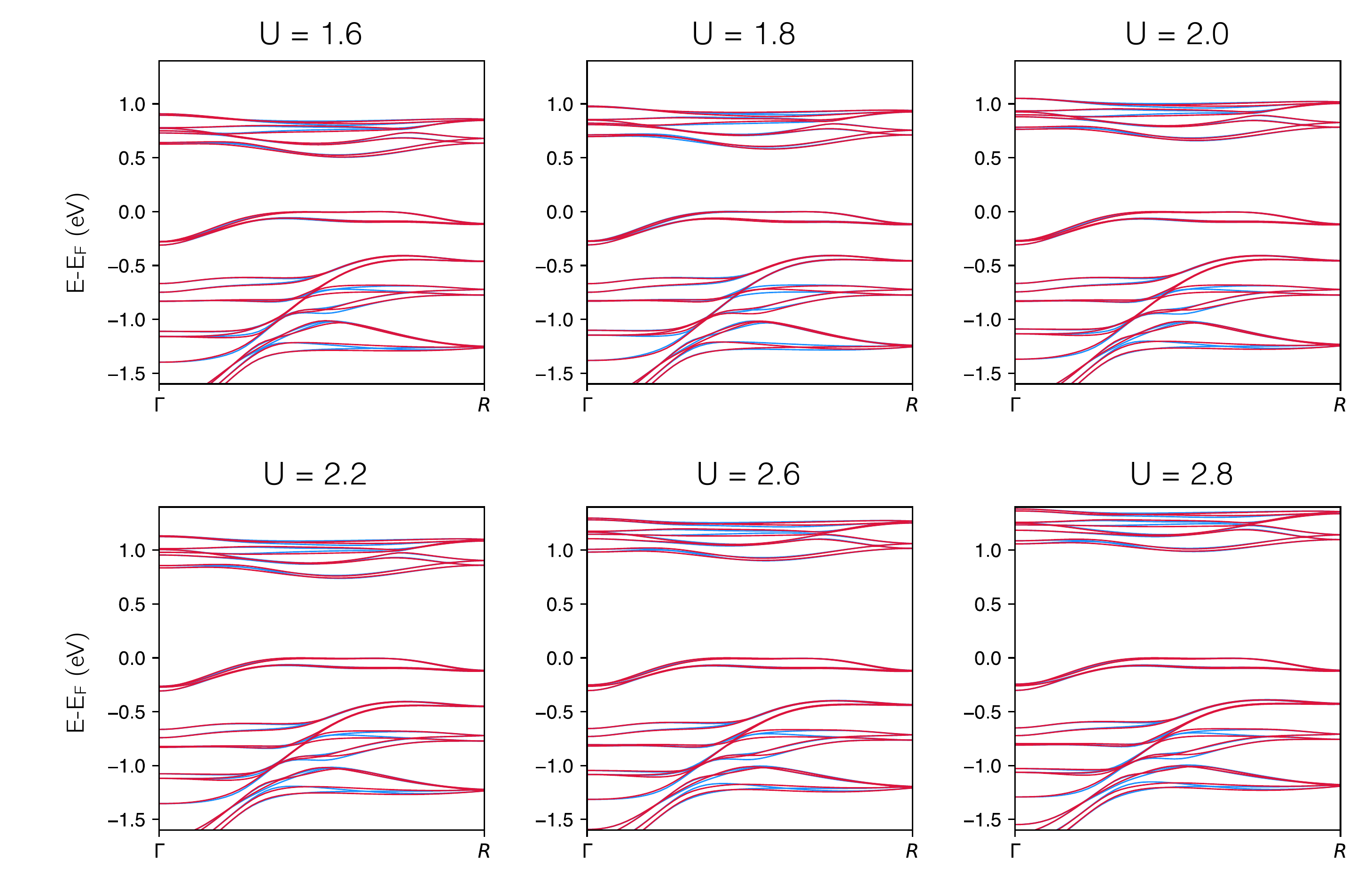}
    \caption{Hubbard $U$ dependent electronic band structures for the G-ALM phase without spin-orbit coupling. The different colors denote the spin-up and spin-down polarized bands.}
    \label{BS_noSOC}
\end{figure}


\begin{figure}[!htb]
    \centering
    \includegraphics[scale=0.24]{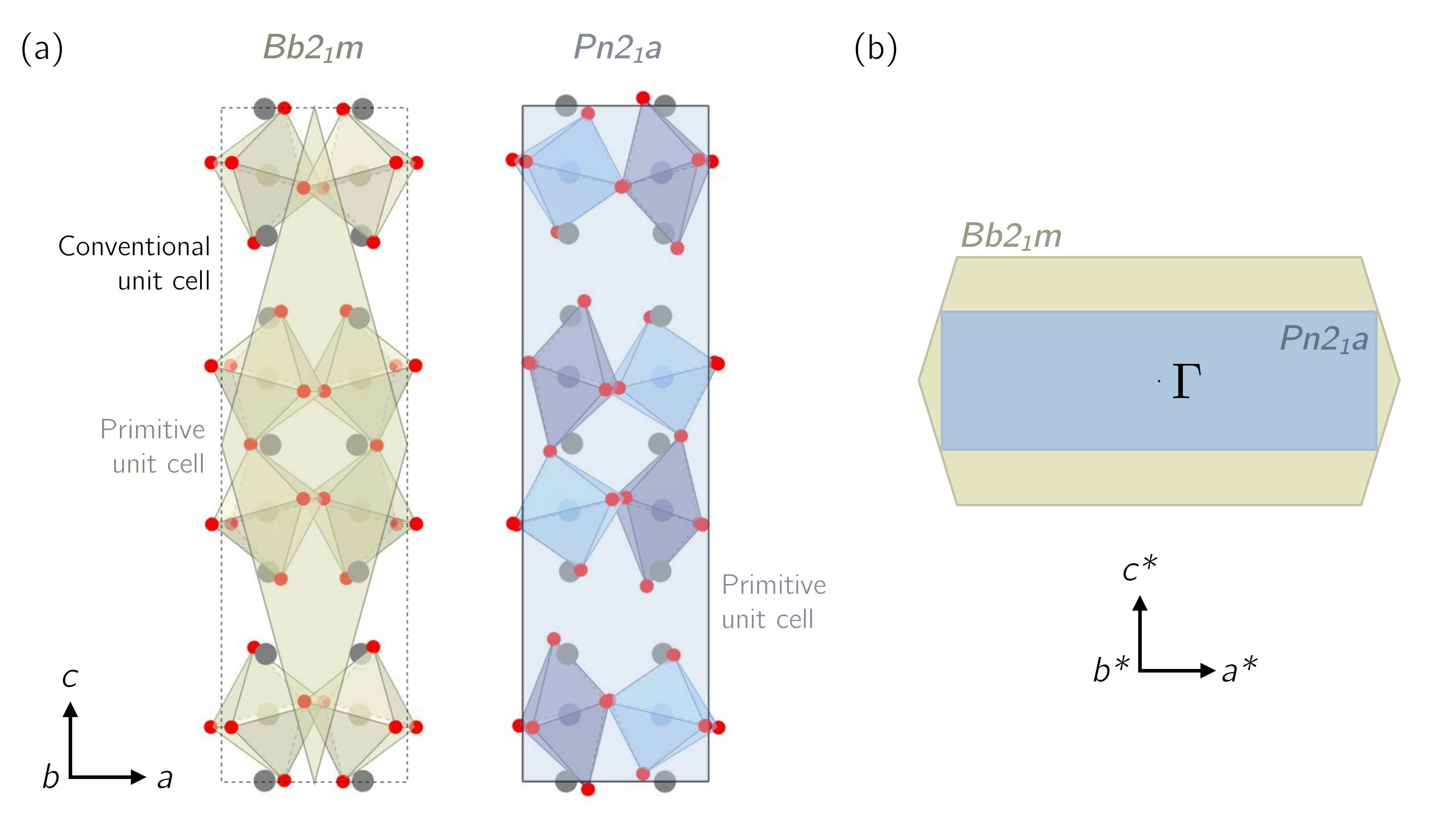}
    \caption{(a) Illustration of the primitive unit cells (shaded region) for the two space groups. The primitive lattice along the crystallographic $b$ direction remains unaffected by the loss of $B$-centering operation, while the unit cell doubles in the plane orthogonal to the $b$ axis. The primitive unit cell of \Bb contains two formula units of \cro, while the primitive \Pn structure contains four formula units. (b) Corresponding Brillouin zones for the two space groups. The Brillouin zone along the $b^{*}$ axis is unaffected by the change of the primitive unit cell, while zone-folding is expected in the plane orthogonal to the $b^{*}$ axis.}
    \label{platt}
\end{figure}

\clearpage

\section*{Supplementary Notes}


\subsection{Sample characterization} \label{sample_char}

\subsubsection{Composition analysis for Ti substituted sample}

\begin{figure}[!htb]
    \centering
    \includegraphics[scale=0.3]{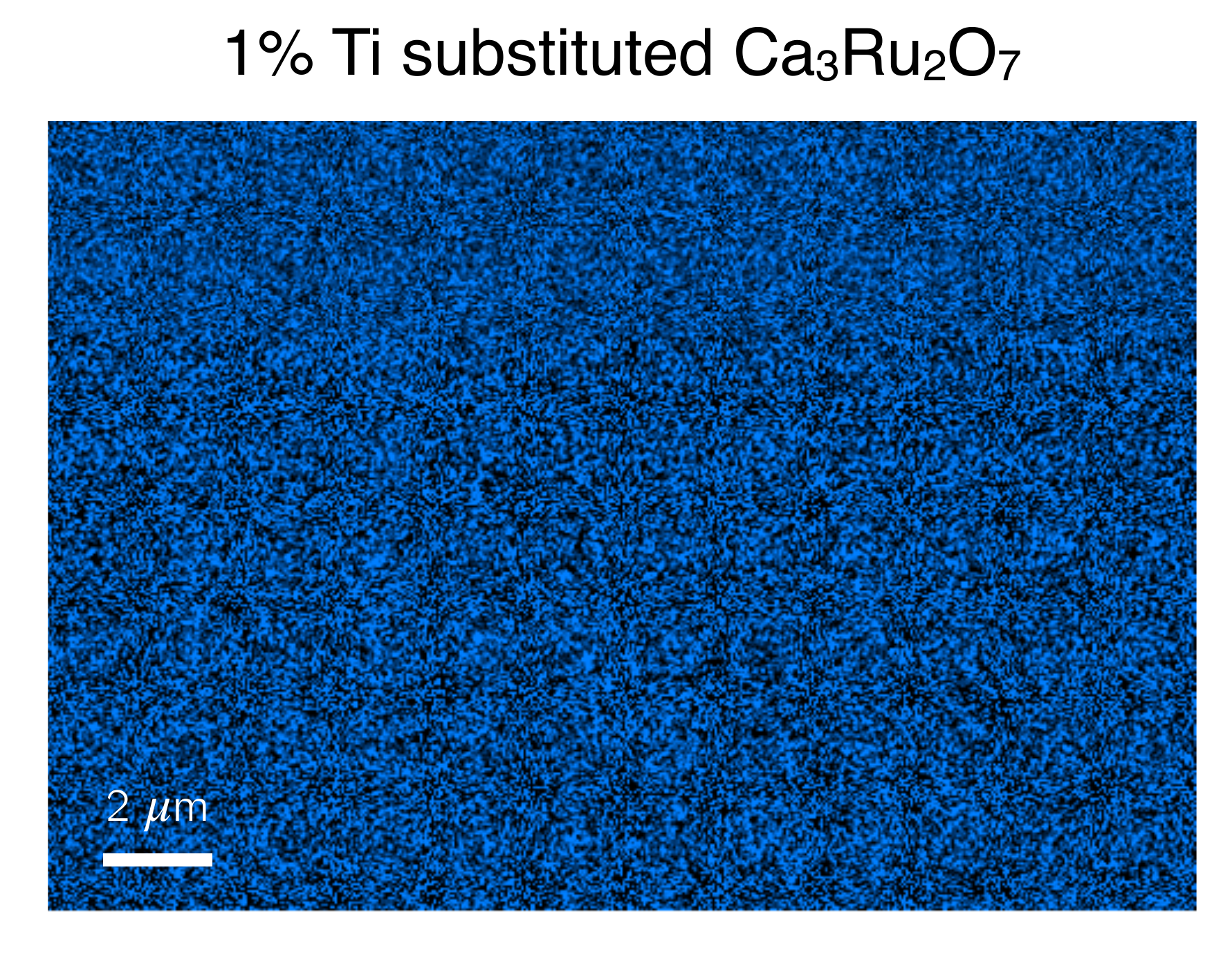}
    \caption{EDS map showing the spatial distribution of Ti atoms for 1\% Ti substitution.}
    \label{edsmap}
\end{figure}

Elemental composition was quantified using energy dispersive X-ray spectroscopy (EDS). EDS spectra were acquired at an accelerating voltage of 15 kV and a probe current of 1.6 nA. Spectra were collected over multiple regions on the sample; the reported atomic percentages represents the spatial average, with uncertainties determined from the standard deviation across the multiple sites. The results are summarized in Table \ref{edstab}. The Ca/(Ru+Ti) ratio $\sim$ 1.5 confirms the stoichiometry of the measured samples. Due to the low energy of the Oxygen K$\alpha$ emission ($\sim$ 0.53 keV), the exact atomic percentages are unreliable. To further confirm the homogeneity of the Ti distribution, EDS mapping was performed and is shown in Fig. \ref{edsmap}. The EDS maps confirm the homogeneous distribution of Ti at micrometer length scales.

\begin{table}[!htb]
\begin{center}
\begin{tabular}{|c|c|}
\multicolumn{2}{c}{\textbf{Atomic percentages (\%)}}     \\
\hline
element           & 1\% Ti substitution  \\
\hline

Ca         & 27.96 $\pm$ 0.12  \\
Ru         & 18.47 $\pm$ 0.09  \\
Ti         & 0.20 $\pm$ 0.03   \\
O          & 53.32 $\pm$ 0.23  \\
\hline
\multicolumn{2}{c}{\textbf{Elemental ratios}}                             \\
\hline
element           & 1\% Ti substitution          \\
\hline
Ca/(Ru+Ti) & 1.497 $\pm$ 0.016 \\
Ti/(Ru+Ti) & 0.011 $\pm$ 0.002 \\
Ru/(Ru+Ti) & 0.989 $\pm$ 0.011 \\
\hline
\end{tabular}
\caption{Quantitative compositional analysis for the 1 \% Ti substituted \cro.}
\label{edstab}
\end{center}
\end{table}

\subsubsection{Lab X-ray characterization of phase purity}

\begin{figure}[!htb]
    \centering
    \includegraphics[scale=0.28]{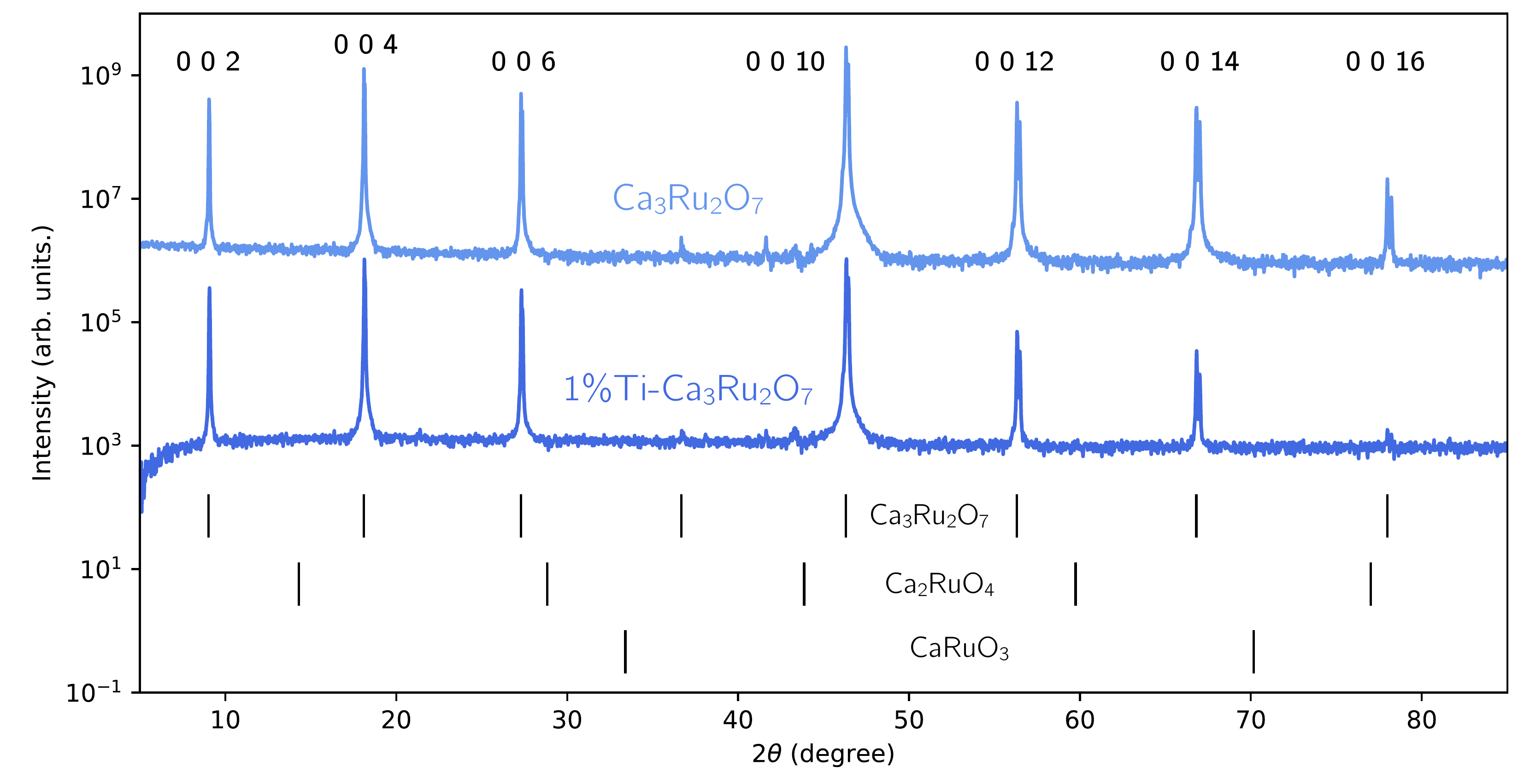}
    \caption{Lab based $\theta - 2 \theta$ X-ray diffraction performed at room temperature using a Cu $K_{\alpha}$ X-ray source. The expected peak positions for \cro and other common impurity phases are marked by dashed lines below the experimental data. }
    \label{th2th}
\end{figure}

The absence of other Ruddlesden-Popper phases was verified using lab based X-ray diffraction. The sample was mounted with the $c$ axis lying along the out-of-plane direction, allowing us to selectively probe the 00$L$ X-ray reflections. Figure \ref{th2th} shows the comparison of the \ticrox ($x = 0, 0.01$) crystals to the expected 00$L$ reflections for CaRuO$_{3}$, Ca$_{2}$RuO$_{4}$ and \cro. Since the ionic radii for Ti$^{+4}$ is close to that of Ru$^{+4}$, the comparison was made with the parent compound. The \ticrox crystals closely match the peaks predicted for \cro, confirming the 327 stoichiometry of the samples.

\subsubsection{Polarized light microscopy}

\begin{figure}[!htb]
    \centering
    \includegraphics[scale=0.3]{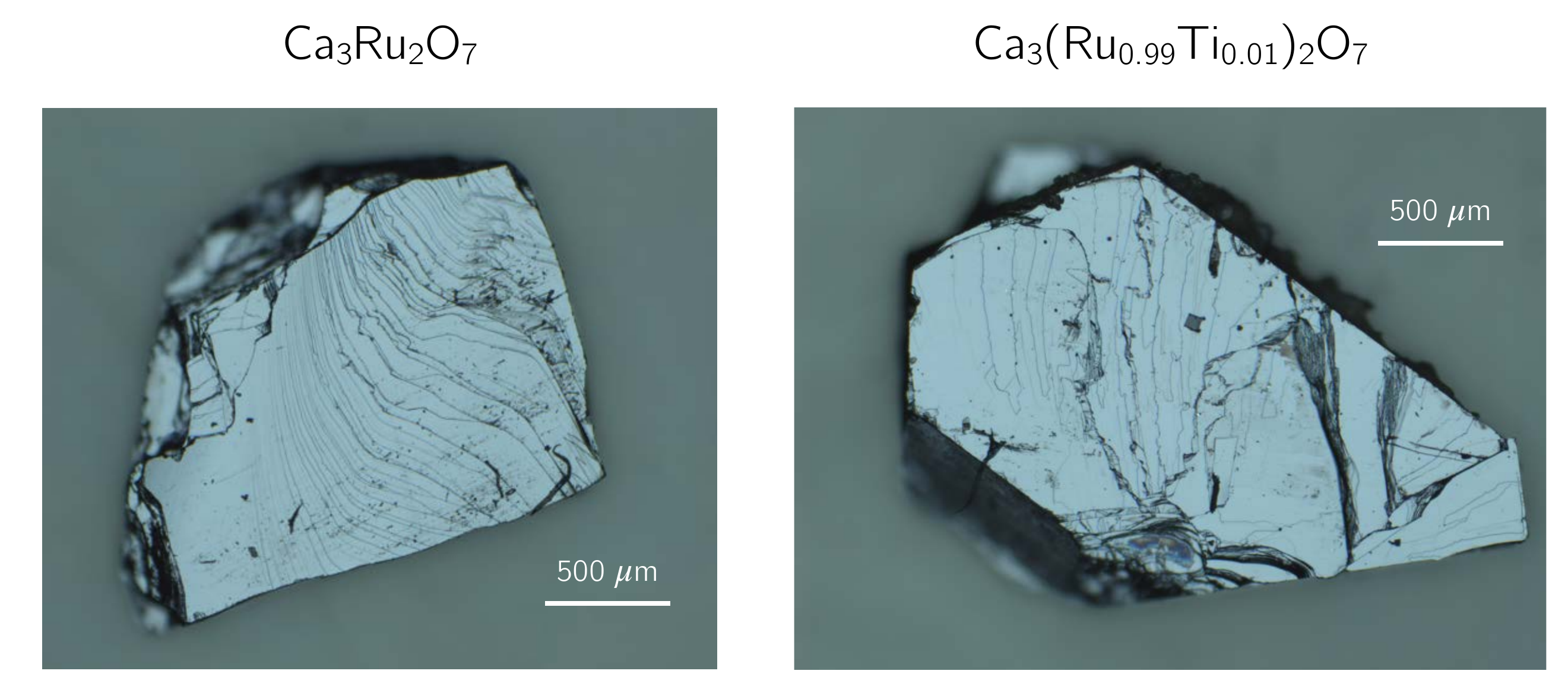}
    \caption{Images of the sample taken under cross-polarized configuration. No visible signs of twinning is observed.}
    \label{plm}
\end{figure}

The presence of twinning was checked using polarized light microscopy under a cross-polarized incident beam and analyzer combination. No clear effect of twin-domains is observed as shown in Fig. \ref{plm}.

\subsubsection{X-ray Laue back-scattering}

\begin{figure}[!htb]
    \centering
    \includegraphics[scale=0.28]{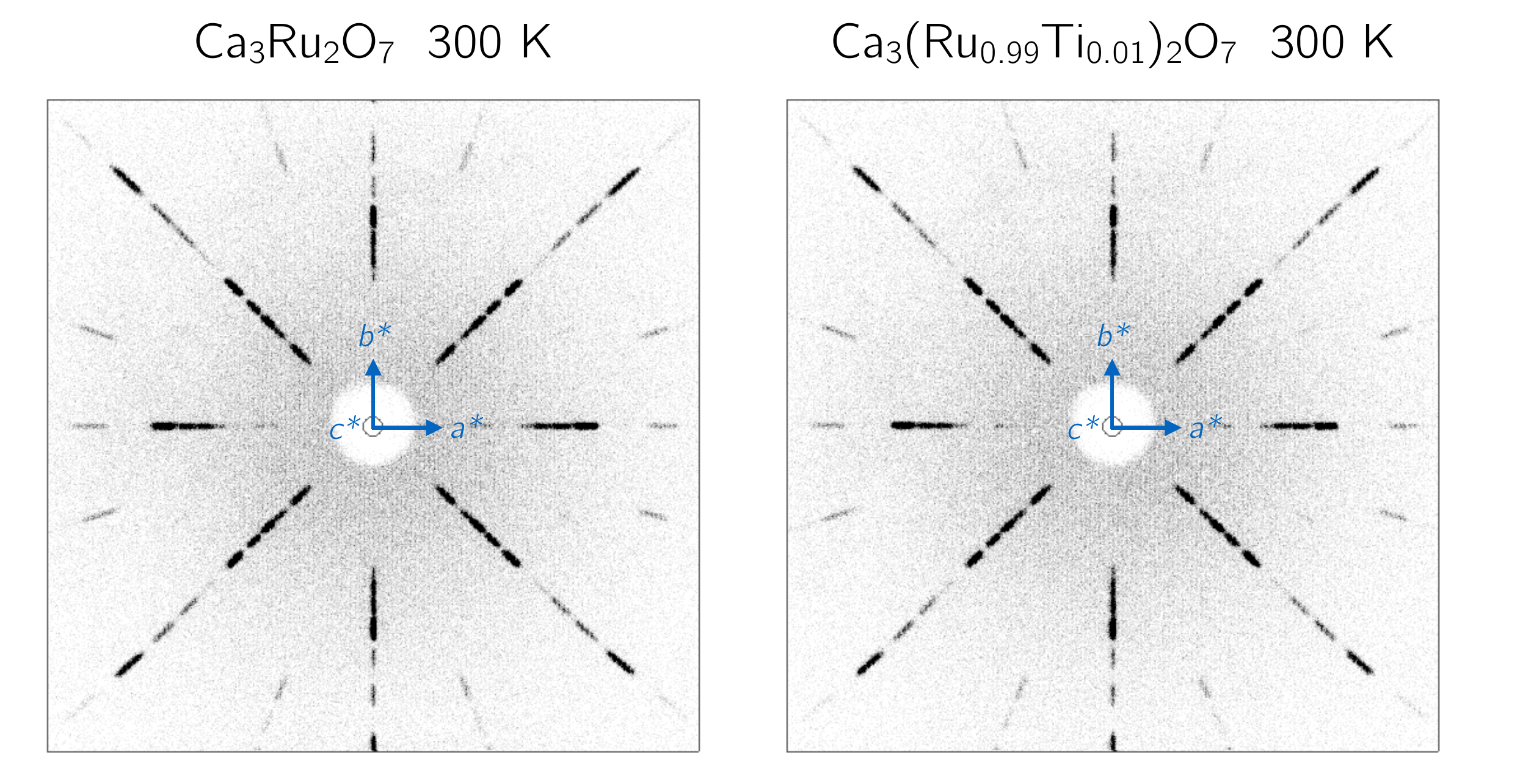}
    \caption{Lab based Laue back-scattering for \cro and \ticro to pre-orient samples and check for twinning.}
    \label{laue}
\end{figure}

To further assess the single domain nature of the grown crystals, Laue back-scattering was performed. We do not identify any visible signatures of twinning in our samples as shown in Fig. \ref{laue}. The samples were also pre-oriented for synchrotron X-ray diffraction using the Laue method. 

\subsubsection{Transport measurements}

\begin{figure}[!htb]
    \centering
    \includegraphics[scale=0.65]{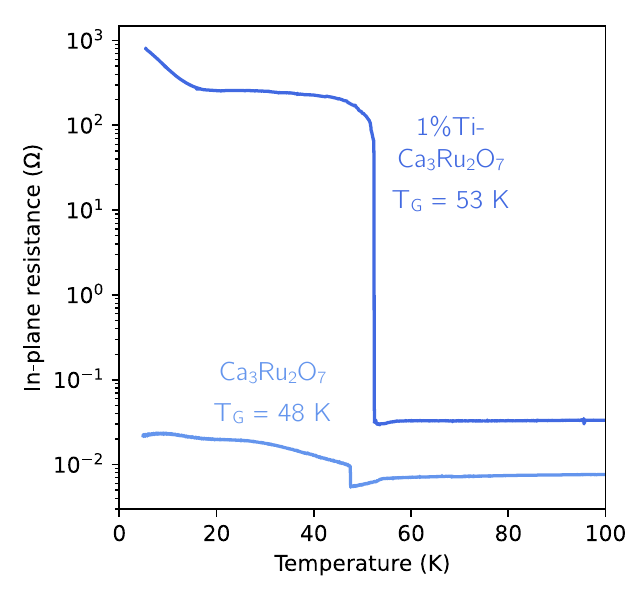}
    \caption{Temperature dependent electrical resistance for the parent \cro as well as the \ticro samples.}
    \label{transport}
\end{figure}

To verify the electronic states, we performed electrical linear transport measurements where the longitudinal resistance along the $b$ axis was measured as in Fig. \ref{transport}. For \cro, the metal-pseudogap transition is observed at $T_{G} = 48$K, while the \ticro sample undergoes a metal-insulator transition at $T_{G}$ = 53 K.

\subsubsection{Magnetometry measurements}

\begin{figure}[!htb]
    \centering
    \includegraphics[scale=0.65]{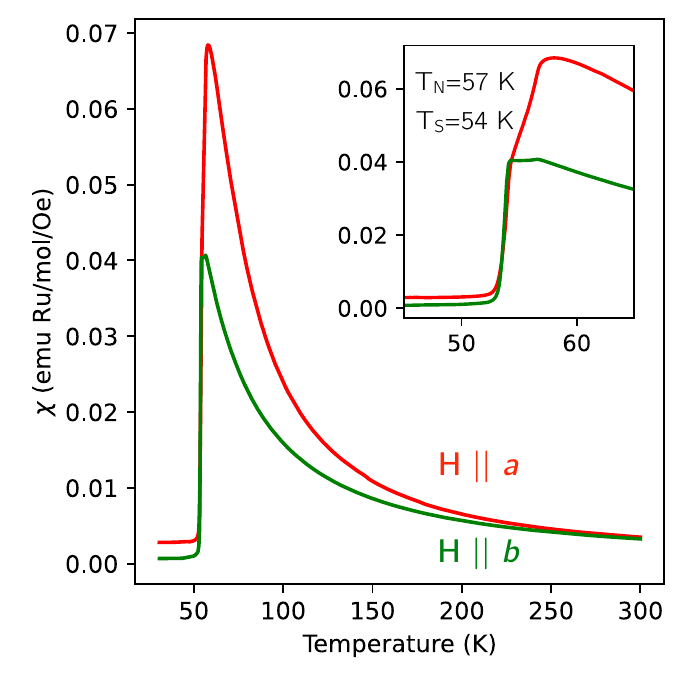}
    \caption{Temperature dependent magnetometry for the \ticro sample.}
    \label{squid}
\end{figure}

Temperature dependent magnetometry measurements were performed to determine the magnetic transition temperatures for the \ticro sample as the magnetic phase transitions are sensitive to the Ti concentration. A magnetic field of 100 Oe was applied along the $a$ and $b$ axes and data was collected under zero-field cooling as shown in Fig. \ref{squid}. The \Neel and spin-reorientation transitions were extracted to be $T_{N} = 57$ K and $T_{S} = 54$ K respectively.  


\subsection{Symmetry adapted distortion mode analysis for a reference independent amplitude bound} \label{samdist}

\begin{figure}[!htb]
    \centering
    \includegraphics[scale=0.25]{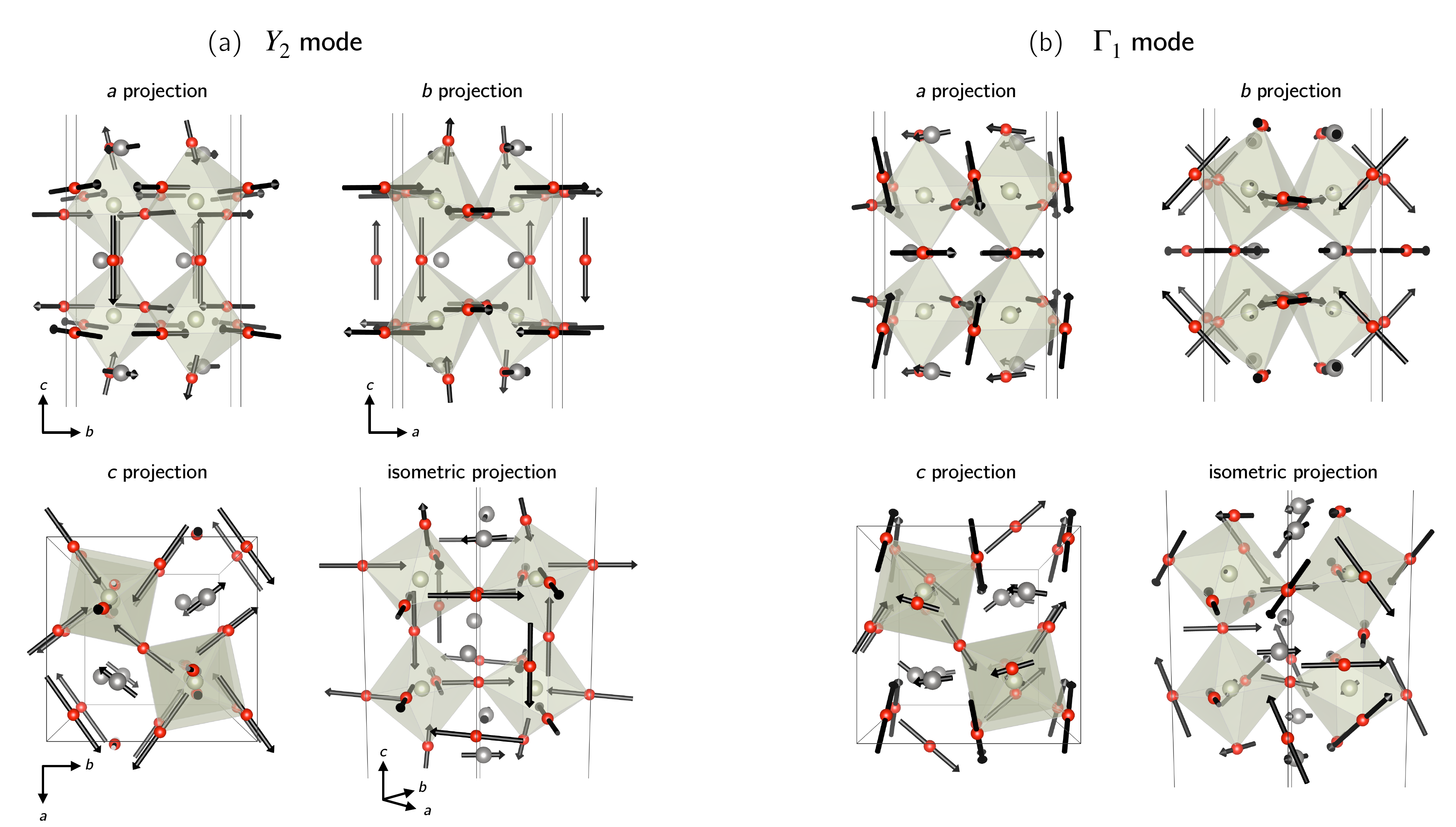}
    \caption{Projections of the (a) $Y_{2}$ and (b) $\Gamma_{1}$ symmetry modes that drives the structural transition from \Bb to \Pn viewed along various crystallographic axes. The $Y_{2}$ mode acts as the primary order parameter for the transition. The arrows represent the atomic displacements and are uniformly scaled for visual clarity.}
    \label{modes}
\end{figure}

A symmetry adapted mode decomposition of the structural transition from \Bb to \Pn performed using AMPLIMODES \cite{amplimodes} shows that the atomic displacements obtained from the DFT relaxed structures can be decomposed into a combination of $Y_{2}$ and $\Gamma_{1}$ modes. We use the $U=1.2$ structure as the reference parent structure (\Bb) while the structures obtained for higher $U$ values are treated as the distorted child structures. The zone boundary $Y_{2}$ mode constitutes the primary symmetry breaking order parameter of the transition and is responsible for the loss of the fractional lattice translation $B$ thereby doubling the primitive unit cell (Fig. \ref{platt}). The zone center $\Gamma_{1}$ mode does not break any additional symmetries beyond what $Y_{2}$ already breaks. The atomic displacement patterns of the two modes are shown in Fig. \ref{modes}. \\

\begin{figure}[!htb]
    \centering
    \includegraphics[scale=0.3]{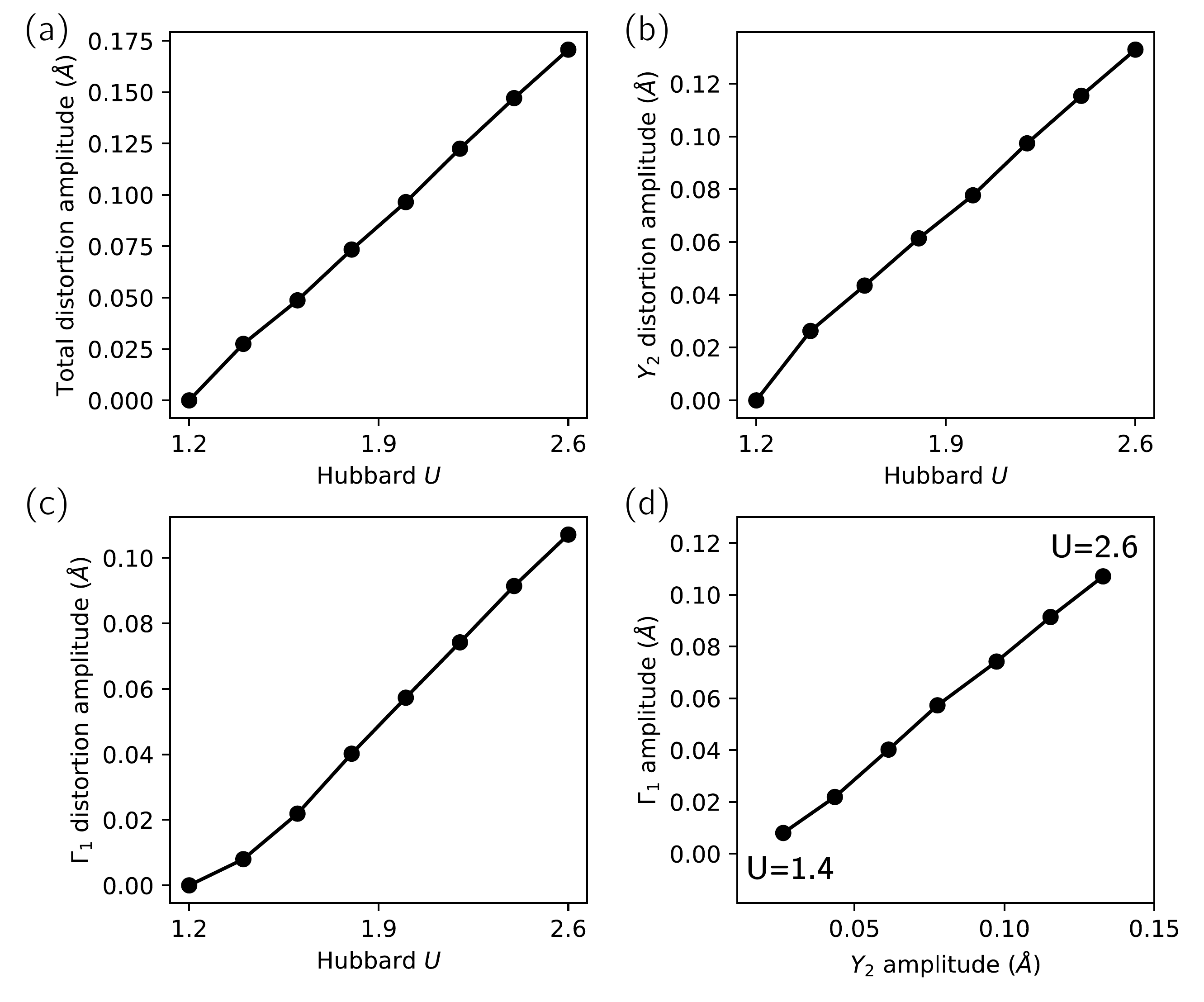}
    \caption{Extraction of the (a) total distortion amplitude and the amplitudes of the (b) $Y_{2}$ and (c) $\Gamma_{1}$ symmetry modes as the Hubbard $U$ is varied. The structure for $U=1.2$ (\Bb) is chosen as the parent structure. The observed linearity supports the use of a linear interpolation scheme between the distorted structures. (d) Relation between the amplitudes of the $Y_{2}$ and $\Gamma_{1}$ modes. The linear dependence justifies the use of a single scalar $\lambda$ to parameterize the distortion as done in the main text.}
    \label{ampli}
\end{figure}

Figs. \ref{ampli}a-c demonstrate that the total distortion amplitude as well as the amplitudes of the $Y_{2}$ and $\Gamma_{1}$ symmetry modes increase smoothly and monotonically with the Hubbard $U$ between $U=1.2$ and $U=2.6$ eV. Thus, the Hubbard $U$ acts as an effective control parameter for the overall magnitude of the distortion. To justify the use of a single scalar parameter $\lambda$ to characterize the distortion amplitude at fixed $U$, we recall the definition given in Sec. III. C, $\vb{r}_{\lambda} = \vb{r}_{B} + \lambda (\vb{r}_{P} - \vb{r}_{B})$ where $\vb{r}_{B}$ and $\vb{r}_{P}$ denote the atomic positions in the \Bb ($U$ = 1.2) and \Pn (reference $U$) unit cells, respectively. Rather than allowing the $Y_{2}$ and $\Gamma_{1}$ mode amplitudes to vary independently, we examine their relative weights across the explored $U$ range. As seen in Fig. \ref{ampli}d, the amplitudes of the $Y_{2}$ and $\Gamma_{1}$ modes are strongly coupled and are linearly related. This justifies the use of a scalar $\lambda$ used to treat the distortion. In addition, the $\lambda$ scaling assumed a linear interpolation between the structures, which is validated in Figs. \ref{ampli}a-c where the amplitudes are linearly related to the Hubbard $U$. While both $U$ and $\lambda$ (defined relative to a specific reference $U$) can be used to independently control the distorted structure, the $Y_{2}$ symmetry mode amplitude is the primary order parameter and represents a physical, reference invariant measure of the distortion. A mapping between the $U$ and $\lambda$ values to the $Y_{2}$ mode amplitude is hence needed. \\

\begin{figure}[!htb]
    \centering
    \includegraphics[scale=0.27]{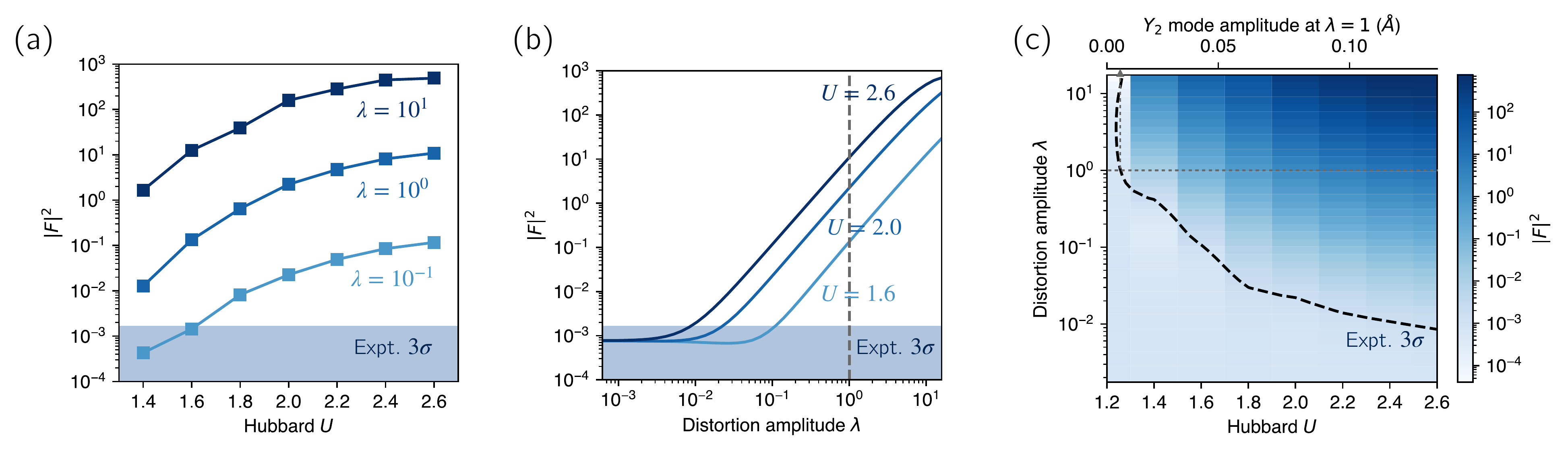}
    \caption{Dependence of the structure factor squared $|F|^{2}$ for the 2 2 7 reflection obtained by varying $U$ and $\lambda$ to control the distortion amplitude. (a) Evolution of $|F|^{2}$ as a function of $U$ for selected $\lambda$ values. (b) Variation of $|F|^{2}$ with $\lambda$ for representative $U$ values. (c) Map of the calculated $|F|^{2}$ as a function of both $U$ and $\lambda$. $\lambda$ is always defined relative to the particular $U$ value. The dashed contour represents the experimental $3 \sigma$ bound and corresponds to the same distorted structure that can be realized by parameterization with multiple $U$ and $\lambda$ combinations. The secondary axis at the top maps the $Y_{2}$ mode distortion to the $U$ value at $\lambda=1$, allowing for a reference-independent physically meaningful bound.}
    \label{lambdaU}
\end{figure}

We now examine how the calculated structure factor squared $|F|^{2}$ for a representative \Pn unique reflection depends on the two descriptors $U$ and $\lambda$. We choose the 2 2 7 reflection for this case. Fig. \ref{lambdaU}a shows the $|F|^{2}$ as a function of $U$ at fixed $\lambda$ values (defined relative to each $U$). For a fixed $\lambda$, $|F|^{2}$ increases monotonically with $U$, with larger distortions (higher $\lambda$) resulting in higher $|F|^{2}$. The complementary cut of $|F|^{2}$ dependence on $\lambda$ for select $U$ values is shown in Fig. \ref{lambdaU}b, reproducing the characteristic increase as seen in Figs. 3c and 5c of the main text and confirming that this behavior is representative across the full $U$ range. Crucially, the derived bounds from the experimental $3 \sigma$ threshold depends on both, $U$ and $\lambda$ (which is defined relative to the particular $U$ value chosen). Fig. \ref{lambdaU}c combines both the $U$ and $\lambda$ dependencies to a single map of $|F|^{2}$ across a range of $U$ and $\lambda$ values. The contours represent a fixed value of $|F|^{2}$ across the ($U$, $\lambda$) plane and essentially describe the same distorted structure. In other words, different reference structures (obtained at different $U$) scaled by different amplitudes $\lambda$ can reproduce the same net distortion and hence the calculated $|F|^{2}$. The dashed contour in Fig. \ref{lambdaU}c marks the experimental $3 \sigma$ threshold. For any $U$ value, the corresponding $\lambda$ value on this contour is the upper bound our measurement places on the distortion amplitude of the reference structure obtained at that $U$. The region above and to the right of the contour is excluded by the absence of a detectable signal in our data. \\

To obtain a bound on the physical $U$-independent order parameter amplitude, we use the top axis of Fig. \ref{lambdaU}c which gives the $Y_{2}$ mode amplitude at $\lambda=1$ for each $U$. Since the contour represents identically distorted structures, the appropriate $Y_{2}$ mode amplitude characterizing that structure can be read directly from the top axis at the $U$ value where the contour crosses $\lambda=1$. This allows us to re-express our experimentally derived bounds directly in terms of the $Y_{2}$ mode amplitude, removing the dependence of the quoted bound on the chosen reference structure. For the parent compound, the $3 \sigma$ threshold (as shown in Fig. \ref{lambdaU}c) translates to a $Y_{2}$ amplitude bound of around 0.008 \AA, independent of the particular $U$ used to generate the reference \Pn structure. The same method can be applied to other \Pn unique reflections to obtain similar physically meaningful order parameter bounds. 


\subsection{Verification of the parent \Bb space group through second harmonic generation} \label{shgsec}

Our X-ray measurements are designed to test the \Pn distortion by targeting specific reflections that distinguish it from the parent \Bb phase. We do not, however, directly determine the crystal symmetry of the parent as this requires access to a larger set of reflections than accessible with the diffraction geometry and energy. The structural symmetry of \cro and \ticrox has been previously determined by several independent groups in both powder and single crystal samples using X-ray and neutron probes \cite{yoshida05, resonantxray, dashwood23, rashba24, nscat, cycloid2, petkov23}. All structural studies so far assign \cro to the space group \Bb (number 36), which have also been supported by DFT calculations \cite{spatialticro, lda}. However, the DFT study by Puggioni et al. \cite{puggioni20} suggested a possible symmetry lowering transition to \Pn (space group 33) at low temperatures, whose structure remains experimentally unconfirmed for more than half a decade since. This low symmetry structure enables altermagnetism in \cro, and combined with its polar character, allows for electric field and strain control of altermagnetic functionality. Hence, our diffraction search for the low temperature \Pn structure was a targeted test of the DFT prediction and not an exhaustive symmetry determination. \\

\begin{figure}[!htb]
    \centering
    \includegraphics[scale=0.3]{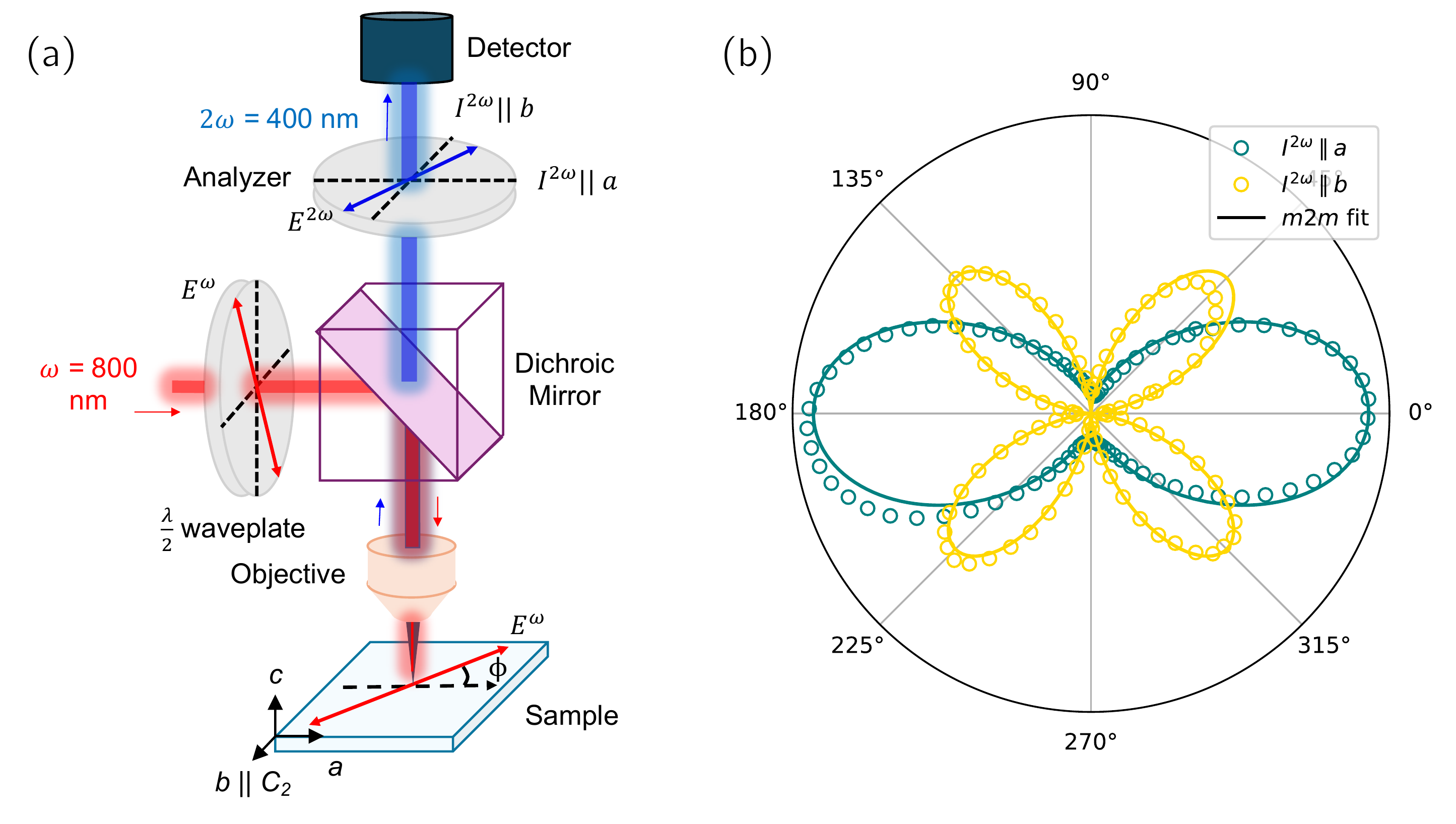}
    \caption{(a) Schematic of the second harmonic generation (SHG) setup. The $\lambda/2$ waveplate continuously changes the polarization of the incident light on the sample (denoted by $\phi$) while an analyzer with pass axis along the crystallographic $a$ and $b$ directions allows for the selection of the particular polarization of the generated second harmonic light. (b) SHG intensity as a function of $\phi$ for the two analyzer configurations. The solid line denotes the fit assuming $m2m$ point group.}
    \label{shg}
\end{figure}

While all our measured reflections are consistent with the \Bb space group (number 36), we identify alternate space groups $Bbm2$ (40) and $Bbmm$ (63) that satisfy the same reflection conditions as that of \Bb. As a result, diffraction alone cannot conclusively determine the space group in this case. Complementary second harmonic generation (SHG) measurements can be used to differentiate between the candidate structures since they show different responses depending on the point group as well as the measurement geometry. Since the sample surface normal is parallel to $c$ axis ($a$ and $b$ axes lie on the surface plane), $Bbm2$ whose 2-fold polar axis is oriented along the $c$ direction is not expected to produce any SHG response at normal incidence. In contrast, \Bb is expected to yield an SHG response under similar measurement conditions. Following the procedure outlined in Ref. \cite{lei18}, we perform SHG measurements at room temperature and find normal incidence SHG signal that agrees with $m2m$ point group as shown in Fig. \ref{shg}. The SHG polarimetry, combined with the limited reflection set from our synchrotron X-ray measurements, clearly points towards the \Bb structure as the parent crystalline phase. Detailed derivations of the fitting expressions may be found in \cite{lei18}.


\subsection{Ferroelastic twinning artifacts and identification of twin-insensitive \Pn unique reflection conditions} \label{multidomain}

\begin{figure}[!htb]
    \centering
    \includegraphics[scale=0.3]{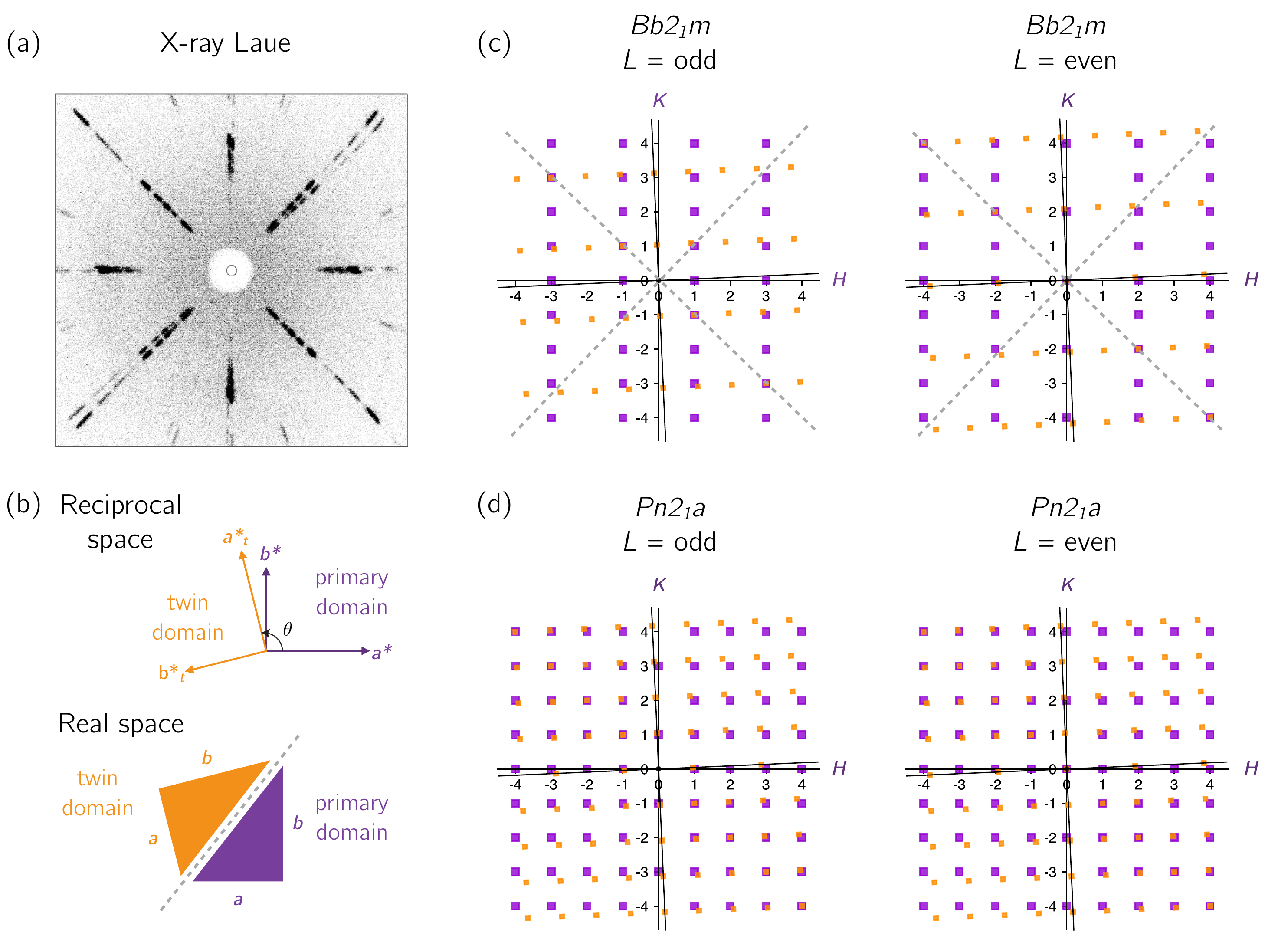}
    \caption{(a) X-ray Laue pattern of a $ab$ twinned crystal. (b) Geometric relation between the two domains in real and reciprocal space. $HK$ cuts illustrating the symmetry allowed reflections for (c) \Bb and (d) \Pn space groups overlaid with the twin contribution. The colors represent the reflections coming from the primary (purple) and the twin (orange). The diagonal grey dashed lines represent $|H| = |K|$.}
    \label{md_RSM}
\end{figure}

Owing to its ferroelastic nature, \cro exhibits a strong propensity to form structural domains to minimize its free energy. These polar domains appear as $\sim$90\de twins in the $ab$ plane and 180\de anti-phase domains along the in-plane $ab$ and out-of-plane $c$ directions \cite{lei18, stone19}. The $ab$ plane twinning is easily visualized by X-ray Laue back-scattering as shown in Fig. \ref{md_RSM}a while the 180\de anti-phase domains show no X-ray contrast but instead disrupt the long-range crystalline order and lead to peak broadening in the reciprocal space. Fig. \ref{md_RSM}b shows the schematic real space domain configuration consistent with the X-ray Laue. The crystallographic axes of the twin are rotated by $\theta$ with respect to the axes of the primary domain where $\theta$ is related to the in-plane lattice parameters and is given by $\theta = 180^{\circ} - 2 \tan^{-1}(a/b)$. The rotation of the real space crystallographic axes implies a rotation of the reciprocal space directions as illustrated in Fig. \ref{md_RSM}b. The presence of such twin domains can, in some cases, impede our X-ray survey for signals associated with the \Pn distortion as contributions from the twin domains may introduce additional signal. Although care was taken to select crystals free of twinning using polarized light microscopy and lab based X-ray Laue (Supplementary Note \ref{sample_char} sections 3,4), our high flux synchrotron X-ray measurements revealed a small residual fraction of twin domains ($\approx$ 0.3\%). \\  

To illustrate how the twin domains can complicate the identification of signatures associated with the \Pn distortion, we plot the $HK$ reciprocal space cuts for the two candidate space groups as shown in Figs. \ref{md_RSM}c,d. The purple points represent the symmetry allowed reflections for the primary domain, which is superimposed with the orange points representing the reciprocal lattice points from the twin domain obtained by rotating the reciprocal lattice. Consider the reflections $HKL$ where $|H| \neq |K|$ such as is the case for the reflections $1 \ 0 \ 1$ and $0 \ \bar{1}\ 1$; the former is allowed in both the \Bb and \Pn structures, while the latter is permitted only in \Pn. Due to the similarity of the in-plane lattice constants $a \approx b$, the reciprocal space in the vicinity of $0 \ \bar{1} \ 1$ may contain intensity arising from the allowed $1 \ 0 \ 1$ reflection of a twin domain, thereby obscuring the unambiguous identification of a true \Pn signal. In other words, the two twin domains combine to give a ``twin integrated systematic absence'' that is distinct from those of either domains individually. For a general reflection $HKL$ with $|H| \neq |K|$, this twin integrated systematic absence can render a reflection forbidden in the primary domain alone allowed once the twinning is accounted for. \\

In contrast, for reflections $HKL$ satisfying $|H| = |K|$ as marked by the dashed line in Fig. \ref{md_RSM}c, the twin-related contribution arises from $K \bar{H} L$. Both these reflections satisfy identical selection rules, and hence are either simultaneously allowed or simultaneously forbidden. Put differently, for reflections where $|H| = |K|$, the twin integrated systematic absence is identical to the systematic absence for the individual domains. Consider the specific case of $1 \bar{1} L$ reflections, which for the \Bb structure is allowed for odd $L$ and forbidden for even $L$, and is permitted for all $L$ in \Pn. Thus, any observed peak at even $L$ cannot be an artifact of twinning and must correspond to a genuine symmetry breaking \Pn distortion. Hence reflections satisfying $|H| = |K|$ are a robust test of the \Pn distortion even in the presence of the residual multidomain character of our crystals, and we thus restrict our analysis to reflections satisfying this criterion. \\  

\begin{figure}[!htb]
    \centering
    \includegraphics[scale=0.23]{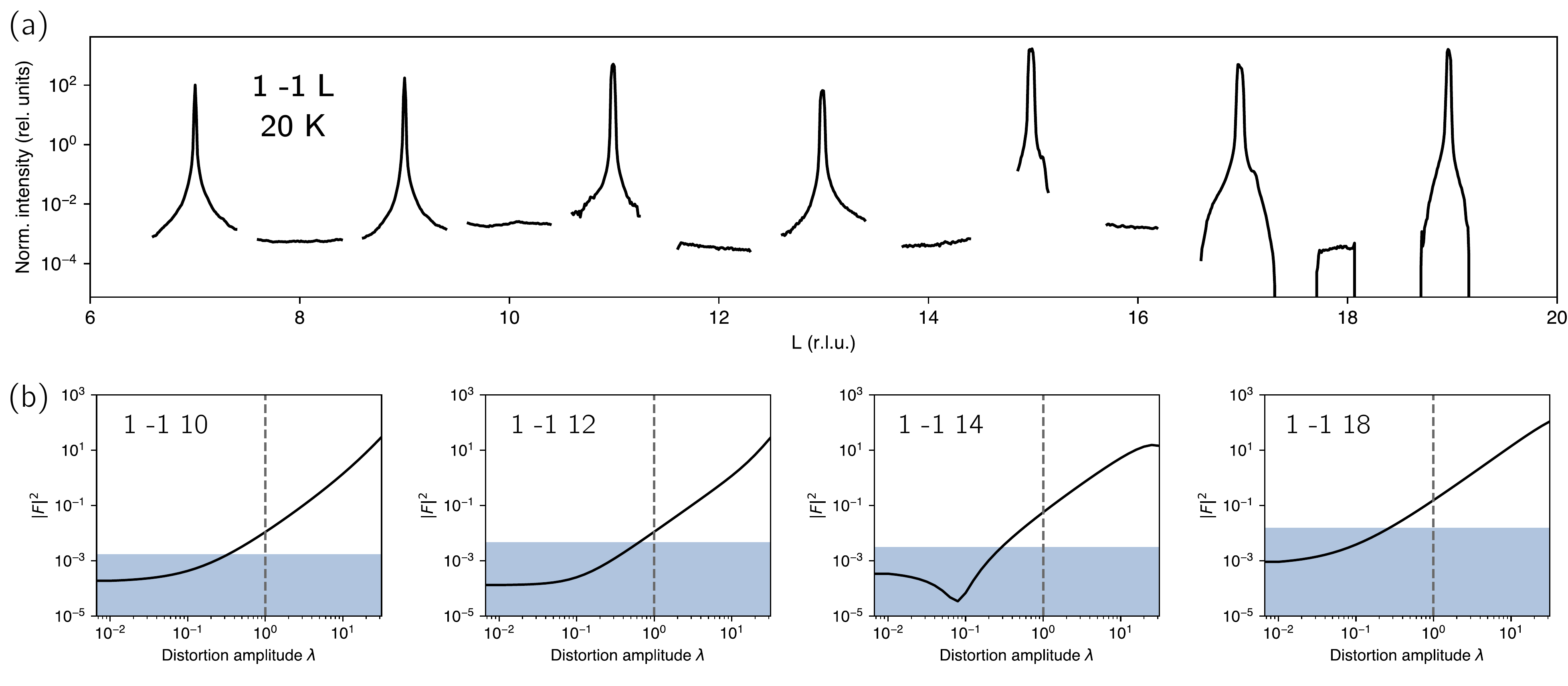}
    \caption{(a) Several additional reflections of the type $1 \ \bar{1} \ L$ measured on \cro at 20 K. The $1 \ \bar{1} \ 2n$ reflections are forbidden for \Bb even in the presence of twinning but allowed for \Pn. No indication of any intensity higher than the background is observed at the \Pn unique reflection conditions. (b) The distortion amplitude estimated from the null signal (horizontal shaded region) is smaller than the value predicted by DFT (vertical line). The distortion bounds for 1 $\bar{1}$ 8 and 1 $\bar{1}$ 16 are presented in the main text Fig. 3c. Although the distortion is expected to lower the symmetry in general, we attribute the non-monotonic behavior for the 1 $\bar{1}$ 14 reflection to an ``accidental symmetry'' at an intermediate distortion amplitude, where the structure transiently acquires a higher symmetry than the parent \Bb phase, resulting in the observed local minimum. }
    \label{11L_all}
\end{figure}

Fig. \ref{11L_all}a shows the $L$ profiles of reflections belonging to the $1 \ \bar{1} \ L$ family spanning a large $L$ range from 7 to 19. Sharp Bragg peaks are observed at all odd $L$ ($L = $ 7, 9, 11, 13, 15, 17, 19), while no measurable intensity above the background is detected at any of the \Pn unique even $L$ positions ($L = $ 8, 10, 12, 14, 16, 18). This extended survey is consistent with the absence of any \Pn signature reported in the main text (Fig. 2b) and substantially broadens the twin unambiguous reflection set across which no evidence for the \Pn phase is observed. \\

\begin{figure}[!htb]
    \centering
    \includegraphics[scale=0.44]{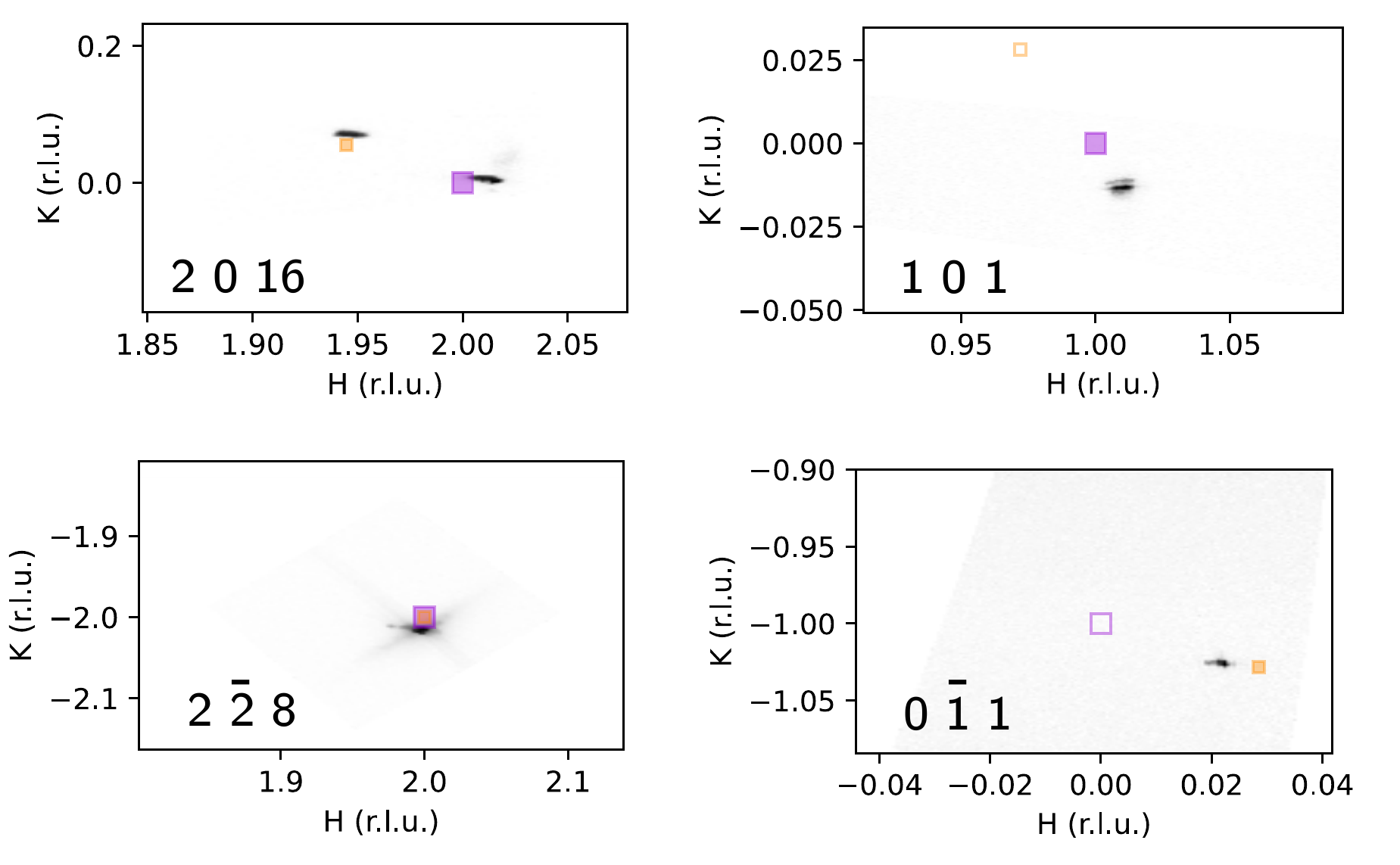}
    \caption{Experimental $HK$ reciprocal space cuts at 20 K. The purple and orange squares represent the calculated peak positions of the primary domain and the twin domain respectively. The hollow squares for the $1 \ 0 \ 1$ and $0 \  \bar{1} \ 1$ reflection conditions represent that the particular reflection is not symmetry allowed by the \Bb structure.}
    \label{md_expt}
\end{figure}

Having established the absence of a \Pn signal along a twin unambiguous reflection line, we now attempt to directly test the twin domain attribution for the specific case of the $0 \ \bar{1} \ 1$ reflection discussed earlier. We compare the experimentally measured peak position at 20 K to the calculated positions of both the primary reflection and its twin domain counterpart, marked by the purple and orange squares respectively in Fig. \ref{md_expt}. The reciprocal space is indexed with respect to the primary domain, and the calculation of the twin domain peak positions is based on the experimental in-plane lattice parameters. Concretely, a primary domain peak at $H \vb{a}^{*} + K \vb{b}^{*} = (H,K)$ might have an additional peak due to the twin at $K \vb{a}^{*}_{t} - H \vb{b}^{*}_{t} = K (\cos \theta \vb{a}^{*} + \frac{b}{a} \sin \theta \vb{b}^{*}) - H (- \frac{a}{b} \sin \theta \vb{a}^{*} + \cos \theta \vb{b}^{*}) = (K \cos \theta + H \frac{a}{b} \sin \theta) \vb{a}^{*} + (K \frac{b}{a} \sin \theta - H \cos \theta) \vb{b}^{*} = (K \cos \theta + H \frac{a}{b} \sin \theta, K \frac{b}{a} \sin \theta - H \cos \theta)$ where $\theta = 180^{\circ} - 2 \tan^{-1}(a/b)$ as defined before. \\  

Both the $2$ $0$ $16$ and the $2 \ \bar{2} \ 8$ reflections are symmetry allowed for \Bb and \Pn structures and are expected to feature an additional contribution in the presence of twinning. The observed $HK$ cuts confirm this, where reciprocal space in the vicinity of $2 \ 0 \ 16$ shows an additional peak from the twin, while the twin contribution at $2 \ \bar{2} \ 8$ overlaps with the primary reflection as suggested by Fig. \ref{md_RSM}. This confirms the accuracy of our orientation matrix and peak position calibration. Further, the $1 \ 0 \ 1$ reflection (symmetry allowed for both \Bb and \Pn) coincides with the calculated position for the primary domain. In contrast, for the $0 \ \bar{1}\ 1$ reflection (forbidden in \Bb but allowed in \Pn), the observed peak does not coincide with the calculated position of the primary domain but instead coincides with the position expected for the twin. In addition, no diffuse scattering is detected at the $0 \ \bar{1} \ 1$ calculated position. This directly demonstrates that the intensity observed at the $0 \ \bar{1} \ 1$ reflection condition originates from the allowed $1 \ 0 \ 1$ reflection of a residual twin domain rather than a true symmetry breaking \Pn distortion. \\

\begin{figure}[!htb]
    \centering
    \includegraphics[scale=0.24]{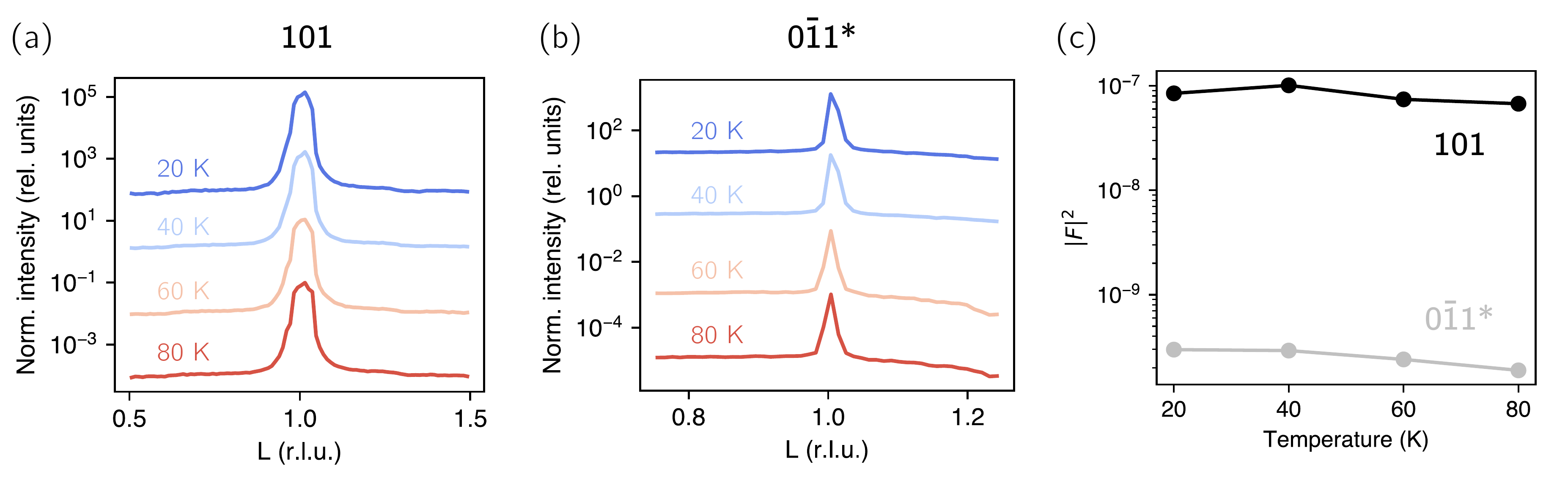}
    \caption{Temperature dependence of the (a) $1 \ 0 \ 1$ and (b) $0 \ \bar{1} \ 1$ reflection conditions. The $1 \ 0 \ 1$ peak is allowed for both \Bb and \Pn structures, while the $0 \ \bar{1} \ 1$ is expected only for the \Pn at low temperatures. However, a peak at the $0 \ \bar{1} \ 1$ reflection condition is observed across 20 K to 80 K due to the multidomain character of the sample. (c) Extracted structure factor squared as a function of temperature for the two reflection conditions show similar behavior, with no anomaly around the expected transition temperature, confirming the extrinsic origin of the measured signal. The ratio of the squared structure factors $|F|^{2}_{101}/|F|^{2}_{0\bar{1}1}$ is related to the domain fraction, which is estimated to be $\approx$ 0.3 \%.}
    \label{md_temp}
\end{figure}

The attribution of the intensity at $0 \  \bar{1} \ 1$ to a twin domain is further corroborated by the temperature dependence of the peak intensities observed at the $1 \ 0 \ 1$ and $0 \  \bar{1} \ 1$ reflection conditions as shown in Fig. \ref{md_temp}. An order parameter associated with the \Pn transition at $T_{S} = 48$ K is expected to show a critical onset below this temperature with no intensity at temperatures higher than $T_{S}$. Instead, the intensity at $0 \  \bar{1} \ 1$ shows a weak temperature dependence between 20 K and 80 K, closely tracking the intensity of the always allowed $1 \ 0 \ 1$ reflection. The $HK$ maps, along with the lack of an anomaly at $T_{S}$ provide independent and consistent evidence that the $0 \  \bar{1} \ 1$ is of extrinsic origin. We thus conclude that no genuine diffraction signature of the \Pn phase is observed across all our measurements. 


\subsection{Relative contributions of various atoms towards X-ray intensity} \label{F2Ca}

\begin{figure}[!htb]
    \centering
    \includegraphics[scale=0.6]{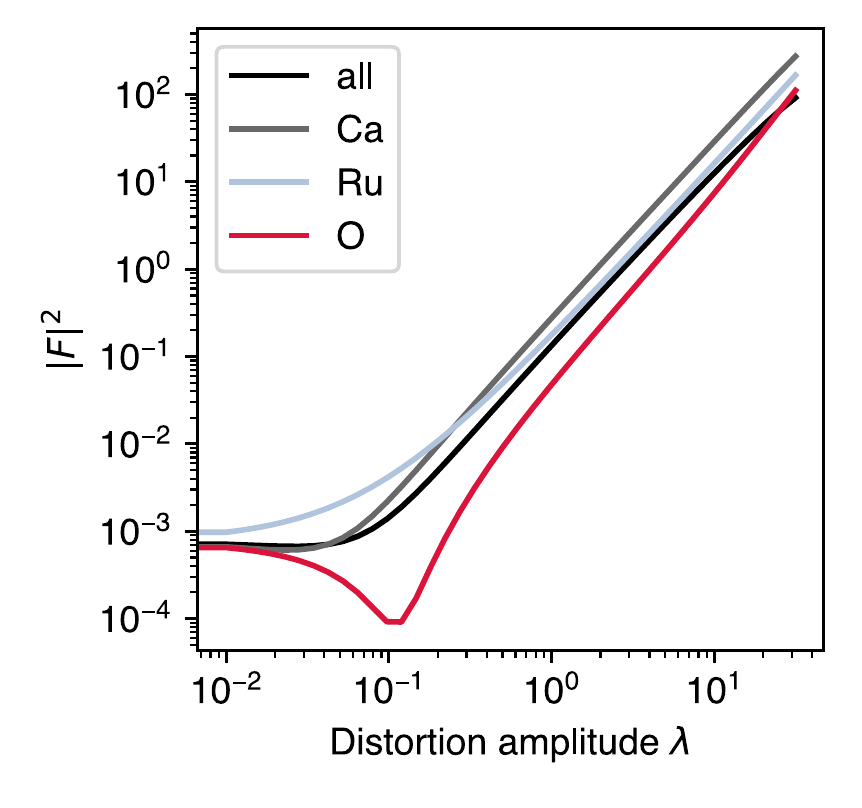}
    \caption{Variation of the square of the structure factor for the 2 2 7 reflection calculated as a function of the distortion amplitude $\lambda$ when only the mentioned atoms are allowed to move.  The Ca atom motions contribute most strongly to the X-ray intensities.}
    \label{CaF2}
\end{figure}

 While generating the scaled structures, we can selectively turn off the displacements of atoms, keeping their coordinates fixed to the \Bb structure for the structure factor calculations. This is necessary to estimate their relative contributions to X-ray squared structure factor. In the following, we scale the Ca, Ru, and O atoms one at a time, to determine which atom contributes most strongly to the squared structure factor and hence X-ray intensity. As shown in Fig. \ref{CaF2}, we find the X-ray intensities are dominated by the Ca atom displacements. \\

 These calculations are also performed for the usual case where all atoms are allowed to displace and is included for comparison. The intensity is given by

\begin{equation}
    I_{\vb{G}} \propto |F_{\vb{G}}|^{2} = |\sum_{j} f_{X} e^{-i \vb{r_{j}} \cdot \vb{G}}|^{2}
\end{equation}

where $j$ labels all unique sites in the unit cell, $f_{X}$ is the atomic form factor of the element $X$ ($=$ Ca, Ru, O) at site $j$ and $\vb{G} = H a^{*} + K b^{*} + L c^{*}$. Since the contribution of each atomic species to the squared structure factor comes with a complex phase (dependent on the position of the atom $\vb{r}_{j}$ and the reflection $\vb{G}$ being probed), the total squared structure factor is not linearly dependent on the contributions coming from each atomic species. 

\subsection{Phase diagram of Ti-substituted \cro} \label{Tiphasediagram}

\begin{figure}[!htb]
    \centering
    \includegraphics[scale=0.3]{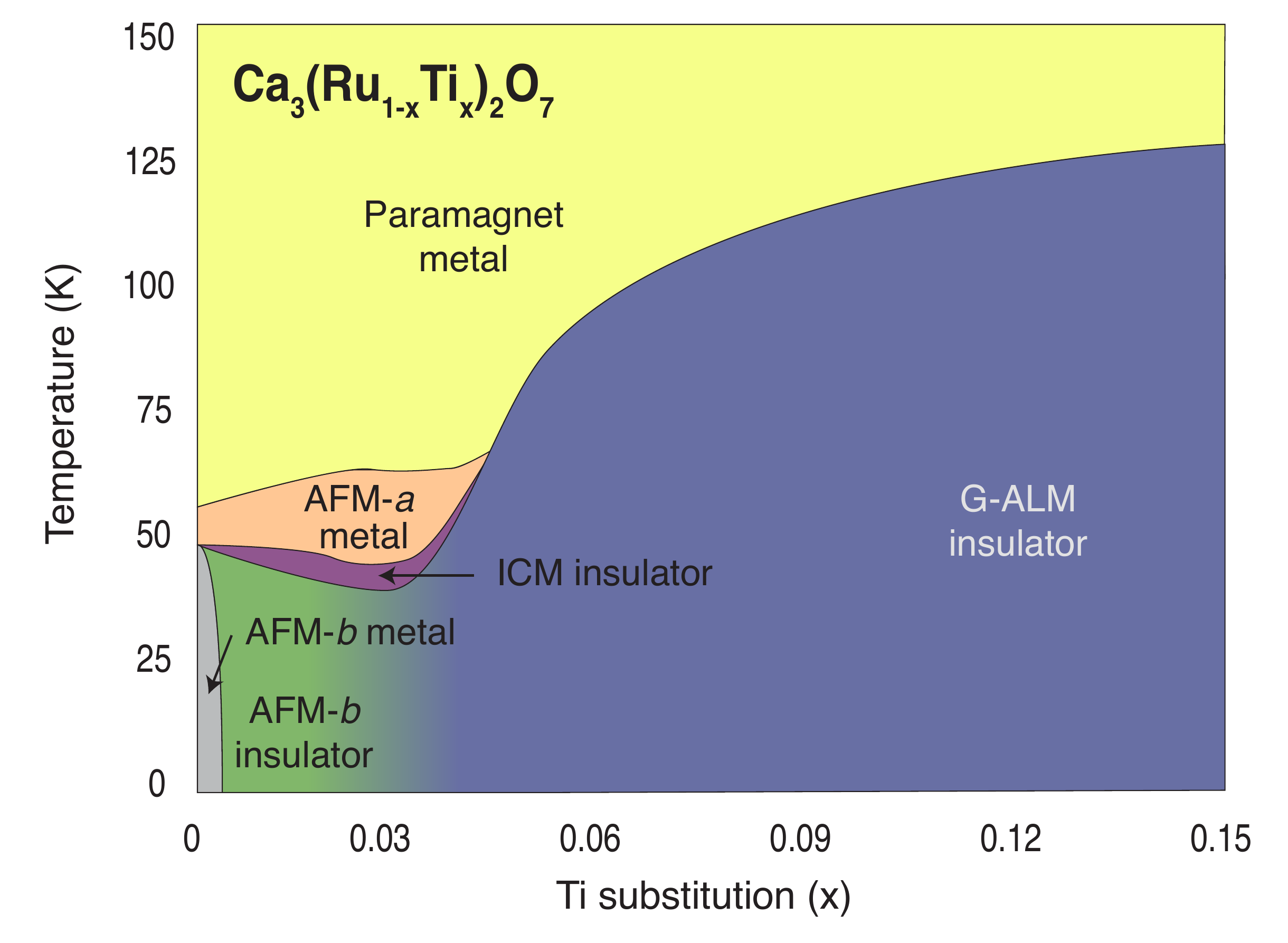}
    \caption{Electronic and magnetic phases of \cro with Ti substitution. Figure adapted from \cite{peng16}. Descriptions of the various phases may be found in the supporting text.}
    \label{phase_diagram}
\end{figure}

With dilute Ti substitution at the Ru site, the electronic ground state changes from the quasi two-dimensional metal for \cro to a Mott insulator with as little as 0.3\% Ti. The magnetic ground state also evolves from the AFM-$b$ magnetic structure in the parent compound to a G-AFM magnetic ordering that emerges at around 3\% Ti concentration. In the G-AFM phase, the spins are coupled antiferromagnetically along the in-plane and out-of-plane directions, and are oriented at an angle $\sim$30\de with respect to the $b$ axis, and $\sim$60\de relative to the $a$ and $c$ axes. The magnetic transition from AFM-$b$ to G-AFM proceeds through an intermediate phase separated regime occurring between 2\% and 4\% Ti, where the majority AFM-$b$ phase at 2\% Ti concentration is progressively replaced with a large volume fraction of the G-AFM phase at 4\% Ti. Within this window, the relative fractions of the two magnetic structures are highly sensitive to the Ti concentration. A summary of the various ground states in \ticrox is shown in Fig. \ref{phase_diagram}. The G-AFM magnetic ordering in the \Bb crystal lattice makes it a polar altermagnet (G-ALM) and spans a large region of the phase diagram. On increasing temperature, the Mott insulating ground state transitions to an AFM-$a$ metallic state for Ti concentrations below 5\%. This is accompanied by the appearance of an incommensurate magnetic structure just below the metal-insulator transition bridging the high temperature AFM-$a$ and low temperature AFM-$b$/G-AFM phases. For higher concentrations (Ti $>$ 5\%), the transition proceeds directly from the low temperature insulating G-ALM phase to a metallic paramagnet at higher temperatures. \\

\subsection{DFT calculations considering the two magnetic structures AFM-$b$ and G-AFM} \label{dft_sm}

\begin{figure}[!htb]
    \centering
    \includegraphics[scale=0.9]{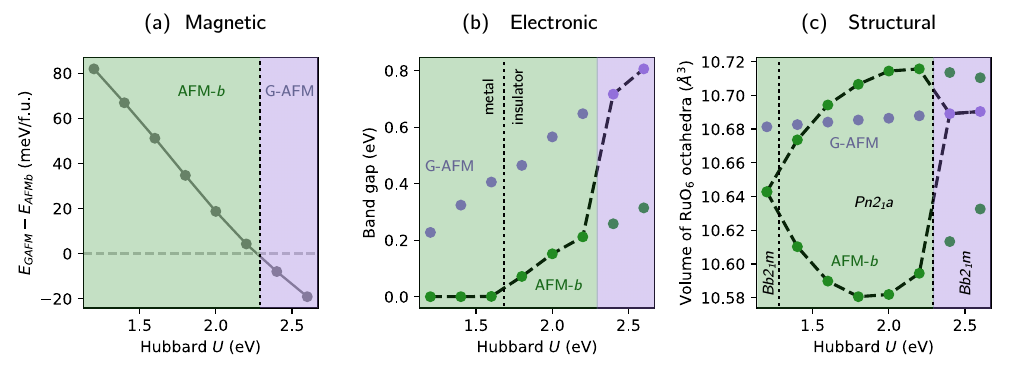}
    \caption{Hubbard $U$ dependence of the (a) magnetic phase stability of AFM-$b$ and G-AFM phases, (b) band gap and (c) volume of \ruo octahedra. Based on the most energetically favorable magnetic structure (shaded background), the band gap and space group undergo a metal-insulator and structural phase transitions.}
    \label{dft_phases}
\end{figure}

DFT+$U$ calculations were performed for two different magnetic structures, AFM-$b$ and G-AFM. The electronic band structures and the relaxed crystal structures were calculated for each of the magnetic phases independently as shown by the distinct traces for the two magnetic orderings in Fig. \ref{dft_phases}. Depending on the most stable magnetic configuration, the band gap and crystal structure undergoes a change denoted by the solid black line.

\subsection{Quantitative relation between X-ray peak maxima and integrated intensity} \label{profile}

Since the Bragg peak detection criterion is defined by the peak intensity $I_{\text{max}}$ exceeding the $3 \sigma$ background level, while the structure factor is related to the integrated intensity $I_{\text{intg}}$ (peak volume), a method connecting these two peak properties is required. \\

Consider a 3 dimensional Gaussian peak profile given by 
\begin{equation}
     I(H, K, L) = I_{0} \exp \left( -\left[ \frac{(H - H_{0})}{2 \sigma_{H}^{2}} + \frac{(K - K_{0})}{2 \sigma_{K}^{2}} + \frac{(L - L_{0})}{2 \sigma_{L}^{2}} \right] \right)
 \end{equation} 
The intensity maximum is $I_{\text{max}} = I_{0}$ occurring at $(H_{0}, K_{0}, L_{0})$ while the integrated intensity (integrated over the whole 3D space) is $I_{\text{intg}} = I_{0} (2 \pi)^{3/2} \sigma_{H} \sigma_{K} \sigma_{L}$. The full-width at half-maximum (FWHM) $\Gamma$ is related to the standard deviation $\sigma$ by $\Gamma = 2 \sqrt{2 \ln(2)} \sigma$. This gives $I_{\text{intg}} = \frac{1}{8} \left( \frac{\pi}{\ln 2} \right)^{3/2} I_{0} \Gamma_{H} \Gamma_{K} \Gamma_{L} = \eta I_{\text{max}} \Gamma_{H} \Gamma_{K} \Gamma_{L}$ with $\eta = 1/8 \cdot (\pi/\ln 2)^{3/2} = 1.206$. \\

Similarly, for a 3D normalizable Lorentzian expressed by
\begin{equation}
    I(H, K, L) = I_{0} \frac{8}{\pi^{2} \Gamma_{H} \Gamma_{K} \Gamma_{L}} \left[ 1 + \left( \frac{H-H_{0}}{2 \Gamma_{H}} \right)^{2} + \left( \frac{K-K_{0}}{2 \Gamma_{K}} \right)^{2}  + \left( \frac{L-L_{0}}{2 \Gamma_{L}} \right)^{2}  \right]^{-2}
\end{equation}
the integrated intensity $I_{\text{intg}} = \eta I_{\text{max}} \Gamma_{H} \Gamma_{K} \Gamma_{L}$ with $\eta = \pi^{2}/8 = 1.233$. 


\subsection{Validity of the assumption of equal reciprocal space peak widths for \Bb and \Pn peaks} \label{fwhm}
 
A reduced real-space correlation length broadens the diffraction peaks and modifies the relationship between the maximum intensity of the peak and the integrated intensity. Thus, clear distinction between the sample properties such as the coherent real-space volume and volume fraction and the experimental observables such as peak widths and intensities is needed. For this, we consider a simple but representative diffraction model of our structure. \\

\begin{figure}[]
  \centering
  \includegraphics[scale=0.23]{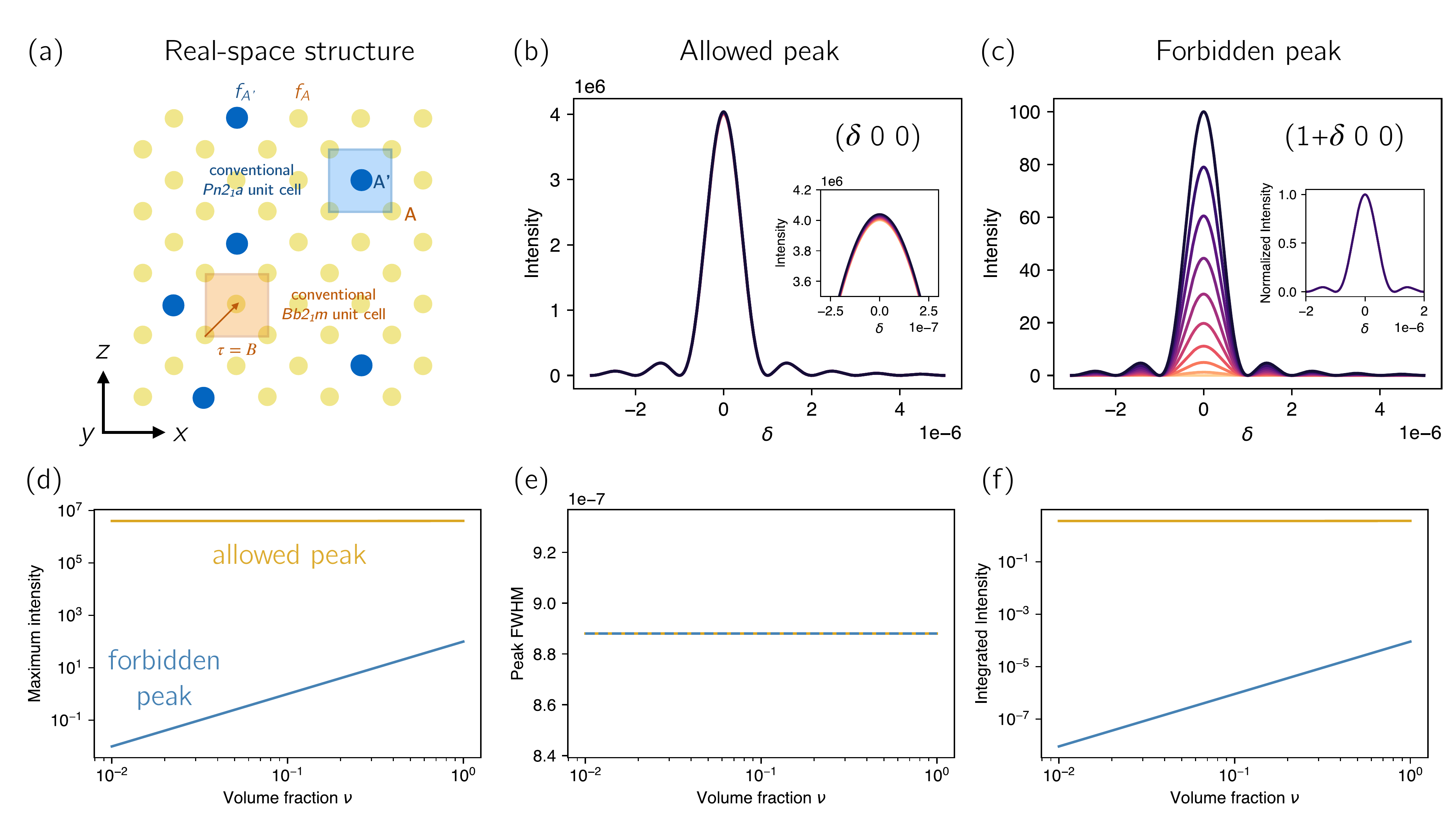}
  \caption{A diffraction model for random ``impurities'' introduced coherently in a lattice. (a) $XZ$ projection of a 3D  cubic lattice. The impurities are denoted by A$^{\prime}$ while the original sites are denoted by A. $f_{A}, f_{A^{\prime}}$ denotes the respective form factors. Linecuts along the $H$ direction near the (b) 0 0 0 allowed peak and (c) 1 0 0 forbidden peak. The inset of (b) shows a zoom-in near the peak while the inset of (c) plots the normalized intensities. The different colors represent different volume fractions $\nu$. The model parameters are $\Delta=0.01$, $N_{x}=10^{6}$, $f_{A}=1$. Volume fraction dependence of the (d) maximum intensity, (e) FWHM and (f) integrated intensities.}
  \label{model}
\end{figure}

We consider a 3D cubic lattice with lattice constant set to unity. Assume we have $N_{c} = N_{x}N_{y}N_{z}$ cells and every cell has two sites: (i) a basis A at $\vb{R} = (x,y,z)$ ($x,y,z$ are integers) with form factor $f_{A}$ and (ii) a site at $\vb{R}+\vb{\tau}$ where $\vb{\tau} = (0.5, 0, 0.5)$ which is occupied by a basis A$^{\prime}$ with probability $\nu$ and by A with probability $1-\nu$ independently for every cell including along the $y$ direction. Thus, the average volume fraction of the A$^{\prime}$ sites is $\nu$. Such a setup, as shown in Fig. \ref{model}a, describes the \Pn ($\equiv$ A$^{\prime}$) phase coherently embedded in the \Bb ($\equiv$ A) matrix. Since the \Pn distortion is small, $f_{A^{\prime}} \approx f_{A}$. Let $\Delta = (f_{A^{\prime}} - f_{A})/f_{A} \ll 1$. The disorder averaged form factor at the second site is $\langle f_{\tau} \rangle = \nu f_{A^{\prime}} + (1-\nu)f_{A} = f_{A}(1+\nu \Delta)$. The diffracted amplitude $A(\vb{q}) = \sum_{\vb{R}} e^{-i 2 \pi \vb{q} \cdot \vb{R}} \left[ f_{A} + \langle f_{\tau}\rangle e^{-i 2 \pi \vb{q} \cdot \vb{\tau}} \right]$ where $\vb{q} = (h, k, l)$ and the diffracted intensity is given by $I(\vb{q}) = \frac{1}{N_{c}} \langle |A(\vb{q})|^{2} \rangle$ where $\langle \cdot \rangle$ denotes the disorder averaging. \\

Similar to a mean-field like approach, we can write the amplitude $A(\vb{q}) = \bar{A}(\vb{q}) + \delta A(\vb{q})$ where $\langle \delta A(\vb{q}) \rangle = 0$ and $\bar{A} = \langle A(\vb{q}) \rangle$. The intensity, which is proportional to $\langle |A(\vb{q})|^{2} \rangle$, can then be written as $\langle |A(\vb{q})|^{2} \rangle = |\bar{A}(\vb{q})|^{2} + \langle | \delta A(\vb{q})|^{2} \rangle$ where the first term arises due to the average order in the lattice (Bragg scattering) while the second term comes from the disorder induced by the A$^{\prime}$ site (diffuse scattering). The diffuse scattering term simplifies to becomes $I_{\text{diffuse}} = N_{c} f_{A}^{2} \Delta^{2} \nu (1-\nu)$ and has no $\vb{q}$ dependence. We now focus on the Bragg part $|\bar{A}(\vb{q})|^{2}$. The Bragg amplitude $\bar{A}$ 
\begin{align*}
 \bar{A} & = \sum_{\text{sites}} e^{-i 2 \pi (hx + ky + lz)} \left[ f_{A} + \langle f_{\tau} \rangle e^{-i 2 \pi \tau \cdot (h k l)} \right] \\
 & = \sum_{x,y,z} e^{-i 2 \pi (hx + ky + lz)} \left[ f_{A} + f_{A}(1 + \nu \Delta) e^{-i \pi (h+l)} \right] \\
 & = f_{A} \left[ 1 + (1 + \nu \Delta) e^{-i \pi (h+l)} \right] \sum_{x=0}^{N_{x}}e^{-i 2\pi h x} \cdot \sum_{y=0}^{N_{y}}e^{-i 2\pi k y} \cdot \sum_{z=0}^{N_{z}}e^{-i 2\pi l z} \\
 & = f_{A} \left[ 1 + (1 + \nu \Delta) e^{-i \pi (h+l)} \right] \left( e^{-i \pi h (N_{x}-1)} \frac{\sin(N_{x}\pi h)}{\sin(\pi h)} \right) \\
 & \ \ \ \ \  \times \left( e^{-i \pi k (N_{y}-1)} \frac{\sin(N_{y}\pi k)}{\sin(\pi k)} \right) \left( e^{-i \pi l (N_{z}-1)} \frac{\sin(N_{z}\pi l)}{\sin(\pi l)} \right)
\end{align*}

\begin{align*}
   |\bar{A}|^{2} & = f_{A}^{2} \left| 1 + (1 + \nu \Delta) e^{-i \pi (h+l)} \right|^{2} \frac{\sin^{2}(N_{x}\pi h)}{\sin^{2}(\pi h)} \frac{\sin^{2}(N_{y}\pi k)}{\sin^{2}(\pi k)} \frac{\sin^{2}(N_{z}\pi l)}{\sin^{2}(\pi l)}  \\
 & = f_{A}^{2} \left[ 1 + (1 + \nu \Delta)^{2} + 2(1 + \nu \Delta) \cos(\pi (h+l)) \right] \frac{\sin^{2}(N_{x}\pi h)}{\sin^{2}(\pi h)} \frac{\sin^{2}(N_{y}\pi k)}{\sin^{2}(\pi k)} \frac{\sin^{2}(N_{z}\pi l)}{\sin^{2}(\pi l)} 
\end{align*}

The Bragg intensity is finally $I_{\text{Bragg}} = |\bar{A}|^{2}/N_{c}$. This satisfies the same selection rules as in the manuscript i.e. $h+l$ = even corresponds to an allowed peak, while $h+l$ = odd corresponds to a forbidden peak. Now consider a reciprocal space line cut $q = (h_{0} + \delta, 0, 0)$ where $h_{0} = 0$ is an allowed peak while $h_{0} = 1$ is a forbidden peak. The peak profile along this direction can be calculated as shown in Fig. \ref{model}b,c for various volume fractions $\nu$. From this, we can extract the $\nu$ dependence of the maximum peak intensity, FWHM and the integrated intensity as plotted in Fig. \ref{model}d-f. The forbidden peak intensity increases monotonically with the volume fraction $\nu$ as expected while the allowed peak intensity shows a weak $\nu$ dependence, decreasing for higher $\nu$ (inset of Fig. \ref{model}b). In contrast, the FWHM for both the allowed and forbidden peaks are equal to each other and insensitive to $\nu$. The only assumptions used in the model are that the A$^{\prime}$ sites are distributed randomly without any correlation and are coherent with the lattice i.e. induces no strain. Based on the above model, assuming $\Gamma_{H} \Gamma_{K} \Gamma_{L}$ to be equal for both the allowed as well as the forbidden peaks is a valid assumption since our DFT calculations constrain the unit-cell parameters as $U$ is varied. We note that in the coherent regime discussed above, the forbidden peak intensity scales as $I \propto \nu^{2}$ whereas if the \Pn is incoherent, the intensity $I \propto \nu$. If the A$^{\prime}$ sites are correlated, then the diffuse part of the intensity $\langle | \delta A(\vb{q})|^{2} \rangle$ picks up a contribution at the reciprocal lattice points while the Bragg part remains unaffected. This diffuse scattering manifests as a broad peak whose width scales inversely with the correlation length. The volume fraction of \Pn on the other hand is an independent sample property that cannot be estimated from our current measurements without determining the amplitude of the \Pn distortion. 


\subsection{Effect of partial volume fraction conversion on distortion bounds} \label{partialvol}

Given our DFT calculations constrain the unit-cell parameters as $U$ is varied, we treat the partial volume fraction scenario in the coherent (i.e. no lattice strain) regime. In such a case, the integrated intensity $I_{\text{intg}} = \nu^{2} |F|^{2}$ where $\nu$ is the volume fraction of the phase is related to the maximum intensity $I_{\text{max}}$ through the correlation volume $V_{HKL} = \eta \Gamma_{H}\Gamma_{K}\Gamma_{L}$ as $I_{\text{intg}} = I_{\text{max}}V_{HKL}$. For a \Pn unique reflection condition where no clear peak is observed, we limit the $I_{\text{max}}$ to $3 \sigma$, which translates to $\nu^{2} |F(\lambda)|^{2} \geq 3 \sigma V_{HKL}$. For small distortions $\lambda$, $|F(\lambda)|^{2} \sim F_{1} \lambda^{2}$. This constraints $\lambda$ to $\lambda_{\text{max}} = \sqrt{3 \sigma V_{HKL}/(F_{1} \nu^{2})}$ i.e. $\lambda_{\text{max}} \sim \lambda_{1}/\nu$ where $\lambda_{1}$ is the distortion amplitude assuming an entire volume fraction conversion from \Bb to \Pn. Hence, reduced volume fraction conversion weakens the bound on $\lambda$ by increasing it. For example, if only 50\% of the initial \Bb transitions into \Pn, then the bound on $\lambda$ doubles. Measured intensity at a \Bb allowed reflection condition does not enable obtaining an estimate of $\nu$ since the integrated intensity comes from both \Bb and \Pn as $|\nu F_{P} + (1 - \nu)F_{B}|^{2} \approx |F_{B}|^{2}$ as $F_{P} \approx F_{B}$. \\

\begin{figure}[!htb]
  \centering
  \includegraphics[scale=0.27]{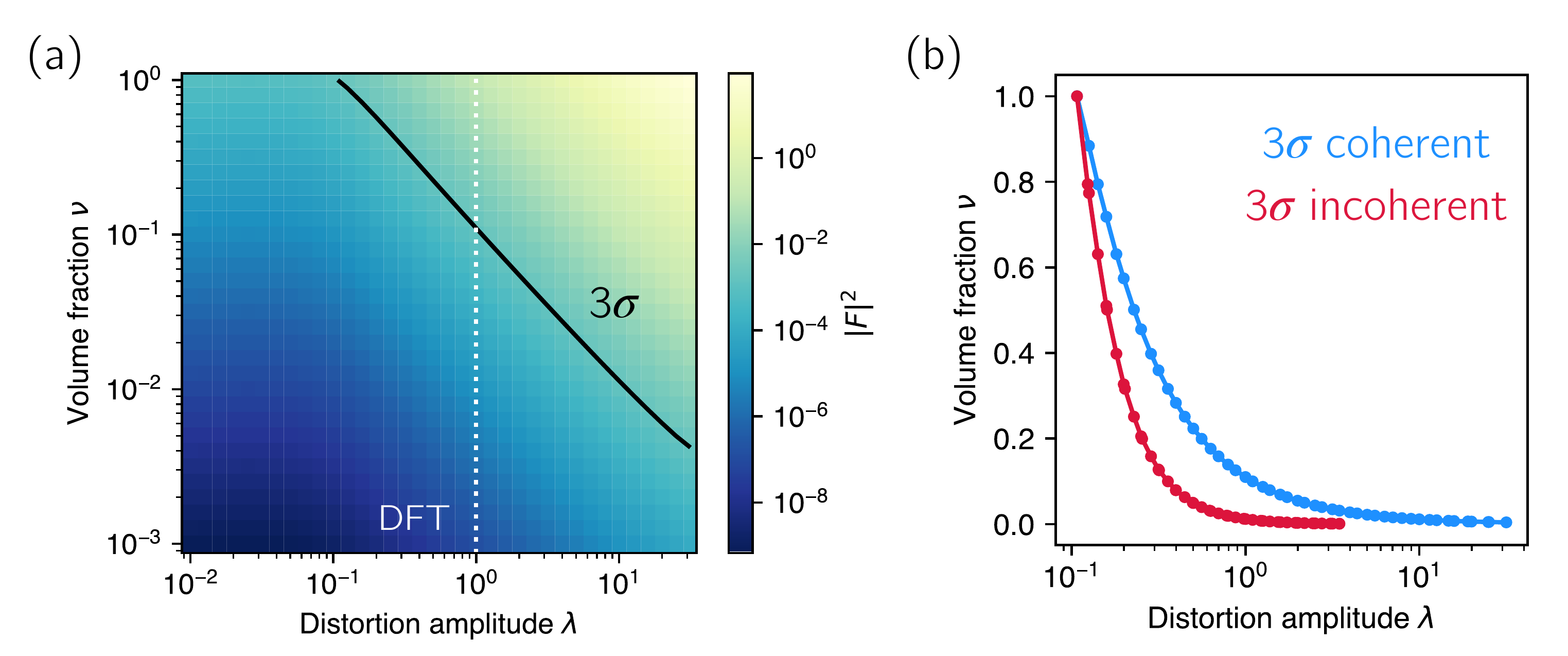}
  \caption{(a) Variation of the structure factor squared $|F|^{2}$ for the 2 $\bar{2}$ 7 reflection parameterized by the distortion amplitude $\lambda$ (defined with respect to the \Pn structure obtained for $U$ = 1.6 eV) and the volume fraction $\nu$ in the coherent regime. The solid black line denotes the $3\sigma$ value obtained from our experiments while the dotted white line refers to the DFT predicted distortion. (b) For constant $|F|^{2}$, $\lambda \sim 1/\nu$ for coherent scattering and $\lambda \sim 1/\sqrt{\nu}$ if incoherent. }
  \label{volfrac}
\end{figure}

Fig. \ref{volfrac}a shows the dependence of the squared structure factor for the 2 $\bar{2}$ 7 reflection as a function of the distortion amplitude and volume fraction. The experimental $3\sigma$ value for this reflection is also plotted as a contour. Assuming a volume fraction of $\nu = 1$ gives $\lambda_{\text{max}} = 0.12$. If we relax this assumption, the $\lambda$ bound increases. Assuming the distortion amplitude predicted by DFT is realized experimentally, then only $\nu \lesssim 0.1$ i.e. less than 10\% of the sample transitions into the \Pn phase. In the incoherent regime, the integrated intensity $I_{\text{intg}} = \nu |F|^{2}$ and $\lambda_{\text{max}} \sim \lambda_{1}/\sqrt{\nu}$. In this case, $\nu \lesssim 0.01$ if the DFT predicted amplitude is physically realized. In both the cases, such a small volume fraction of \Pn is not enough to explain the nonlinear transport signal which is a bulk sensitive probe. \\

The key signature of the \Pn phase from nonlinear transport comes from $\sigma_{bbb}$ which is symmetry-forbidden for \Bb while is allowed for \Pn. If we imagine the partial volume fraction conversion of \Bb to \Pn results in the formation of \Pn clusters in a \Bb matrix, then the situation is similar to ``metallic'' ($\sigma_{bbb} \neq 0$) patches in an ``insulating'' ($\sigma_{bbb} = 0$) matrix. Strictly speaking, this is valid only for the nonlinear conduction channel while the sample remains metallic for both \Bb and \Pn. Assuming a percolation like scenario for the nonlinear conductivity, the percolation threshold in this case is $\nu \approx 0.31$ in 3D \cite{percolation3} and $\nu \approx 0.31-0.47$ for quasi 2D transport \cite{percolation2}. Site percolation on a simple square lattice is assumed. In either case, the minimum volume fraction of \Pn to allow the observation of a bulk nonlinear transport signal is $\nu \approx 0.31$, which increases our $\lambda$ bound by $1/\nu \approx 3.2$ ($1/\sqrt{\nu} \approx 1.8$) times to $\lambda \approx$ 0.22 ($\lambda \approx$ 0.38) in the (in)coherent case. In both cases, the value still remains below the distortion predicted by DFT ($\lambda = 1$) and the probable assumption of a $\sim$31\% \Pn phase does not change our conclusion that PBEsol + $U$ over-stabilizes the \Pn phase. We however note that a macroscopic nonlinear response does not necessarily require geometrical percolation of the \Pn regions and the percolation scenario presented above should be treated as an illustrative case rather than a strict bound on the \Pn volume fraction.  \\


\clearpage


\end{document}